\documentclass[]{jfm} 

\usepackage{graphicx}
\usepackage{newtxtext}
\usepackage{newtxmath}
\usepackage{natbib}
\usepackage{hyperref}
\usepackage{subfigure}
\hypersetup{
    colorlinks = true,
    urlcolor   = blue,
    citecolor  = black,
}

\newcommand{\RomanNumeralCaps}[1]
\linenumbers

\usepackage{subfigure}

\renewcommand{\u}{\boldsymbol{u}}
\renewcommand{\v}{\boldsymbol{v}}
\renewcommand{\a}{\boldsymbol{a}}
\renewcommand{\b}{\boldsymbol{b}}
\renewcommand{\d}{\boldsymbol{d}}
\newcommand{\x}{\boldsymbol{x}}
\newcommand{\X}{\boldsymbol{X}}
\newcommand{\n}{\boldsymbol{n}}
\renewcommand{\t}{\boldsymbol{t}}
\renewcommand{\sf}{\boldsymbol{s_f}}
\newcommand{\sr}{\boldsymbol{s_r}}
\newcommand{\T}{\mathsfbi{T}}
\newcommand{\Ca}{\textit{Ca}}
\newcommand{\Gr}{\textit{Gr}}
\newcommand{\Ma}{\textit{Ma}}
\newcommand{\Oh}{\textit{Oh}}
\newcommand{\Bo}{\textit{Bo}}
\renewcommand{\Pr}{\textit{Pr}}
\renewcommand{\cos}{\text{cos}}
\renewcommand{\sin}{\text{sin}}
\newcommand{\F}{\boldsymbol{F}}

\newcommand{\A}{\boldsymbol{A}}
\newcommand{\B}{\boldsymbol{B}}
\newcommand{\C}{\boldsymbol{C}}
\newcommand{\D}{\boldsymbol{D}}
\newcommand{\E}{\boldsymbol{E}}
\renewcommand{\H}{\boldsymbol{H}}
\newcommand{\K}{\boldsymbol{K}}
\newcommand{\G}{\boldsymbol{G}}
\newcommand{\Phat}{\boldsymbol{\hat{\boldsymbol{P}}}}

\newcommand{\f}{\boldsymbol{f}}
\newcommand{\g}{\boldsymbol{g}}

\title{Generalisation of the Total Linearisation Method to Three-dimensional Free-Surface Flows}

\author{Tyler Benkley\aff{1}
  \corresp{\email{tyler.benkley@epfl.ch}}, 
  S. Deparis \aff{2},
  P. Ricci \aff{3},
 \and A. Mortensen\aff{1}}

\affiliation{\aff{1}Laboratory of Mechanical Metallurgy, EPFL, Switzerland
\aff{2}MATH, EPFL, Switzerland
\aff{3}Swiss Plasma Center, EPFL, Switzerland}

\begin{document}
\maketitle

\begin{abstract}

An iterative Finite Element method predicated on a linearisation of the weak form around a reference configuration is derived for general, three-dimensional, free-surface flows, including systems with moving contact lines. The method is a rigorous generalisation of the Total Linearisation Method that was proposed by \citet{kruyt_total_1988} for two-dimensional flows with contact angles limited to $90^\circ$. In contrast to existing numerical methods for free-surface-flow problems, the present linearisation produces a weak form that is devoid of displacement degrees of freedom in the bulk, thus nearly halving the size of the linear system when compared to standard linearised methods. A novel preconditioner, whose implementation is made possible by the size reduction, is employed to solve the large resulting monolithic Jacobian systems with the Generalised Minimum Residual Method. The proposed method and the preconditioner are shown to be effective on two numerical examples of capillary flows, namely (i) the cylindrical die swell problem, solved both in 3D Cartesian coordinates and under the Ansatz of axisymmetry; and (ii) an enclosed 2D thermo-capillary problem. For the die swell, numerical results are validated by existing experimental results and prior simulations, and confirm both the extension to 3D and theoretical convergence rates. For the thermo-capillary problem, simulations verify earlier calculations. Additional simulations are also carried out for a new range of contact angles made possible by the extension.

\end{abstract}



\section{Introduction} \label{Sec:Introduction}

Free boundary problems (FBPs) \citep{crank_free_1987} are a challenging type of partial differential equation (PDE) where the domain boundary is an unknown that must also be determined as part of the solution. An additional boundary condition (BC) then arises on the free boundary as a result of it being unknown. FBPs have a vast range of applications including, for example, free-surface flows \citep{katopodes_free-surface_2019} in micro-fluidic devices \citep{stone_engineering_2004} or ink-jet printers \citep{wijshoff_drop_2018}, multi-phase fluid flows \citep{brennen_fundamentals_2006} in secondary petroleum recovery \citep{osti_5112525} or aluminium production by electrolysis \citep{lacamera1992magnetohydrodynamics,gerbeau_modeling_2004}, solidification processes \citep{kurz_fundamentals_2023} such as casting of metals \citep{viswanathan_casting_2008} or Laser Powder Bed Fusion in additive manufacturing \citep{chowdhury_laser_2022}, fluid-structure interaction problems \citep{hou_numerical_2012,dowell_modeling_2000}, relevant for instance in the prevention of blood-vessel aneurysms \citep{di_martino_fluidstructure_2001} or in assessing the structural integrity of long-span bridges \citep{elgohary_3d-modeling_2017}, or optimal stopping in mathematical finance \citep{sirjaev_optimal_2006} such as that of American options \citep{rapuch_american_2005}.

Typically, the free boundary's position depends on the value of field variables whose value, in turn, depends on the position of the free boundary, thus, often requiring the need for iteration over the domain. In the last three decades, in the context of fluid dynamics, researchers have focused primarily on time-dependent FBPs. As it stands, the advent of Eulerian diffuse-interface methods has revolutionised this field of study. Techniques such as Volume-of-Fluid \citep{hirt_volume_1981,mohan_volume_2024,popinet_accurate_2009}, the Level-Set method \citep{osher_fronts_1988,sharma_level_2015}, the Phase-Field method \citep{cahn_free_1958,kim_phase-field_2012}, coupled with the Continuum Surface Model \citep{brackbill_continuum_1992}, have endowed computer simulations with the ability to handle complicated interface dynamics riddled with breakup and coalescence, without the difficulties encountered in mesh-movement steps in Arbitrary Lagrangian-Eulerian (ALE) methods \citep{scardovelli_direct_1999}. Smoothed-Particle Hydrodynamics \citep{lucy_numerical_1977,monaghan_smoothed_2005}, another diffuse-interface alternative to ALE, eludes the issues posed by topology change by virtue of its mesh-free nature.

While diffuse-interface methods are undoubtedly better equipped to track complex moving interfaces than sharp-interface methods, the latter continue to be needed when precise tracking of a smooth interface is required. Rather than recovering the location of a free surface \textit{a posteriori}, sharp-interface methods explicitly embed its location into boundary conditions (BCs) like the Young-Laplace law, the no-shear stress condition or the kinematic equation, obviating the need for reconstruction and the errors that can ensue therefrom. In sharp-interface methods, these BCs are all seamlessly integrated as natural BCs into one fully-coupled finite-element weak form. The sharp nature of the interface thus implies that surface-tension effects need not be smeared across the free surface as they are in diffuse methods, where capillary forces are usually approximated via the widely-used Continuum Surface Model \citep{brackbill_continuum_1992}. This can, for example, lead to spurious eddy currents when surface-tension effects are large, as explained by \citet{scardovelli_direct_1999}. 

Sharp-interface methods are also inherently well-suited to the solution of steady problems. Indeed, zeroing partial derivatives with respect to time in the Navier-Stokes equations and reframing the kinematic equation as an impermeability equation makes it possible to deploy the entire tool-kit of numerical techniques for nonlinear system solution. This opens the door to using, for instance, Newton or quasi-Newton methods and thus achieving quadratic convergence if ensuing linear systems can be inverted. By contrast, in order to incorporate the smoothed free-interface variable, diffuse-interface methods typically rely on special interface-advection equations, whose solution is usually decoupled from the Navier-Stokes equations by virtue of their transient nature. As a result, for the solution of steady problems, the equations in transient form are generally evolved from an initial state with the aim of converging, in time, to a steady solution. Convergence rates are therefore at the mercy of problem dynamics, which may entail a restrictive Courant-Friedrichs-Lewy condition and, in turn, extra computations to reach a solution.

We devote here our attention to sharp-interface methods and address problems in need of high-resolution tracking of an unknown smooth interface free of topological change. Previously tractable only in simple configurations, sharp-interface methods for FBPs saw large progress in the latter part of the $20^\text{th}$ century with the advent of increased computing power. In the 1970s, first numerical solutions of FBPs on dam flow, stellar evolution, viscous laminar jet flow were reported \citep{cryer_technical_nodate,reddy_finite_1978}. 
The first technique to be used, known as the Trial Method, involves deliberately relaxing one BC on the free boundary. Then, the original FBP is treated as a fixed-domain PDE in the absence of the relaxed BC, and solved using standard numerical methods designed for PDEs over a fixed domain. Having thus solved the relaxed problem over a guessed domain, using the relaxed BC, one formulates a new guess of the domain so as to execute an ALE scheme. As in a Picard process, one thus iterates on the domain with the aim of converging to the free-boundary solution. The fixed-point nature of the algorithm implies that its convergence, when observed, is usually linear in the number of unknowns \citep{cuvelier_numerical_1990}. The main advantage of this method is that it is very simple to implement with an existing PDE solution method. On the other hand, it is not always obvious which BC must be relaxed, nor how to formulate a new and converging free-boundary guess given a relaxed solution. Numerical experiments by \citet{silliman_separating_1980} show that convergence is affected, in a given problem, by the values of its dimensionless groups. A poor choice of the relaxed BC can furthermore slow down or even impede convergence altogether (see \citet{cuvelier_numerical_1990} for an illustration where a Trial Method is shown to be divergent on a simple one-dimensional FBP). This pathology has also been reported by \citet{slikkerveer_implicit_1996} in free-surface flows dominated by surface tension. Although a remedy was discovered to make this specific case numerically stable \citep{slikkerveer_implicit_1996}, there are no established guidelines for more general problems. As a result, it is not always possible to predict the outcome of a Trial Method.

To overcome the limitations of Trial Methods, more sophisticated iterative schemes predicated on Newton iteration over a nonlinear, fully-coupled finite element weak form were introduced in the 1980s \citep{saito_study_1981,ruschak_method_1980,ettouney_finite-element_1983,kruyt_total_1988}. These methods rely on the hypothesis that, given a reference configuration $\Omega_0\subset \mathbb{R}^3$ (in the parlance of solid mechanics) of the appropriate topology, a sufficiently smooth displacement field $\d:\Omega_0\rightarrow \mathbb{R}^3$ exists such that the reference configuration can be mapped by a translation $\d$ to the unknown fluid domain $\Omega \subset \mathbb{R}^3$, also known as the deformed configuration
\begin{equation} \label{Eq:DisplacementHypothesis}
    \Omega = \{\x\in\mathbb{R}^3:\x(\X) = \X + \d(\X),\,\forall \X\in \Omega_0 \}.
\end{equation}
With this assumption, the weak form of the Navier-Stokes equations, which is typically integrated over $\Omega$ and $\partial \Omega$, can be parametrised by the reference configuration. Integrals, initially not tractable for such a problem within the traditional FEM framework, become computable when parametrised by a prescribed domain $\Omega_0$. The parametrisation of integration measures, differential operators and tangent/normal vectors then gives rise to nonlinear terms dependent on $\d$. As such, the FEM weak form becomes a coupled system of equations encompassing both the field variables of the Navier-Stokes equations and $\d$. Under the additional hypothesis that $\d$ is small enough, Newton-Raphson's algorithm converges quadratically in the number of iterations to a solution of the FBP. Two possibilities arise in between Newton iterations: one may either deform the reference configuration as prescribed by the FEM solution and update the mesh such that $\Omega_0$ converge to $\Omega$ (the ALE approach); or one may retain the same reference configuration and solve the entire problem in the parametrised domain (the Eulerian approach).

One complication of these parametrised formulations is the presence of $\d$ in volume integrals. Indeed, this calls for a description of displacement within $\Omega$ even though displacement is not governed by a physical law therein. The involvement of $\d$ in integrals over the bulk of $\Omega_0$ is furthermore known to be important, because ignoring the dependency on $\d$ can compromise second-order convergence rates \citep{saito_study_1981}. To deal with this, one approach is to supply an analytical relation for displacement within the bulk, thus defining interior degrees of freedom (DOFs) as a function of free surface DOFs. This can easily be achieved when the problem permits a description of the free surface by a height function, e.g. $\d = d_y(x) \boldsymbol{e_y}$ \citep{saito_study_1981,cuvelier_thermocapillary_1986}, but calls for special care when dealing with rotating menisci, as explained by \citet{saito_study_1981}. Another more general approach to obtain a description of $\d$ is to solve an artificial PDE for displacement in $\Omega_0$ \citep{christodoulou_discretization_1992}. This has become routine in fluid-structure interaction simulations \citep{fernandez_newton_2005,hou_numerical_2012,deparis_numerical_2004,crosetto_parallel_2011,tenderini_reduced_2024} and endows the user with a simple method to cope with the dependency of integrals on values of $\d$ in the bulk of $\Omega_0$. Then, upon solution of each iteration, displacement values computed at vertices can be used to move mesh vertices in ALE schemes. Nodal values of field variables are then displaced together with mesh vertices, thus enabling simultaneous iteration over the free-boundary and field variables. This mesh-movement technique, known as Elliptic Mesh Generation \citep{thompson_boundary-fitted_1982}, sidesteps the need for re-meshing in-between iterations. 

Pseudo-solid models whereby $\d$ is subject to Cauchy's equations of motion have been successful in several FBPs including 2D solidification of a pure metal \citep{sackinger_newtonraphson_1996}, 2D transient free-surface flows \citep{sackinger_newtonraphson_1996}, 3D steady laminar jets and coating flows \citep{cairncross_finite_2000,baer_finite_2000}, and have served as the foundation of the open-source software GOMA \citep{schunk_goma_2013}. Other Elliptic Mesh Generation variants have ensued from this approach in recent years \citep{fraggedakis_discretization_2017,anthony_sharp_2023}. 

Despite its success, the elliptic extension of displacement from the free surface to the bulk is not without drawbacks, as it significantly increases the monolithic linear system size. In 3D incompressible fluid flow simulations, in which the number of unknowns is already very large, the addition of $\d$, a 3D vector field, nearly doubles the problem size. Even more problematic is the poor conditioning of the monolithic system \citep{cairncross_finite_2000}, which can cripple iterative solvers like GMRES altogether in situations where direct solution is no longer viable. It is said that the \textit{distinguishing conditions}, i.e. the additional BCs that must be introduced for $\d$ \citep{schunk_iterative_2002}, not native to the free surface flow but merely to the artificial PDE, contribute negatively to the condition number \citep{cairncross_finite_2000}. A plethora of approaches, such as stabilised equal-order velocity-pressure fields, heavy-duty algebraic preconditioners like incomplete LU factorisation with fill, or reordering strategies \citep{schunk_iterative_2002}, have been attempted to address this concern. Of all these approaches, even the most successful one, namely the stabilised FEM approach, sometimes fails to sufficiently improve the condition number of the monolithic system for iterative solution to be possible \citep{rao_3d_nodate}. Even when it does succeed, the convergence rates of the FEM approximation are inferior to those of the Taylor-Hood elements \citep{taylor_numerical_1973}, which comply with the Ladyzhenskaya–Babuska–Brezzi (LBB) condition \citep{brezzi_existence_1974}. Segregated algorithms or artificial transient solutions are an alternative, but they compromise the quadratic convergence rate of Newton's algorithm when they do converge, and increase the amount of computational work required to compute a solution. 

We aim in this manuscript to address the aforementioned issues by building on a Newton approach entitled the Total Linearisation Method (TLM) that was initially proposed by \citet{kruyt_total_1988}. It is an ALE technique that distinguishes itself from others by a subtle, additional linearisation step in which functions $f:\Omega\rightarrow \mathbb{R}$ are reformulated as:
\begin{equation} \label{Eq:TLMAdditionalStep}
    f(\x(\X)) = f(\X + \d(\X)) = f(\X) + \d\cdot \bnabla f(\X) + \mathcal{O}(\|\d\|^2).
\end{equation}
When employed in volume integrals, by virtue of the divergence theorem, this simple step rids the FEM weak form of all bulk terms involving $\d$.

As such, when employing this additional step in the FEM weak form, $\d$ appears only on the free surface in the final linearised equations which remain exact everywhere up to first order in displacement. It is therefore not necessary to solve for fictitious DOFs of $\d$ in the bulk of $\Omega_0$ when doing so for field variables and free-surface DOFs of $\d$. As such, if Elliptic Mesh Generation is implemented to prescribe motion of mesh vertices, $\d$ is extended from the interface to the bulk of $\Omega_0$ via the solution of an artificial PDE in a step that is decoupled from the FBP. This drastically reduces the size of the FBP, and avoids the adverse effects on the conditioning of the monolithic system that are posed by the artificial equation and its distinguishing conditions, while simultaneously sidestepping the need for re-meshing. 

Although they did not use Elliptic Mesh Generation, \citet{kruyt_total_1988} were successful in applying this ALE technique to the 2D die-swell problem. In this specific case, they constrained the displacement field to be directed normal to the free surface in all locations. This reduces the displacement variable $\d$ to a scalar field that lives only the interface, thus minimising the number of DOFs of this variable in FBPs. The computation of the Jacobian of the weak form encoded in the linearisation has also led to success of this method in linear stability analysis \citep{abubakar_linear_2022}.

While the TLM addresses the issue of bulk-extended displacement, its original form, as posed by \citet{kruyt_total_1988}, is not sufficiently general for the description of free-surface flows with moving contact lines. Indeed, the restriction of the displacement field to the normal direction prohibits motion of a contact line along the wetted, rigid surface unless the contact angle happens to be $90^\circ$. The calculus of \citet{kruyt_total_1988} is thus predicated on this limiting assumption and was furthermore only carried out in two dimensions. Another consequence of Relation (\ref{Eq:TLMAdditionalStep}), not yet noted, is the presence of an additional order of differentiation on $f$. While this is not a problem in volume integrals by virtue of the divergence theorem, on 3D surfaces, more complex vector calculus is required in order to reduce higher-order derivatives, therefore making generalisation of the calculus of \citet{kruyt_total_1988} from 2D to 3D flows nontrivial. 

We propose here an extension of the TLM approach of \citet{kruyt_total_1988} that addresses these limitations. By not restricting the displacement field to the normal direction, and with the use of 3D vector calculus, we derive mathematically a Newton-Raphson, ALE algorithm capable of solving 3D, time-dependent, capillary free-surface flows with moving contact lines. Our formalism yields a linearised weak form devoid of displacement DOFs in the bulk, which significantly reduces the computational size of the fully-coupled system with respect to existing Newton methods. Moreover, when the displacement field is expressed in coordinates of a basis of vectors normal and tangent to the free surface, only the normal component of displacement remains in surface integrals. This reduces the computational size of our problem even further, and bypasses the need for the \textit{distinguishing conditions} that are said to be detrimental to conditioning \citep{cairncross_finite_2000}. In a second, decoupled step, Elliptic Mesh Generation is implemented in order to compute displacement DOFs via an artificial PDE. In addition, in order to cope with the large system size of the FBP, a physics-based, block preconditioner that enables the use of LBB-compliant elements is derived for the iterative solution of the monolithic linear system in steady flows. 

Our extension of TLM, together with our preconditioner, is implemented in Fenics Legacy and Fenicsx \citep{logg_dolfin_2010} with the mixed-dimensional PDE package \citep{daversin-catty_abstractions_2021} and PETSc \citep{petsc-user-ref,petsc-web-page} in two steady, flow problems, namely (i) the 3D die-swell problem, modeled both in 3D Cartesian coordinates and axisymmetric cylindrical coordinates for comparison; (ii) a 2D thermo-capillary moving contact-line problem, originally solved by \citet{cuvelier_thermocapillary_1986}, where two sides of a container are subject to different temperatures with both Grashof and Marangoni convection. In both examples, a range of simulations are observed to be consistent with results from the literature.

This contribution is structured as follows: in Section \ref{Sec:Problem}, a general problem statement of free-surface flows is given in conjunction with its weak form; then, in Section \ref{Sec:Linearisation}, the FEM system is linearised, upon parametrisation by a reference configuration, by virtue of 3D surface calculus. Linearisation steps nonessential to the understanding of the reader are reserved for Appendices \ref{appA} and \ref{appB}. Together with the linearised weak form, an algorithm predicated on Elliptic Mesh Generation is given to solve FBPs. In Section \ref{Sec:Preconditioner}, the physics-based preconditioner is introduced together with its inversion algorithm. In Section \ref{Sec:Simulations}, the method is employed to solve the aforementioned die-swell and thermo-capillary problems. 

\section{Problem Definition} \label{Sec:Problem}

\subsection{Physical Problem}

We begin by giving a general definition of the free-surface flows that are treated here. Let us define $\Omega(\tau)\in\mathbb{R}^3$ for $\tau\in(0,T)$ as a fluid volume. Its boundary can be decomposed into two disjoint sets $\partial \Omega(\tau) = \Sigma_f(\tau) \cup \Sigma_r(\tau)$ where $\Sigma_f(\tau)$ denotes the free surface and $\Sigma_r(\tau)$ denotes rigid surfaces. The fluid is assumed to be Newtonian and undergoes incompressible flow. Free surfaces are governed by surface tension in a quiescent atmosphere. Hence, the time-dependent FBP is governed by the following the problem.

\begin{figure}
    \centering
    \includegraphics{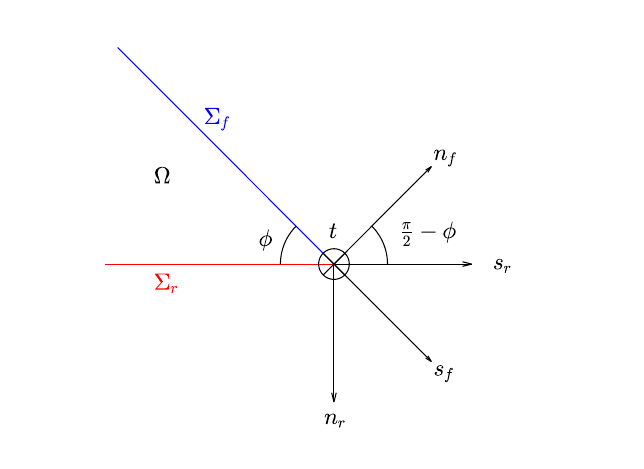}
    \caption{Orthogonal bases at the contact line; $\t$ is parallel to the contact line; $\phi$ is the contact angle}
    \label{fig:ContactLine}
\end{figure}

\begin{equation} \label{Eq:StrongForm}
    \begin{split}
    &\text{Find $(\u,p,\Sigma_f,p_a)\in V\times Q\times S \times \mathbb{R} $ such that} \\
    &\begin{cases}
        \rho ( \frac{\partial\u}{\partial \tau} + (\u\cdot \bnabla)\u) = -\bnabla p + \mu \nabla^2\u, \text{ in $\Omega(\tau)$ for $\tau\in(0,T)$} \\
        \bnabla \cdot \u = 0, \text{ in $\Omega(\tau)$ for $\tau\in(0,T)$} \\
        \frac{d\x}{d\tau} \cdot \n = \u\cdot\n, \text{ on $\Sigma_f(\tau)$ for $\tau\in(0,T)$} \\
        \n\times (\T \n) = 0, \text{ on $\Sigma_f(\tau)$ for $\tau\in(0,T)$} \\
        \n\cdot (\T \n) = -\gamma(\bnabla_S \cdot \n) - p_a, \text{ on $\Sigma_f(\tau)$ for $\tau\in(0,T)$} \\
        \sf\cdot \sr = \cos(\phi), \text{ on $\partial \Sigma_f(\tau) \cap \partial \Sigma_r(\tau)$ for $\tau\in(0,T)$}\\
        \text{ + ICs, BCs on $\u$ on $\Sigma_r$ and global constraint}
    \end{cases}
    \end{split}
\end{equation}
where $\u$, $p$, $p_a$ and $\T$ are velocity, pressure, atmospheric pressure and the Newtonian stress tensor; $V$ and $Q$ are suitable function spaces for velocity and pressure; $S$ is a suitable space of surfaces for $\Sigma_f$; $\n$, $\sf$ and $\sr$ are respectively (i) the outward-pointing unit normal vector to $\Sigma_f$, (ii) the unit vector tangent to $\Sigma_f$ and normal to $\partial \Sigma_f$ and (iii) the unit vector tangent to $\Sigma_r$ and normal to $\partial \Sigma_r$ as sketched in Figure \ref{fig:ContactLine}; $\rho$, $\mu$, $\gamma$ and $\phi$ are fluid density, dynamic viscosity, surface tension and the contact angle (see Figure \ref{fig:ContactLine}), respectively; and $\bnabla_S = \bnabla - \n\frac{\partial }{\partial n}$ is the surface gradient. In certain cases, an additional global constraint may accompany the system in order to fix atmospheric pressure or the fluid volume. We define the following set of variables in order to non-dimensionalise Problem (\ref{Eq:StrongForm}).

\begin{align}
x^* := \frac{x}{L} && \tau^* := \frac{U\tau}{L} && \u^* := \frac{\u}{U} && p^* := \frac{Lp}{\mu U} && p_a^* := \frac{L p_a}{\gamma} && \T^* := \frac{L \T}{\mu U}
\end{align}
All domains, function spaces, differential operators, etc. are redefined accordingly. In what follows, the $*$ superscript is omitted so as to lighten the notation. All terms are thus considered to be dimensionless in the rest of the manuscript. Applying the standard Galerkin method produces the following weak form.
\begin{equation} \label{Eq:WeakForm}
    \begin{split}
    &\text{Find $(\u,p,\Sigma_f, p_a)\in V\times Q\times S \times \mathbb{R} $ such that} \\
    &\begin{cases}
        \begin{split}
        \int_{\Omega(\tau)} \{\Rey \, &\v\cdot \frac{\partial\u}{\partial \tau} + \Rey\, \v\cdot (\u\cdot \bnabla)\u  + \T : \bnabla \v \} dV \\
        = &\int_{\Sigma_r(\tau)} \v\cdot (\T \n) dS - \frac{1}{\Ca}\int_{\Sigma_f(\tau)} (\bnabla_S \cdot \v + p_a(\v\cdot \n))dS\\
        &+ \frac{1}{\Ca}\int_{\partial \Sigma_f(\tau)} \{\text{cos}(\phi)(\v\cdot \sr)+\text{sin}(\phi)(\v\cdot \n_r)\} dl , \text{ $\forall v\in V$, for $\tau\in(0,T)$}
        \end{split}\\
        \int_{\Omega(\tau)} (\bnabla \cdot \u) q dV = 0, \text{ $\forall q \in Q$, for $\tau\in(0,T)$} \\
        \int_{\Sigma_f(\tau)} \chi (\frac{d\x}{dt} - \u)\cdot \n dS = 0, \text{ $\forall \chi \in M$, for $\tau\in(0,T)$} \\
        \text{ + ICs, BCs on $\u$ on $\Sigma_r(\tau)$ and global constraint}
    \end{cases}
    \end{split}
\end{equation}
where $M$ is a space of scalar functions defined on $\Sigma_f$, $\Rey = \frac{\rho U L}{\mu}$ is the Reynolds number, and $\Ca = \frac{\mu U}{\gamma}$ is the Capillary number. The weak form of the linear momentum equation was obtained by applying integration by parts followed by the divergence theorem for 3D surfaces \citep{weatherburn_differential_1927}, which is given below.

\begin{equation} \label{Eq:DivergenceTheorem}
    \int_{\Sigma_f} (\bnabla_S\cdot \n)(\v\cdot\n)\,dS = \int_{\Sigma_f} (\bnabla_S\cdot \v)\,dS - \int_{\partial \Sigma_f} (\v\cdot\sf) \,dS
\end{equation}

\subsection{Parametrisation by Reference Configuration}

As mentioned in Section \ref{Sec:Introduction}, the requisite integration with respect to $\Omega(\tau)$ for Galerkin approximation is intractable in Equation System (\ref{Eq:WeakForm}). As a result, given a reference domain $\Omega_0\subset\mathbb{R}^3$ and time instant $\tau \in (0,T)$, and under the hypothesis that a displacement field $\d$ satisfying
\begin{equation} \label{Eq:DisplacementHypothesisTime}
    \Omega(\tau) = \{ \x\in\mathbb{R}^3 : \x(\X,\tau) = \X + \d(\X,\tau),\,\forall \X\in\Omega_0  \}\text{ in a neighborhood of $\tau $}
\end{equation}
exists, we parametrise integrals by the reference domain $\Omega_0$. Note that this equation is a generalisation of (\ref{Eq:DisplacementHypothesis}) to time-dependent problems. The following identities hold for functions, differential operators, the normal vector and integration measures:
\begin{equation} \label{Eq:Parametrization}
    \begin{cases}
        f(\x,\tau) = f(\X + \d(\X,\tau),\tau) \text{, in $\Omega_0$}\\
        \bnabla_{\x} = \F^{-T}\bnabla_{\X} \text{, in $\Omega_0$} \\
        \n = \frac{\F^{-T}\n_0}{\| \F^{-T}\n_0 \|} \text{, over $\partial \Omega_0$}  \\
        \t = \frac{\F \t_0}{\|\F\t_0\|} \text{, on $\partial \Sigma_{f,0}\cap \partial \Sigma_{r,0}$} \\
        \int_{\Omega(\tau)}\,.\,dV = \int_{\Omega_0} \,.\,\text{det}(\F)\,dV\\
        \int_{ \Sigma_f(\tau)}\,.\,dS = \int_{ \Sigma_{f,0}}\,.\,\|\F^{-T}\n_0\|\text{det}(\F)  \,dS \\
        \int_{\partial \Sigma_f(\tau)} \,.\, dl = \int_{\partial \Sigma_{f,0}} \,.\, \|\F\t_0 \| dl\\
    \end{cases}
\end{equation}
where $\F(\X,\tau) = \boldsymbol{I} + \bnabla_{\X}\d(\X,\tau)$ is the deformation gradient and $\t$ is the unique unit tangent vector oriented positively along the curve $\partial\Sigma_f \cap \partial \Sigma_r$, as shown in Figure \ref{fig:ContactLine}. Note that $\bnabla_{\x}$ designates the gradient in the Lagrangian coordinate system while $\bnabla_{\X}$ designates the gradient in the reference-configuration coordinate system. Moreover, the subscript 0 denotes that a variable pertains to the reference configuration. Parametrisation by $\Omega_{0}$ and, thus, replacement of terms by the above relations, leads to an FEM weak form involving $\d$ in bulk integrals. As explained in Section \ref{Sec:Introduction}, numerous methods such as those of \citet{sackinger_newtonraphson_1996,cairncross_finite_2000,baer_finite_2000}, couple the parametrised weak form with a necessarily artificial equation for $\d$ such that the latter's DOFs in the volume are uniquely defined. This enables one to overcome the dependency of the parametrised Navier-Stokes equations on $\d$ by closing the mathematical problem; however, the approach has the disadvantage of requiring one to solve simultaneously the physical laws of the system at hand and the artificial equation. This non-physical problem augmentation nearly doubles the FBP's size with respect to that of the Navier-Stokes equations for a fixed domain and can have dire consequences on the conditioning of ensuing linear systems \citep{cairncross_finite_2000,schunk_iterative_2002}. In the next Section, we extend the aforementioned TLM, that is, the method of \citet{kruyt_total_1988} that was developed for the solution of the 2D die-swell problem, to a method that is capable of solving general free-surface problems and circumvents the issues posed by the above non-physical system augmentation. 

\section{Linearisation of Free-Surface-Flow Problems} \label{Sec:Linearisation}

In this Section, free-surface-flow Problem (\ref{Eq:WeakForm}) is linearised, upon parametrisation by the reference configuration $\Omega_0$, around a solution guess $(\u_0,p_0,\d_0,p_{a,0})$. In ALE algorithms, at each iteration, the reference configuration constitutes the current guess of the domain. Upon parametrisation of integrals by $\Omega_0$, the displacement field that fulfills this requirement is therefore the identically-zero vector field. 

In what follows the '$\approx$' sign denotes that terms of second order or higher have been neglected. Under the hypothesis that $\d$ is small enough, Relations (\ref{Eq:Parametrization}) can be approximately reformulated as:
\begin{equation} \label{Eq:LinearizationRelations}
    \begin{cases}
        f(\x,\tau) \approx f(\X,\tau) + \d(\X,\tau)\cdot \bnabla_{\X}f(\X,\tau) \text{, in $\Omega_{0}$}\\
        \bnabla_{\x} \approx \{\boldsymbol{I} - (\bnabla_{\X}\d)^T\}\bnabla_{\X} \text{, in $\Omega_{0}$} \\
        \n \approx \n_0 - \n_0 \cdot (\bnabla_S \d) \text{, on $\partial \Omega_{0}$} \\
        \t \approx \t_0 + (\frac{d\d}{dl}\cdot \n_0)\n_0 \text{, on $\partial \Sigma_{f,0}\cap \partial \Sigma_{r,0}$} \\
        \int_{\Omega(\tau)}\,.\,dV \approx \int_{\Omega_0} \,.\,(1+\bnabla_{\X}\cdot \d)\,dV\\
        \int_{ \Sigma_f(\tau)}\,.\,dS \approx \int_{ \Sigma_{f,0}}\,.\,(1+\bnabla_S\cdot \d)  \,dS \\
        \int_{\partial \Sigma_f(\tau)} \,.\, dl \approx \int_{\partial \Sigma_{f,0}} \,.\, (1+\frac{d\d}{dl}\cdot \t_0) dl\\
    \end{cases}
\end{equation}

In what follows, so as to lighten the notation, the subscript 0 will be omitted from normal and tangent vectors and from differential operators, as these will be associated to the relevant integration measure. The subscript 0 will however continue to be used in order to designate the reference configuration $\Omega_0$ and current guess, $(\u_0,p_0,0,p_{a,0})$, around which the weak form is linearised.

As explained in Section \ref{Sec:Introduction}, the main advantage of TLM is the ability to remove $\d$ from volume integrals as demonstrated below:
\begin{equation} \label{Eq:VolumeIntegrals}
\begin{split}
    \int_{\Omega(\tau)} f dV &\approx \int_{\Omega_0} \{ f + \d\cdot \bnabla f + f(\bnabla\cdot\d)  \}dV = \int_{\Omega_0} \{ f + \bnabla\cdot(f\d)  \}dV \\
    &= \int_{\Omega_0} f\,dV + \int_{\partial \Omega_{0}} f(\d\cdot \n)\,dS
\end{split}
\end{equation}
This relation will be used throughout the linearisation.

\subsection{Momentum Equation}

Starting with the time derivative of velocity, terms emanating from the momentum equation are linearised one by one.
\begin{equation}
    \begin{split}
        \int_{\Omega(\tau)} \Rey\, \v\cdot \frac{\partial\u}{\partial \tau} dV \approx &\int_{\Omega_0} \Rey\, \v\cdot \frac{\partial\u}{\partial \tau} dV + \int_{ \Sigma_{f,0} } \Rey\,  (\d\cdot \n)\, (\v\cdot \frac{\partial\u}{\partial \tau}) dS \\
    \approx &\int_{\Omega_0} \Rey\, \v\cdot \frac{\partial\u}{\partial \tau} dV + \int_{ \Sigma_{f,0} } \Rey\, (\d\cdot \n)(\v\cdot \frac{\partial\u_0}{\partial \tau}) dS 
    \end{split}
\end{equation}
The advection part of the momentum Equation (\ref{Eq:WeakForm}) becomes
\begin{equation}
\begin{split}
    \int_{\Omega(\tau)} \Rey\, \v\cdot (\u\cdot \bnabla)\u dV \approx &\int_{\Omega_0} \Rey\, \v\cdot (\u\cdot \bnabla)\u dV + \int_{ \Sigma_{f,0} } \Rey\,  (\d\cdot \n) \v\cdot (\u\cdot \bnabla)\u dS \\
    \approx &\int_{\Omega_0} \Rey\, \v\cdot \{ (\u_0\cdot \bnabla)\u + (\u\cdot \bnabla)\u_0 - (\u_0\cdot \bnabla)\u_0 \} dV \\
    &+ \int_{ \Sigma_{f,0} } \Rey\, (\d\cdot \n) \v\cdot (\u_0\cdot \bnabla)\u_0 dS
\end{split}
\end{equation}
In the last step, a Taylor series in velocity is expanded up to first order. In the next step, the stress term, also found on the left-hand-side of the momentum equation, is linearised to become
\begin{equation} \label{Eq:StressTensorLinearization}
\begin{split}
    \int_{\Omega(\tau)} \T : \bnabla \v dV &\approx \int_{\Omega_0} \T : \bnabla \v dV + \int_{\Sigma_{f,0}} (\d\cdot \n) \T : \bnabla \v dS
\end{split}
\end{equation}
The tensor inner product on the free surface can be reformulated as follows.
\begin{equation}
    \begin{split}
        \T:\bnabla \v &= (-p\mathsfbi{I} + \bnabla \u + (\bnabla \u)^\text{T}):\bnabla \v \\
        &= -p (\bnabla_S \cdot \v) - p (\frac{\partial \v}{\partial n}\cdot \n) + (\bnabla \u + (\bnabla \u)^\text{T}):\bnabla \v
    \end{split}
\end{equation}
By virtue of the no-shear BC and following a gradient basis change, the tensor product between the strain rate tensor and the test function gradient is simplified.
\begin{equation}
    \begin{split}
        (\bnabla \u + (\bnabla \u)^\text{T}):\bnabla \v &= (\bnabla \u +\bnabla \u)^\text{T}):(\bnabla_S \v + \n \frac{\partial \v}{\partial n}) \\
        &= (\bnabla\u + (\bnabla\u)^\text{T}):\bnabla_S \v + 2 (\n\cdot\frac{\partial \u}{\partial n})(\n\cdot\frac{\partial \v}{\partial n}) + \text{O}(\|\d\|) \\
    \end{split}
\end{equation}
Therefore, the surface integral in Equation (\ref{Eq:StressTensorLinearization}) becomes
\begin{equation} \label{Eq:LinearizedStress}
\begin{split}
    \int_{\Sigma_{f,0}} (\d\cdot \n) \T : \bnabla \v dS &\approx \int_{\Sigma_{f,0}} (\d\cdot \n) \{ \T:\bnabla_S \v + (\n\cdot \T \n)(\n\cdot \frac{\partial \v}{\partial n}) \} dS \\
    &\approx \int_{\Sigma_{f,0}} (\d\cdot \n) \{\T_0:\bnabla_S \v + (\n\cdot \T_0 \n)(\n\cdot \frac{\partial \v}{\partial n}) \} dS
\end{split}
\end{equation}
Moreover, stress terms integrated over the rigid surface $\Sigma_r$ are linearised. For simplicity, in what follows, it is assumed that all subsets of $\Sigma_r$ upon which contact lines may move, are flat. As a result, the normal vector $\n$ is unchanged along $\Sigma_{r,0}$ which implies in turn that $\bnabla_S \n = 0$. Furthermore, since the contact line may not be displaced into $\Sigma_r$, $\d\cdot\n_r = 0$ on $\Sigma_{r,0}$.
\begin{equation} \label{Eq:RigidIntegrals}
    \begin{split}
        \int_{\Sigma_r(\tau)} \v\cdot (\T \n) dS &\approx \int_{\Sigma_{r,0}} (\v+(\d\cdot \bnabla_S)\v) \cdot \{( \T + (\d\cdot \bnabla_S)\T ) \n\} (1+\bnabla_S\cdot \d) dS \\
        &\approx \int_{\Sigma_{r,0}}\{ \v\cdot (\T \n)(1+\bnabla_S \cdot \d) +(\d\cdot \bnabla_S)(\v\cdot (\T \n))\} dS \\
        &\approx \int_{\Sigma_{r,0}}\{ \v\cdot (\T \n) + \bnabla_S \cdot (\d \v \cdot (\T \n))\} dS \\
        &\approx \int_{\Sigma_{r,0}} \v\cdot \T \n dS + \int_{\partial \Sigma_{r,0}} (\d\cdot \sr) \, \v\cdot (\T_0 \n_r)) dl
    \end{split}
\end{equation}
In the last step, the divergence Theorem for Surfaces (\ref{Eq:DivergenceTheorem}), is used. 

\subsection{Continuity Equation}

No additional terms arise on the continuity equation since $\bnabla\cdot \u = \mathcal{O}(\|\d\|)$ on $\Sigma_{f,0}$.
\begin{equation} \label{Eq:LinearizedContinuity}
\begin{split}
    \int_{\Omega(\tau)} (\bnabla \cdot \u)q dV &\approx \int_{\Omega_0} (\bnabla \cdot \u)q dV + \int_{\Sigma_{f,0}} (\d\cdot \n)(\bnabla \cdot \u)q dS \\
     &\approx \int_{\Omega_0} (\bnabla \cdot \u)q dV = 0
\end{split}
\end{equation}

\subsection{Kinematic Equation}

In the next few steps, the velocity term is linearised.
\begin{equation} \label{Eq:KinematicEquationStart}
    \begin{split}
        \int_{\Sigma_f(\tau)} (\u\cdot \n)\chi dS \approx \int_{\Sigma_{f,0}} &\{ (\u + (\d\cdot \bnabla)\u)\cdot (\n-\n\cdot (\bnabla_S\d) )\} (\chi+(\d\cdot \bnabla_S)\chi)(1+\bnabla_S\cdot \d) dS \\
        \approx \int_{\Sigma_{f,0}} &\{  (\u\cdot \n)\chi + \chi \n\cdot (\d\cdot \bnabla)\u - \chi \n\cdot (\u\cdot \bnabla_S)\d\\
        &+ (\u\cdot \n)(\d\cdot \bnabla_S)\chi + \chi(\u\cdot \n)(\bnabla_S\cdot \d) \} dS
    \end{split}
\end{equation}
We note that since test function $\chi$ is defined only on over $\Sigma_{f,0}$, its linear expansion is
\begin{equation}
    \chi \approx \chi + \d\cdot\bnabla_S\chi.
\end{equation}
Identity (\ref{Eq:AppendixIdentity}) from Appendix \ref{appA} is applied in the subsequent step.
\begin{equation}
    \begin{split}
        \chi \n \cdot  \{(\d\cdot \bnabla)\u - (\u\cdot \bnabla_S )\d\} = &\bnabla_S\cdot (-\chi \n\times(\u\times \d)) + (\d\cdot \n)(\u\cdot \bnabla_S)\chi \\
        &- (\u\cdot \n)(\d\cdot \bnabla_S)\chi - \chi(\u\cdot \n) (\bnabla_S\cdot \d) + \chi (\d\cdot \n)(\bnabla\cdot \u)
    \end{split}
\end{equation}
Applying incompressibility, the velocity term present in the kinematic equation simplifies to 
\begin{equation}
    \begin{split}
        \int_{\Sigma_f(\tau)} (\u\cdot \n)\chi dS \approx &\int_{\Sigma_{f,0}} \{  (\u\cdot \n)\chi + (\d\cdot \n)(\u\cdot \bnabla_S)\chi \} dS \\
        &- \int_{\partial \Sigma_{f,0}}\chi(\n_f\times(\u\times \d))\cdot \sf \,dl
    \end{split}
\end{equation}
By virtue of the basis transformation at the contact line (see Figure \ref{fig:ContactLine}),
\begin{equation} \label{Eq:CLBases}
    \begin{cases}
        \sf = \cos(\phi) \sr + \sin(\phi) \n_r \\
        \n_f = \sin(\phi) \sr - \cos(\phi) \n_r \\
    \end{cases} \text{ on $\partial \Sigma_f$}
\end{equation}
we can expand and simplify the vector products 
\begin{equation}
    \begin{split}
        \sf\cdot (\n_f\times(\u\times \d)) = &(\u\cdot \sf)(\d\cdot \n_f) - (\u\cdot \n_f)(\d\cdot \sf) \\
        \approx &\{(\u\cdot \sr)\cos(\phi) + (\u\cdot \n_r)\sin(\phi) \}(\d\cdot \sr)\sin(\phi) \\
        & - \{(\u\cdot \sr)\sin(\phi) - (\u\cdot \n_r)\cos(\phi)  \}(\d\cdot\sr)\cos(\phi) \\
        = &(\u\cdot\n_r)(\d\cdot \sr)
    \end{split}
\end{equation}
where in the first step, we employ the fact that $\d\cdot \n_r = 0$ on $\Sigma_{r,0}\cap\Sigma_{f,0}$. We are left with the subsequent velocity-term linearisation.
\begin{equation} \label{Eq:KinematicEquationVelocityTerm}
    \begin{split}
        \int_{\Sigma_f(\tau)} (\u\cdot \n)\chi dS &\approx \int_{\Sigma_{f,0}} \{  (\u\cdot \n)\chi + (\d\cdot \n)(\u\cdot \bnabla_S)\chi \} dS - \int_{\partial \Sigma_{f,0}}\chi(\u_0\cdot\n_r)(\d\cdot\sr) \,dl
    \end{split}
\end{equation}
Note that, apart from incompressibility, the exact same steps can be applied when replacing $\u$ by $\frac{d \x}{d\tau}$ in Equation (\ref{Eq:KinematicEquationStart}). In this case, the first-order Taylor expansion 
\begin{equation}
    \frac{d\x(\X,\tau)}{d\tau} = \frac{d}{d\tau}( X + (\d\cdot\bnabla)X ) = \frac{\partial \d}{\partial \tau}
\end{equation} 
is exact. This produces
\begin{equation} \label{Eq:KinematicEquationDisplacementTerm}
    \begin{split}
        \int_{\Sigma_f(\tau)} (\frac{d\x}{d\tau}\cdot \n)\chi dS &\approx \int_{\Sigma_{f,0}} \{  (\frac{\partial\d}{\partial\tau}\cdot \n)\chi + (\d\cdot \n)(\frac{\partial\d}{\partial\tau}\cdot \bnabla_S)\chi + \chi(\d\cdot\n)(\bnabla\cdot\frac{\partial\d}{\partial\tau}) \} dS
    \end{split}
\end{equation}
where the fact that $\frac{\partial \d}{\partial \tau}\cdot \n_r = 0$ on $\partial \Sigma_{f,0}$ is employed to zero the curve integral. The linearised kinematic equation is taken to be the difference between Equations (\ref{Eq:KinematicEquationVelocityTerm}) and (\ref{Eq:KinematicEquationDisplacementTerm}).

\subsection{Dynamic Equation}

We begin with the atmospheric pressure term. The linearisation of $p_a(\v\cdot \n)$ is identical to that of $\chi(\u\cdot \n)$, save for the fact that $\bnabla p_a = 0$ and that incompressibility is not imposed on test function $\v$. Thus, the linearisation yields
\begin{equation} \label{Eq:LinearizedAtmosphericPressure}
\begin{split}
    \int_{\Sigma_f(\tau)} p_a(\v\cdot \n) dS \approx \int_{\Sigma_{f,0}} &\{ p_a(\v\cdot \n) + p_{a,0}(\d\cdot \n)(\bnabla_S\cdot \v) + p_{a,0}(\d\cdot \n)(\frac{\partial \v}{\partial n}\cdot \n)\} dS \\
    &- \int_{\partial \Sigma_{f,0}} p_{a,0} (\v\cdot\n_r)(\d\cdot\sr) dl
\end{split}
\end{equation}
In the next step, the surface divergence of $\v$ along $\Sigma_f$ is linearised using Relation (\ref{Eq:SurfaceDivergenceLinearizationFinal}) of Appendix \ref{appB}.
\begin{equation} \label{Eq:LinearizedDynamicEquation}
    \begin{split}
        \int_{\Sigma_f(\tau)} (\bnabla_S\cdot \v) dS \approx \int_{\Sigma_{f,0}} &\{ \bnabla_S \cdot \v + (\d\cdot \n) (\n\cdot \frac{\partial \v}{\partial n} + \bnabla_S\cdot \v )(\bnabla_S\cdot \n) \\ & - (\d\cdot \n)(\bnabla_S\n)^T:\bnabla_S\v + (\bnabla_S\v\cdot \n)\cdot \bnabla_S(\d\cdot \n) \} dS \\
        + \int_{\partial \Sigma_{f,0}} &\{(\d\cdot \n_f)(\sf\cdot \frac{\partial \v}{\partial n_f})  + (\d\cdot \sf) (\bnabla_S \cdot \v) \}dl
    \end{split}
\end{equation}
We now linearise the curve integral involving the contact angle in Equation (\ref{Eq:WeakForm}). Recalling that $\Sigma_{r,0}$ is assumed to be locally flat, $\{ \sr ,\t\}$ forms an orthogonal basis of $\Sigma_{r,0}$ and the rules of linearisation for a 2D curve apply:
\begin{equation}
    \begin{cases}
        dl \approx (1+\frac{d\d}{dl}\cdot \t) dl \\
        \frac{d}{dl} \approx (1-\frac{d\d}{dl}\cdot \t)\frac{d}{dl} \\
        \sr \approx \sr - (\frac{d\d}{dl}\cdot \sr)\t \\
        \t \approx \t + (\frac{d\d}{dl}\cdot \sr)\sr \\
        \n_r = \n_r \\
    \end{cases}
\end{equation}
This gives rise to
\begin{equation} \label{Eq:LineIntegralDevelopment}
\begin{split}
    \int_{\partial \Sigma_f(\tau)} \{\cos(\phi)(\v\cdot \sr&) + \sin(\phi)(\v\cdot\n_r)\} dl \\
    \approx \int_{\partial \Sigma_{f,0}}& \{\cos(\phi)(\sr - (\frac{d\d}{dl}\cdot \sr)\t) + \sin(\phi)\n_r\}\cdot \{\v+(\d\cdot\bnabla)\v\}(1+\frac{d\d}{dl}\cdot \t) dl \\
    \approx \int_{\partial \Sigma_{f,0}}& \{ (\cos(\phi)\sr + \sin(\phi)\n_r)\cdot (\v+(\d\cdot\bnabla)\v + \v (\frac{d\d}{dl}\cdot \t) ) \\
    &- \cos(\phi)(\frac{d\d}{dl}\cdot\sr)(\v\cdot\t) \} dl \\
\end{split}
\end{equation}
Note that, on $\partial \Sigma_{f,0}$, $\sf = \cos(\phi)\sr + \sin(\phi)\n_r + \mathcal{O}(\|\d\|)$. In addition, having assumed here that $ \d\cdot \n_r = 0$ on $\partial\Sigma_{f,0}$ and $\frac{d\n_r}{dl} = 0$, one obtains that $\frac{d\d}{dl}\cdot \n_r = 0$ on $\partial \Sigma_{f,0}$. This implies in turn that 
\begin{equation}
    -\cos(\phi)(\frac{d\d}{dl}\cdot \sr) = \frac{d\d}{dl}\cdot (-\cos(\phi)\sr - \sin(\phi)\n_r) \approx - \frac{d\d}{dl}\cdot \sf, \text{ on $\partial\Sigma_{f,0}$}
\end{equation}
This allows one to rewrite Equation (\ref{Eq:LineIntegralDevelopment}) as
\begin{equation} \label{Eq:LinearizedLineIntegral}
\begin{split}
    \int_{\partial \Sigma_f(\tau)} \{\cos(\phi)(\v\cdot \sr&) + \sin(\phi)(\v\cdot\n_r)\} dl \\
    \approx \int_{\partial \Sigma_{f,0}}& \{ (\cos(\phi)\sr + \sin(\phi)\n_r)\cdot \v+\sf\cdot(\d\cdot\bnabla)\v + (\v\cdot\sf) (\frac{d\d}{dl}\cdot \t) ) \\
    &- (\frac{d\d}{dl}\cdot\sf)(\v\cdot\t) \} dl \\
    = \int_{\partial \Sigma_{f,0}}& \{ (\cos(\phi)\sr + \sin(\phi)\n_r)\cdot \v+\sf\cdot(\d\cdot\bnabla)\v + (\v\times \frac{d\d}{dl})\cdot \n_f \} dl \\
\end{split}
\end{equation}
where the identity 
\begin{equation} \label{Eq:VectorIdentity}
(\boldsymbol{a}\times \boldsymbol{b})\cdot \n_f = (\boldsymbol{a}\cdot \sf)(\boldsymbol{b}\cdot \t) - (\boldsymbol{a}\cdot\t)(\boldsymbol{b}\cdot\sf)
\end{equation}
is used. We now take the sum of first-order terms in integrals over $\Sigma_{f,0}$ in Equations (\ref{Eq:LinearizedDynamicEquation}) and (\ref{Eq:LinearizedLineIntegral}):
\begin{equation}
    \begin{split}
        - (\d\cdot \n_f)(&\sf\cdot\frac{\partial\v}{\partial n_f}) - (\d\cdot\sf)(\bnabla_S\cdot\v) + \sf\cdot(\d\cdot\bnabla)\v + (\v\times\frac{d\d}{dl})\cdot \n_f \\
        &= -(\d\cdot\sf)(\frac{d\v}{dl}\cdot\t) + (\d\cdot\t)(\frac{d\v}{dl}\cdot\sf) + (\v\times \frac{d\d}{dl})\cdot \n_f \\
        &= (\v\times \frac{d\d}{dl})\cdot \n_f + (\frac{d\v}{dl}\times \d)\cdot \n_f \\
        &= \frac{d}{dl}\{ (\v\times\d)\cdot\n_f \} - (\v\times \d)\cdot \frac{d\n_f}{dl}
    \end{split}
\end{equation}
where in the first step, an expansion of the surface gradient $\bnabla_S = \t\frac{d}{dl} + \sf\frac{\partial}{\partial s_f}$ leads to the canceling-out of two terms. Moreover, in the second step, Identity (\ref{Eq:VectorIdentity}) is reused, followed by product differentiation. Note that $\int_{\partial\Sigma_{f,0}} \frac{d}{dl}\{ (\v\times\d)\cdot\n_f \} dl = 0$ by Stokes' Theorem since $\partial \Sigma_{f,0}$ is a closed curve by definition. Additionally, we expand the following term:
\begin{equation}
    \begin{split}
        (\v \times \d)\cdot \frac{d\n_f}{dl} = &\{ (\v\cdot\t)(\d\cdot\n_f) - (\v\cdot\n_f)(\d\cdot\t) \} \underbrace{(\frac{d\n_f}{dl}\cdot \sf)}_{=:A} \\
        &+ \underbrace{\{(\v\cdot\n_f)(\d\cdot\sf) - (\v\cdot\sf)(\d\cdot\n_f)\}}_{=:B}(\frac{d\n_f}{dl}\cdot \t)
    \end{split}
\end{equation}
Once again, using Basis Change (\ref{Eq:CLBases}), we observe that
\begin{equation}
    \begin{split}
        A &=  (\cos(\phi)\sr + \sin(\phi)\n_r) \cdot \frac{d}{dl}(\sin(\phi)\sr - \cos(\phi)\n_r) \\
        &= \sin(\phi) (\cos(\phi)\sr + \sin(\phi)\n_r) \cdot \frac{d\sr}{dl} \\
        &= \sin^2(\phi) (\n_r \cdot \frac{d\sr}{dl}) \\
        &= 0
    \end{split}
\end{equation}
due to the facts that $\frac{d\n_r}{dl} = 0$ and $\frac{d\sr}{dl} = (\frac{d\sr}{dl}\cdot\t)\t$, since curve $\partial \Sigma_{f,0}$ lies in a plane generated by $\{\sr,\t\}$. 

We also have that
\begin{equation}
    \begin{split}
        B =  &\{\v\cdot(\sin(\phi)\sr - \cos(\phi)\n_r)\}\,\{\d\cdot(\cos(\phi)\sr + \sin(\phi)\n_r)\} \\
        &- \{\v\cdot(\cos(\phi)\sr + \sin(\phi)\n_r)\}\,\{\d\cdot(\sin(\phi)\sr-\cos(\phi)\n_r)\} \\
        = &\v\cdot(\sin(\phi)\sr - \cos(\phi)\n_r)(\d\cdot\sr)\cos(\phi) \\
        & - \v\cdot(\cos(\phi)\sr + \sin(\phi)\n_r)\,(\d\cdot\sr)\sin(\phi) \\
        = & - (\v\cdot\n_r)(\d\cdot\sr)
    \end{split}
\end{equation}
where $\d\cdot\n_r = 0$ is imposed in the second step, followed by straightforward computations. In light of Equations (\ref{Eq:LinearizedAtmosphericPressure}), (\ref{Eq:LinearizedDynamicEquation}) and (\ref{Eq:LinearizedLineIntegral}), the dynamic-equation terms may \textit{in fine} be linearised as follows:
\begin{equation} \label{Eq:LinearizedDynamicEquationResult}
    \begin{split}
        - \int_{\Sigma_f(\tau)} &\{\bnabla_S\cdot \v + p_a(\v\cdot\n) \} dS + \int_{\partial \Sigma_f(\tau)} \{\cos(\phi)(\v\cdot \sr) + \sin(\phi)(\v\cdot\n_r)\} dl \\
        \approx - \int_{\Sigma_{f,0}} &\{ \bnabla_S \cdot \v + p_a(\v\cdot \n) + (\d\cdot \n) (\n\cdot \frac{\partial \v}{\partial n} + \bnabla_S\cdot \v )(\bnabla_S\cdot \n) \\
        &- (\d\cdot \n)(\bnabla_S\n)^T:\bnabla_S\v
        + (\bnabla_S\v\cdot \n)\cdot \bnabla_S(\d\cdot \n) \\
        &+ p_{a,0}(\d\cdot \n)(\bnabla_S\cdot \v) + p_{a,0}(\d\cdot \n)(\frac{\partial \v}{\partial n}\cdot \n)  \} dS \\
        + \int_{\partial \Sigma_{f,0}} &\{ \cos(\phi)(\v\cdot\sr) + \sin(\phi)(\v\cdot\n_r) + (p_{a,0} + \frac{d\n_f}{dl}\cdot\t) (\v\cdot\n_r)(\d\cdot\sr) \} dl
    \end{split}
\end{equation}

\subsection{Global Constraint}

The global constraint is generally a supplementary point-wise or integral equation on the physical system that serves (i) to accompany the added DOF, namely atmospheric pressure in the case of open capillary flows, or (ii) to constrain the volume in the event of enclosed flows. This global constraint is problem-dependent and must be defined \textit{ad hoc}. 

In the case of open flows, on the outlet, we typically impose zero tangent velocity on the function space, and zero normal stress. This circumvents pressure indeterminateness but calls for an additional DOF, that is atmospheric pressure, which defines the pressure reached after the surface-tension-induced jump across the interface. The accompanying global constraint for the latter is taken to be the Young-Laplace equation evaluated on the intersection of the outlet with the free surface, which is otherwise omitted from the weak form due to a Dirichlet BC on tangent velocity. Mathematically speaking,

\begin{equation}
    \int_{\partial\Sigma_f(\tau)\cap\partial\Sigma_r(\tau)} (\t\cdot \frac{d\n_f}{dl} + p_a)\,dl = 0
\end{equation}
which yields the linearised equation
\begin{equation}
    \int_{\partial\Sigma_{f,0}\cap\partial\Sigma_{r,0}} \{ \t\cdot \frac{d\n_f}{dl} + p_a - (\t\cdot \frac{d\n_f}{dl})^2 (\d\cdot \n_f) - (\t \cdot \frac{d\n_f}{dl}) \frac{d}{dl}(\d\cdot\t) \} \,dl \approx 0
\end{equation}
In the case of enclosed flows, atmospheric pressure is no longer needed in the weak formulation since a normal-stress BC is imposed only on the free surface. As such, the Young-Laplace equation fixes the pressure reference. Instead of a Young-Laplace constraint, a global volume constraint must then be employed in order to fix the input fluid volume $|\Omega|$.
\begin{equation}
    \int_{\Omega(\tau)} \,dV = |\Omega|
\end{equation}
which is linearised as
\begin{equation}
    \int_{\partial \Omega_0} (\d\cdot\n)\,dS = |\Omega| - \int_{\Omega_0}\,dV
\end{equation}
and is employed as a constraint on displacement together with a Lagrange multiplier present in the kinematic equation.

\subsection{Global Linearised System}

A linearised version of the general free-surface-flow System (\ref{Eq:WeakForm}) can now be expressed where all integrals have been linearised with respect to a reference solution $(\u_0,p_0,0,p_{a,0})$. Note that, by virtue of the dynamic equation, the sum $\int_{\Sigma_{f,0}}(\d\cdot \n)(\n\cdot \T_0\n + \frac{1}{\Ca}(\bnabla_S\cdot \n))(\n \cdot \frac{\partial \v}{\partial n})dS \approx 0 $ that stems from Equations (\ref{Eq:LinearizedStress}), (\ref{Eq:LinearizedAtmosphericPressure}) and (\ref{Eq:LinearizedDynamicEquation}), is second-order in displacement and is thus excluded.

\begin{equation} \label{Eq:LinearizedWeakForm}
    \begin{split}
    &\text{Find $(\u,p,\d,p_a)\in V\times Q\times M^3 \times \mathbb{R} $ such that for $\tau\in (0,T)$} \\
    &\begin{cases}
        \begin{split}
            \int_{\Omega_0} \{ \Rey\,\v\cdot\frac{\partial\u}{\partial \tau} + \Rey\,&\v\cdot(\u_0\cdot\bnabla)\u + \Rey\,\v\cdot(\u\cdot\bnabla)\u_0 + \T:\bnabla\v \} \,dV \\
            + \frac{1}{\Ca}\int_{\Sigma_{f,0}}p_a &(\v\cdot \n)\,dS
            - \int_{\Sigma_{r,0}}\v\cdot(\T\n_r)\,dS - \int_{\partial \Sigma_{r,0}} (\d\cdot \sr)\v\cdot (\T_0\n_r)\,dl \\
            + \int_{\Sigma_{f,0}} (\d\cdot\n&)\{ \Rey\,\v\cdot \frac{\partial \u}{\partial \tau} +\Rey\,\v\cdot (\u_0\cdot\bnabla_S)\u_0 + \T_0:\bnabla_S\v \}\,dS \\
            + \frac{1}{\Ca}\int_{\Sigma_{f,0}}\{ (&\d\cdot\n)(\bnabla_S\cdot\n)(\bnabla_S\cdot\v) - (\d\cdot\n)(\bnabla_S\n)^T:\bnabla_S\v \\
            + &(\bnabla_S\v\cdot\n)\cdot \bnabla_S(\d\cdot\n) + p_{a,0}(\d\cdot\n)(\bnabla_S\cdot\v) \} dS \\
            + \frac{1}{\Ca} &\int_{\partial\Sigma_{f,0}} (p_{a,0}+\frac{d\n_f}{dl}\cdot\t)(\v\cdot\n_r)(\d\cdot\sr)\,dl \\
            = & \int_{\Omega_0} \Rey\,\v\cdot (\u_0\cdot\bnabla)\u_0\,dV - \frac{1}{\Ca} \int_{\Sigma_{f,0}} (\bnabla_S\cdot \v)\,dS \\
            &+ \frac{1}{\Ca}\int_{\partial \Sigma_{f,0}}( \cos(\phi)(\v\cdot\sr) + \sin(\phi)(\v\cdot\n_r) )\,dl, \text{ $\forall \v\in V$}
        \end{split} \\
        \\
        \int_{\Omega_0} (\bnabla \cdot \u)q dV = 0,\text{ $\forall q\in Q$} \\
        \\
        \begin{split}
        \int_{\Sigma_{f,0}} \{  (\frac{\partial \d}{\partial \tau}\cdot\n)&\chi + (\d\cdot \n)(\frac{\partial\d}{\partial\tau}\cdot \bnabla_S)\chi + \chi(\d\cdot\n)(\bnabla\cdot\frac{\partial\d}{\partial\tau}) \}dS \\
        = &\int_{\Sigma_{f,0}} (\u\cdot \n)\chi  + (\d\cdot \n)(\u_0\cdot \bnabla_S)\chi \} dS \\
        &- \int_{\partial\Sigma_{f,0}} \chi (\u_0\cdot \n_r)(\d\cdot\sr) dl = 0, \text{ $\forall \chi\in M$} 
        \end{split} \\
        \\
        + \text{ ICs, BCs on $\u$ on $\Sigma_{r,0}$, BCs for $\d$ and global constraint}
    \end{cases}
    \end{split}
\end{equation}
We remark that, when it comes to displacement, only its normal component $\d\cdot\n$ is present in integrals over $\Sigma_{f,0}$. Tangent components of $\d$ are present only in integrals over $\partial \Sigma_{f,0}$. In fact, only component $\d\cdot\sr$ is involved in curve integrals. An optimal method of solution is therefore to solve for scalar fields $\alpha = \d\cdot \n$ over $\Sigma_{f,0}$ and $\beta = \d\cdot\sf$ over $\partial \Sigma_{f,0}$ while imposing $\d\cdot\n_r = 0$ on $\partial \Sigma_{f,0}$. The reduced problem is thus 

\begin{equation} \label{Eq:LinearizedWeakFormReduced}
    \begin{split}
    &\text{Find $(\u,p,\alpha,\beta,p_a)\in V\times Q\times M_\alpha \times M_\beta \times \mathbb{R} $ such that for $\tau\in (0,T)$} \\
    &\begin{cases}
        \begin{split}
            \int_{\Omega_0} \{ \Rey\,\v\cdot\frac{\partial\u}{\partial \tau} + \Rey\,&\v\cdot(\u_0\cdot\bnabla)\u + \Rey\,\v\cdot(\u\cdot\bnabla)\u_0 + \T:\bnabla\v \} \,dV \\
            + \frac{1}{\Ca}\int_{\Sigma_{f,0}}p_a &(\v\cdot \n)\,dS
            - \int_{\Sigma_{r,0}}\v\cdot(\T\n_r)\,dS \\
            - \int_{\partial \Sigma_{r,0}} (\sin(&\phi)\alpha + \cos(\phi)\beta)\v\cdot (\T_0\n_r)\,dl \\
            + \int_{\Sigma_{f,0}} \alpha\{ \Rey\,&\v\cdot \frac{\partial \u}{\partial \tau} +\Rey\,\v\cdot (\u_0\cdot\bnabla_S)\u_0 + \T_0:\bnabla_S\v \}\,dS \\
            + \frac{1}{\Ca}\int_{\Sigma_{f,0}}\{ &\alpha(\bnabla_S\cdot\n)(\bnabla_S\cdot\v) - \alpha(\bnabla_S\n)^T:\bnabla_S\v \\
            + &(\bnabla_S\v\cdot\n)\cdot \bnabla_S\alpha + \alpha p_{a,0} (\bnabla_S\cdot\v) \} dS \\
            + \frac{1}{\Ca} &\int_{\partial\Sigma_{f,0}} (p_{a,0}+\frac{d\n_f}{dl}\cdot\t)(\v\cdot\n_r)(\sin(\phi)\alpha + \cos(\phi)\beta)\,dl \\
            = & \int_{\Omega_0} \Rey\,\v\cdot (\u_0\cdot\bnabla)\u_0\,dV - \frac{1}{\Ca} \int_{\Sigma_{f,0}} (\bnabla_S\cdot \v)\,dS \\
            &+ \frac{1}{\Ca}\int_{\partial \Sigma_{f,0}}( \cos(\phi)(\v\cdot\sr) + \sin(\phi)(\v\cdot\n_r) )\,dl, \text{ $\forall \v\in V$}
        \end{split} \\
        \\
        \int_{\Omega_0} (\bnabla \cdot \u)q dV = 0,\text{ $\forall q\in Q$} \\
        \\
        \begin{split}
        \int_{\Sigma_{f,0}} \{  (\frac{\partial \d}{\partial \tau}\cdot\n)&\chi + \alpha(\frac{\partial\d}{\partial\tau}\cdot \bnabla_S)\chi + \chi\,\alpha(\bnabla\cdot\frac{\partial\d}{\partial\tau}) \}dS \\
        = &\int_{\Sigma_{f,0}} (\u\cdot \n)\chi  + \alpha(\u_0\cdot \bnabla_S)\chi \} dS \\
        &- \int_{\partial\Sigma_{f,0}} \chi (\u_0\cdot \n_r)(\sin(\phi)\alpha + \cos(\phi)\beta) dl = 0, \text{ $\forall \chi\in M_\alpha$} 
        \end{split} \\
        \\
        \int_{\partial\Sigma_{f,0}} \{\sin(\phi)\beta - \cos(\phi)\alpha\}\upsilon \,dl = 0,\text{ $\forall \upsilon \in M_\beta$ } \\
        \\
        + \text{ ICs, BCs on $\u$ on $\Sigma_{r,0}$, and global constraint}
    \end{cases}
    \end{split}
\end{equation}
Note that, in cases where $\beta = 0$ on $\partial \Sigma_{f,0}$, i.e. $\phi = \frac{\pi}{2}$, one can restrict $\d$ to the normal direction everywhere on $\Sigma_{f,0}$ as was done by \citet{kruyt_total_1988}.

\subsection{Elliptic Extension of Displacement for Mesh Movement} \label{Sec:EllipticExtension}

Upon solving System (\ref{Eq:LinearizedWeakFormReduced}), one obtains values of $\alpha = \d\cdot \n$ on $\Sigma_{f,0}$ and $\beta = \d\cdot\sf$ on $\partial \Sigma_{f,0}$. In this second, decoupled step, displacement values are extended from $\Sigma_{f,0}$ to the bulk of $\Omega_0$ via the solution of an artificial PDE. The computed values of displacement at mesh vertices are then used to displace the mesh before the next FBP iteration. Nodal values of field variables at a given iteration are displaced together with mesh vertices and serve as the reference configuration in the subsequent iteration.

In order to thus extend $\d$ efficiently, we first compute it in Cartesian coordinates on $\Sigma_{f,0}$. In problems where $\beta = 0$, we find $\d$ that solves
\begin{equation}
    \d = \alpha \n, \text{ on $\Sigma_{f,0}$}
\end{equation}
or in weak form,
\begin{equation} \label{Eq:DisplacementProjectionNoBeta}
    \int_{\Sigma_{f,0}} \boldsymbol{\psi} \cdot \d \,dS = \int_{\Sigma_{f,0}} \alpha(\boldsymbol{\psi}\cdot\n)\,dS,\text{ $\forall \boldsymbol{\psi}\in M_\alpha^3$}
\end{equation}
In cases where $\beta \neq 0$, on $\Sigma_{f,0}$, we solve
\begin{equation} \label{Eq:DisplacementProjectionBeta}
    \begin{cases}
        \d = \underset{\psi\in M_\alpha^2}{\text{argmin}} \|\bnabla_S \psi \|_{L_2(\Sigma_{f,0})}^2 \\
        \text{subject to }\d\cdot\n = \alpha,\text{ on $\Sigma_{f,0}$,} \\
        \d\cdot\sf = \beta,\text{ on $\partial \Sigma_{f,0}$} \\
        \text{and }\d\cdot\t = 0,\text{ on $\partial \Sigma_{f,0}$}
    \end{cases}
\end{equation}
Given that they are confined to the free surface and consequently, small, these basis projection problems are solved via direct solution. In the next step, the displacement field is extended from the free surface to the bulk of $\Omega_0$ via the following elliptic problem.
\begin{equation} \label{Eq:DisplacementStrainMinimization}
    \int_{\Omega_0} (\bnabla\d+(\bnabla\d)^T):\bnabla\boldsymbol{\psi}\,dV = 0,\text{ $\forall \boldsymbol{\psi}\in M_{\d}$}
\end{equation}
This (artificial) strain-minimisation problem is implemented with the aim of mitigating mesh distortion, as this can deteriorate FEM solution quality. Furthermore, it yields a symmetric positive-definite system that can be solved with the fast, preconditioned conjugate gradient algorithm. Upon solution, displacement values at vertices are used to move the mesh. In cases where the mesh is significantly deformed and distorted, it may be wise to remesh. Note that it would also be possible to solve the strain-minimisation problem together with the two constraints $\d\cdot\n = \alpha$ and $\d\cdot\sf = \beta$ in one step; however, this requires the solution of a saddle-point problem which is symmetric indefinite and, therefore, slower to solve than Equations (\ref{Eq:DisplacementProjectionBeta}) and (\ref{Eq:DisplacementStrainMinimization}).

\section{Preconditioning} \label{Sec:Preconditioner}

In the current Section, after giving the abstract weak form and discretised block formulation, we provide a preconditioner that is adequate for the solution of System (\ref{Eq:LinearizedWeakFormReduced}) by GMRES in the problems solved herein. 

\subsection{Block Formulation for Steady Flows}

The abstract formulation of the weak form is defined for steady flows where the volume $\Omega\subset \mathbb{R}^3$ is independent of time and $\frac{\partial \d}{\partial \tau} = \frac{\partial \u}{\partial \tau} = 0$. Equation System (\ref{Eq:LinearizedWeakFormReduced}) is then equivalent to

\begin{equation} \label{Eq:AbstractLinearizedWeakForm}
    \begin{split}
    &\text{Find $(\u,p,\alpha,\beta,p_a)\in V\times Q\times M_\alpha\times M_\beta\times\mathbb{R} $ such that } \\
    &\begin{cases}
        a(\u,\v) + b(\v,p) + c(\alpha,\v) + l(\beta,\v) + z(p_a,\v) = F(\v), \text{ $\forall \v\in V$}\\
        b(\u,q) = 0,\text{ $\forall q\in Q$} \\
        e(\u,\chi) + h(\alpha,\chi) + n(p_a,\chi)= 0,\,\text{ $\forall \chi\in M_\alpha$} \\
        j(\alpha,\upsilon) + w(\beta,\upsilon) = 0,\text{ $\forall \upsilon\in M_\beta$} \\
        k(\alpha,\xi) + g(p_a,\xi) = G(\xi),\text{ $\forall \xi\in\mathbb{R}$} \\
    \end{cases}
    \end{split}
\end{equation}
where $a$, $b$, $c$, $l$, $z$, $e$, $h$, $n$, $k$, $g$, $j$, $w$ are the appropriate bilinear forms defined from System (\ref{Eq:LinearizedWeakFormReduced}). $F$ and $G$ are the ensuing linear forms. As mentioned in Section \ref{Sec:Linearisation}, a global constraint is required in free-surface flows to close the mathematical problem. In open flows, atmospheric pressure does so, while in enclosed flows, a Lagrange multiplier associated to an input volume constraint carries out this task. For the sake of generality, the formulation in Equation (\ref{Eq:AbstractLinearizedWeakForm}) accommodates both types of fluid flow problems and refers to the resulting DOF as $p_a$ in both cases. Note that in open flows, $n(p_a,\chi) = 0$ while in enclosed flows, $z(p_a,\v) = 0$ and $n(p_a,\chi) = k(\chi,p_a)$. The preconditioning strategy is nevertheless identical in both cases.

Upon domain discretisation for suitable spaces $V_h\times Q_h\times M_{\alpha,h} \times M_{\beta,h} \times \mathbb{R}$, the linearised block-matrix system is
\begin{equation} \label{Eq:BlockSystem}
    \underbrace{\begin{pmatrix}
        \A & \B^T & \C & \boldsymbol{L} & \boldsymbol{Z} \\
        \B &  0  & 0 & 0 & 0 \\
        \E &  0  & \H & 0 & \boldsymbol{N} \\
        0  &  0  & \boldsymbol{J} & \boldsymbol{W} & 0 \\
        0  &  0  & \K & 0& \G \\
    \end{pmatrix}}_{=:\boldsymbol{M}}
    \begin{pmatrix}
        \boldsymbol{u_h} \\ \boldsymbol{p_h} \\ \boldsymbol{\alpha_h} \\ \boldsymbol{\beta_h} \\ p_{a,h}
    \end{pmatrix}
    = \begin{pmatrix}
        \f \\
        0 \\
        0 \\
        0 \\
        \g \\
    \end{pmatrix}
\end{equation}
where $\A$, $\B$, $\C$, $\boldsymbol{L}$, $\boldsymbol{Z}$, $\E$, $\H$, $\boldsymbol{N}$, $\boldsymbol{J}$, $\boldsymbol{W}$, $\K$ and $\G$ are the matrices emanating from spatial discretisation of Equation (\ref{Eq:AbstractLinearizedWeakForm}) and $\f$, $\g$ are the ensuing right-hand side vectors. In open flows, $\boldsymbol{N} = 0$, while in enclosed flows, $\boldsymbol{Z} = 0$ and $\boldsymbol{N} = \K^T$.

\subsection{Preconditioner for Steady Flows}

As mentioned in Section \ref{Sec:Introduction}, block-matrix systems such as that of Equation (\ref{Eq:BlockSystem}) are inherently ill-conditioned. Iterative methods may thus fail altogether in solving them when coupled only with algebraic preconditioning \citep{schunk_iterative_2002}, as this does not exploit the block structure of the matrix system. 

An ideal block preconditioner is given in the following equation.

\begin{equation} \label{Eq:IdealPreconditioner}
    \boldsymbol{P} = 
    \begin{pmatrix}
        \A & \B^T & \C & \boldsymbol{L} &\boldsymbol{Z} \\
        0 &  -\B\A^{-1}\B^T  & - \B\A^{-1}\C & - \B\A^{-1}\boldsymbol{L} & -\B\A^{-1}\D \\
        0 &  0  & \H + \E\boldsymbol{S}\C & \E\boldsymbol{S}\boldsymbol{L} & \boldsymbol{N} + \E\boldsymbol{S}\boldsymbol{Z} \\
        0 &  0  & \boldsymbol{J} & \boldsymbol{W} & 0 \\
        0 &  0  & \K & 0 & \G
    \end{pmatrix}
\end{equation}
where $\boldsymbol{S} = -\A^{-1} + \A^{-1}\B^T(\B\A^{-1}\B^T)^{-1}\B\A^{-1}$. This preconditioner is ideal in the sense that the condition number of the preconditioned system $\kappa(\boldsymbol{P}^{-1}\boldsymbol{M}) = 1$; it therefore clusters all eigenvalues. However, an explicit representation of the matrix is not readily available given the denseness of the Schur complements in Blocks (2,2) and (3,3). As a result, we propose a computable approximation $\Phat$ that was found to be effective in numerical simulations.

\begin{equation} \label{Eq:ApproximatePreconditioner}
    \Phat = \begin{pmatrix}
        \bar{\A} & \B^T & \C & \boldsymbol{L} & \boldsymbol{Z} \\
        0 &  -\Tilde{\boldsymbol{Q}}  & - \B\bar{\A}^{-1}\C & -\B\bar{\A}^{-1}\boldsymbol{L} & - \B\bar{\A}^{-1}\boldsymbol{Z} \\
        0 &  0  & \H + \E\hat{\boldsymbol{S}}\C & \E \hat{\boldsymbol{S}}\boldsymbol{L} & \boldsymbol{N} + \E \hat{\boldsymbol{S}} \boldsymbol{Z} \\
        0 &  0  & \boldsymbol{J} & \boldsymbol{W} & 0 \\
        0 &  0  & \K & 0 & \G
    \end{pmatrix}
\end{equation}
where $\Tilde{.}$ denotes that a diagonal approximation of the matrix is made, i.e. that off-diagonal components are zeroed, and $\bar{.}$ indicates that a multi-grid algorithm is employed to approximate inversion. $\boldsymbol{Q}$ is the pressure mass matrix whose approximation by a diagonal matrix is known to be spectrally equivalent to the Schur complement of pressure in the limit $\Rey \rightarrow 0$ \citep{elman_finite_2014}. We remark that, in all of our simulations, which are at relatively low Reynolds numbers, this is a suitable approximation. Moreover, matrix $\boldsymbol{S}$ is approximated as
\begin{equation} \label{Eq:SApproximation}
    \hat{\boldsymbol{S}} = -\Tilde{\A}^{-1} + \Tilde{\A}^{-1}\B^T\Tilde{\boldsymbol{Q}}^{-1}\B\Tilde{\A}^{-1}
\end{equation} 
Note that the approximations made in $\Phat$ degrade the eigenvalue clustering property of $\boldsymbol{P}$ and, therefore, compromise its ideality.

\subsection{Preconditioner Application}

At each step of the GMRES, Preconditioner (\ref{Eq:ApproximatePreconditioner}) is applied; in other words, it is inverted so as to solve System
\begin{equation}
        \begin{pmatrix}
        \bar{\A} & \B^T & \C & \boldsymbol{L} & \boldsymbol{Z} \\
        0 &  -\Tilde{\boldsymbol{Q}}  & - \B\bar{\A}^{-1}\C & -\B\bar{\A}^{-1}\boldsymbol{L} & - \B\bar{\A}^{-1}\boldsymbol{Z} \\
        0 &  0  & \H + \E\hat{\boldsymbol{S}}\C & \E \hat{\boldsymbol{S}}\boldsymbol{L} & \boldsymbol{N} + \E \hat{\boldsymbol{S}} \boldsymbol{Z} \\
        0 &  0  & \boldsymbol{J} & \boldsymbol{W} & 0 \\
        0 &  0  & \K & 0 & \G
    \end{pmatrix} \begin{pmatrix}
        \boldsymbol{X_u} \\ \boldsymbol{X_p} \\ \boldsymbol{X_\alpha} \\ \boldsymbol{X_\beta} \\ \boldsymbol{X_{p_a}}
    \end{pmatrix}
     = \begin{pmatrix}
        \boldsymbol{Y_u} \\ \boldsymbol{Y_p} \\ \boldsymbol{Y_\alpha} \\ \boldsymbol{Y_\beta} \\ \boldsymbol{Y_{p_a}}
    \end{pmatrix}
\end{equation}
To this end, an inversion algorithm is given. Note that $\Phat$, when lumping its final three rows and columns into a single block, can be considered as an upper, block-triangular matrix. As such, inversion is required only of block diagonals and can be separated into three steps. 
\begin{equation} \label{Eq:PreconditionerInversionAlgorithm}
\begin{split}
    &\textit{\underline{Inversion Algorithm}} \\
    &\begin{cases}
    \textbf{Step 1:} \\
    \\
        \begin{pmatrix}
    \H + \E\hat{\boldsymbol{S}}\C & \E\hat{\boldsymbol{S}} \boldsymbol{L} & \boldsymbol{N} + \E \hat{\boldsymbol{S}} \boldsymbol{Z} \\
    \boldsymbol{J} & \boldsymbol{W} & 0 \\
    \K & 0 & \G
\end{pmatrix} \begin{pmatrix}
    \boldsymbol{X_\alpha} \\ \boldsymbol{X_\beta} \\ \boldsymbol{X_{p_a}} 
\end{pmatrix} = 
\begin{pmatrix}
        \boldsymbol{Y_\alpha} \\ \boldsymbol{Y_\beta} \\ \boldsymbol{Y_{p_a}}
    \end{pmatrix} \text{ (LU solution)}
    \\
    \\
    \textbf{Step 2:} \\
    \\
    \Tilde{\boldsymbol{Q}} \boldsymbol{X_p} = -\boldsymbol{Y_p} - \B \bar{\A}^{-1} (\C \boldsymbol{X_\alpha} + \boldsymbol{L}\boldsymbol{X_\beta} + \boldsymbol{Z}\boldsymbol{X_{p_a}}) \text{ (inversion with Jacobi preconditioning)}\\
    \\
    \textbf{Step 3:} \\
    \\
    \bar{\A}\boldsymbol{X_u} = \boldsymbol{Y_u} - \B^T \boldsymbol{X_p} - \C \boldsymbol{X_\alpha} - \boldsymbol{L}\boldsymbol{X_\beta} - \boldsymbol{Z}\boldsymbol{X_{p_a}} \text{ (multi-grid inversion with 7 cycles)} \\
    \end{cases}
\end{split}
\end{equation}
In the first step, exact LU inversion can be carried out since the block matrix is relatively small. Indeed, $\alpha$ is defined only on $\Sigma_{f,0}$, $\beta$ on $\partial \Sigma_{f,0}$, while $p_{a}$ is a constant. As for the second step, in the small Reynolds-number limit, Jacobi preconditioning of pressure's mass matrix $\boldsymbol{Q}$ is known to be spectrally equivalent to exact inversion of the Schur complement, $\B\A^{-1}\B^T$ \citep{elman_finite_2014}, and hence, gives a robust approximation. In the third step, multi-grid inversion is employed in order to effect a fast, approximate solution of the convection-diffusion problem.

\section{Numerical Simulations of Steady Flows} 
\label{Sec:Simulations}

In this Section, two numerical examples are solved for the purpose of validating the linearised system given by Equation (\ref{Eq:LinearizedWeakFormReduced}) and testing the preconditioner evoked in Equation (\ref{Eq:ApproximatePreconditioner}).

\subsection{Cylindrical Die Swell} \label{Sec:DieSwellSimulations}

\begin{figure}
    \centering
    \includegraphics{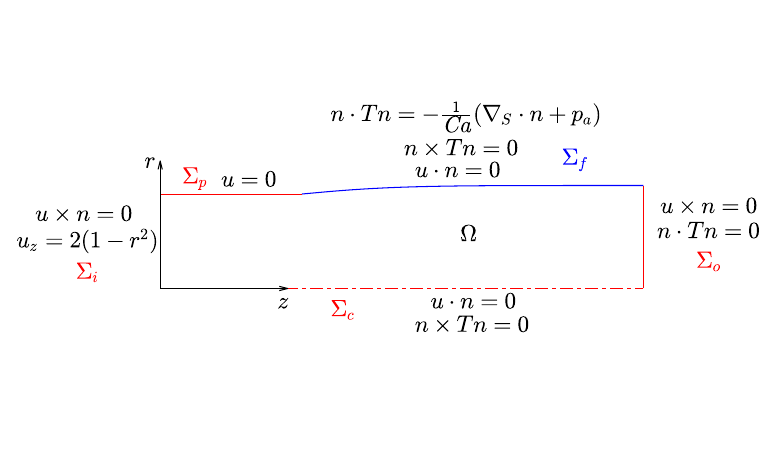}
    \caption{Domain of axisymmetric die swell and boundary conditions}
    \label{fig:CylinderDomain}
\end{figure}

Steady laminar flow out of a sharply-ending circular pipe is modeled here in two distinct simulations. The first is parametrised by cylindrical coordinates with the azimuthal coordinate omitted under the Ansatz of axisymmetry. The second is modeled in 3D Cartesian coordinates. In both simulations, the fluid is assumed to be incompressible, Newtonian, and governed by the Young-Laplace equation along the interface with the atmosphere. The effects of gravity are not considered. 

As in Section \ref{Sec:Linearisation}, the domain boundary $\partial \Omega$ is decomposed into two disjoint sets, $\Sigma_r$ and $\Sigma_f$, where $\Sigma_r = \Sigma_p\cup \Sigma_i \cup \Sigma_c\cup\Sigma_o$ and $\Sigma_f$ is the free surface, as seen in Figure \ref{fig:CylinderDomain}. The fluid enters the numerical domain through $\Sigma_i$, with a Poiseuille flow profile. Up to the die exit, it satisfies the axisymmetry conditions of no-shear and impermeability on the centre boundary $\Sigma_c$ (which is present only in axisymmetric simulations); the no-slip condition is obeyed on the pipe surface $\Sigma_p$. At the outlet $\Sigma_o$, zero normal stress and zero tangential flow are imposed; impermeability, zero shear stress and the Young-Laplace law govern the free surface $\Sigma_f$. Finally, the Young-Laplace law is also imposed on $\partial \Sigma_f\cap \partial \Sigma_r$ as the global constraint. Although the die swell is not a contact-line problem \textit{per se}, the intersection of the free surface with the outlet, $\partial \Sigma_{f}\cap \partial \Sigma_{o}$, is not pinned and therefore, may be likened to a contact line with respective angle $\phi = \frac{\pi}{2}$. 

The model equations in strong form are in Cartesian coordinates with $z$ being the axial direction:
\begin{equation}
    \begin{cases}
        \text{Find $(\boldsymbol{u},p,\Sigma_f,p_a)\in V\times Q\times S \times \mathbb{R}$ such that } \\
        \Rey (\u\cdot \bnabla)\u = -\bnabla p + \nabla^2 \u,\text{ in $\Omega$}\\
        \bnabla \cdot \u = 0,\text{ in $\Omega$} \\
        \u\cdot \n = 0, \n\times \T \n = 0, \n\cdot \T \n = -\frac{1}{\Ca}(\bnabla_S\cdot \n + p_a),\text{ on $\Sigma_f$} \\
        \u = 0,\text{ on $\Sigma_p$} \\
        \u = (0,0,2(1-x^2-y^2)),\text{ on $\Sigma_i$} \\
        \u \cdot \n = 0, \n\times \T \n = 0,\text{ on $\Sigma_c$} \\
        \u\times \n = 0,\n\cdot \T \n = 0,\text{ on $\Sigma_o$} \\
        p_a + \t\cdot \frac{d\n_f}{dl} = 0,\text{ on $\partial \Sigma_f\cap \partial \Sigma_o$}\\
    \end{cases}
\end{equation}
Here, $\Rey =\frac{\rho U R}{\mu}$ where $U$ denotes the mean cross-section velocity and $R$, chosen to be the characteristic length $L$, denotes the pipe radius; the Ohnesorge number is $\Oh = \sqrt{\frac{\Ca}{\Rey}}$. The free surface is pinned at $\partial \Sigma_f \cap \partial \Sigma_p $, implying that $\d = 0$ there, and is constrained to be purely radial at $\partial \Sigma_f\cap \partial \Sigma_o$ since flow is assumed to be fully developed at the domain outlet. This implies that $\beta = \d\cdot \sf = 0$ on $\partial \Sigma_f$, thus reducing the representation of displacement to $\d = \alpha\n_f$ over $\Sigma_f$. 

This leads to the following weak form. Note that that, in the die-swell problem, the integrand in the generalised weak form $(p_{a,0}+\frac{d\n_f}{dl}\cdot\t)\alpha = O(|\alpha|^2)$ on $\partial \Sigma_{o,0}$; it is thus omitted from System (\ref{Eq:DieSwellWeakForm}).  
\begin{equation} \label{Eq:DieSwellWeakForm}
    \begin{split}
    &\text{Find $(\u,p,\alpha,p_a)\in V_h\times Q_h\times M_{\alpha,h} \times \mathbb{R} $ such that} \\
    &\begin{cases}
        \begin{split}
            \int_{\Omega_0} \{ \Rey\,\v\cdot(\u_0\cdot\bnabla)&\u + \Rey\,\v\cdot(\u\cdot\bnabla)\u_0 + \T:\bnabla\v \} \,dV + \frac{1}{\Ca}\int_{\Sigma_{f,0}}p_a(\v\cdot \n)\,dS \\
            + \int_{\Sigma_{f,0}} \alpha \{ \Rey&\,\v\cdot (\u_0\cdot\bnabla_S)\u_0 + \T_0:\bnabla_S\v \}\,dS \\
            + \frac{1}{\Ca}\int_{\Sigma_{f,0}}\{ &\alpha (\bnabla_S\cdot\n)(\bnabla_S\cdot\v) - \alpha (\bnabla_S\n)^T:\bnabla_S\v + (\bnabla_S\v\cdot\n)\cdot \bnabla_S\alpha \} dS \\
            + \frac{1}{\Ca}\int_{\Sigma_{f,0}} &\alpha \, p_{a,0}\,(\bnabla_S\cdot\v)  dS \\
            = & \int_{\Omega_0} \Rey\,\v\cdot (\u_0\cdot\bnabla)\u_0\,dV - \frac{1}{\Ca} \int_{\Sigma_{f,0}} (\bnabla_S\cdot \v)\,dS \\
            &+ \frac{1}{\Ca}\int_{\partial \Sigma_{f,0}\cap \partial \Sigma_{o,0}}v_z\,dl, \text{ $\forall \v\in V_h$}
        \end{split} \\
        \\
        \int_{\Omega_0} (\bnabla \cdot \u)q dV = 0,\text{ $\forall q\in Q_h$} \\
        \\
        \int_{\Sigma_{f,0}} \{  (\u\cdot \n)\chi + \alpha(\u_0\cdot \bnabla_S)\chi \} dS -\int_{\partial\Sigma_{o,0}} \chi u_{0,z}\alpha \,dl = 0, \text{ $\forall \chi\in M_{\alpha,h}$} \\
        \\
        \int_{\partial\Sigma_{f,0}\cap \partial\Sigma_{o,0}} \{ p_a - (\t\cdot \frac{d\n_f}{dl})^2 \alpha \} \,dl = -\int_{\partial \Sigma_{f,0}\cap\partial\Sigma_{o,0}} (\t\cdot \frac{d\n_f}{dl}) \,dl
    \end{cases}
    \end{split}
\end{equation}
 where in both types of simulations, $V_h$ and $Q_h$ form the Taylor-Hood finite element pair \citep{taylor_numerical_1973}, which has been shown to satisfy the inf-sup condition in both Cartesian coordinates \citep{girault_finite_1986} and axisymmetric, cylindrical coordinates \citep{belhachmi_weighted_2006}. The pair $V_h$ and $Q_h$ is used together with quadratic Lagrange elements over $\Sigma_{f,0}$ for $\alpha\in M_{\alpha,h}$, since $\alpha$ is governed by a second-order differential equation, as is $\u$. 
 
 In Cartesian simulations, $\Omega_0$ is discretised into a mesh of second-order, triangular elements. This choice is made in order to avert premature convergence plateaus, which may arise when a $\mathbb{P}^2$-discretisation of displacement prescribes motion of the midpoints of triangle edges in the mesh-movement step, something that is impossible for linear, triangular elements. This choice is in contrast to axisymmetric simulations where the domain is discretised into a mesh of linear, triangular elements since, in practice, convergence is not observed to be hindered by a lack of midpoint movement in the instances exposed here. In the axisymmetric and Cartesian cases, spatial discretisations endow simulations with approximately $10^6$ DOFs and are finer in the vicinity of the stick-slip transition at the pipe exit. 
 
 In axisymmetric simulations, the Ansatz of axisymmetry is exercised so as to reduce $\u$ to the two-dimensional vector field, $(u_r,u_z)$. All differential operators, integration measures and tensor operations are then also recast into cylindrical coordinates; see \citet{gerritsma_spectral_2000} for details.

The initial guess of each simulation domain is a cylindrical free surface of radius $r = 1$. The Stokes-flow result for the cylindrical domain, without enforcing the Young-Laplace law, serves as the starting point for velocity and pressure. Upon calculation of FBP unknowns $(\u,p,\alpha,p_a)$ in a given iteration, the displacement field $\d$ is extended from $\Sigma_{f,0}$ to the bulk via elliptic extension, as explained in Section \ref{Sec:EllipticExtension}. For unknowns $\u$ and $\alpha$, tolerances of $\epsilon = 10^{-4}$ in axisymmetric simulations and $\epsilon = 10^{-5}$ in the Cartesian case are employed as stopping criteria. 

\begin{figure}
    \centering
    \subfigure[]{\includegraphics[width=0.495\linewidth]{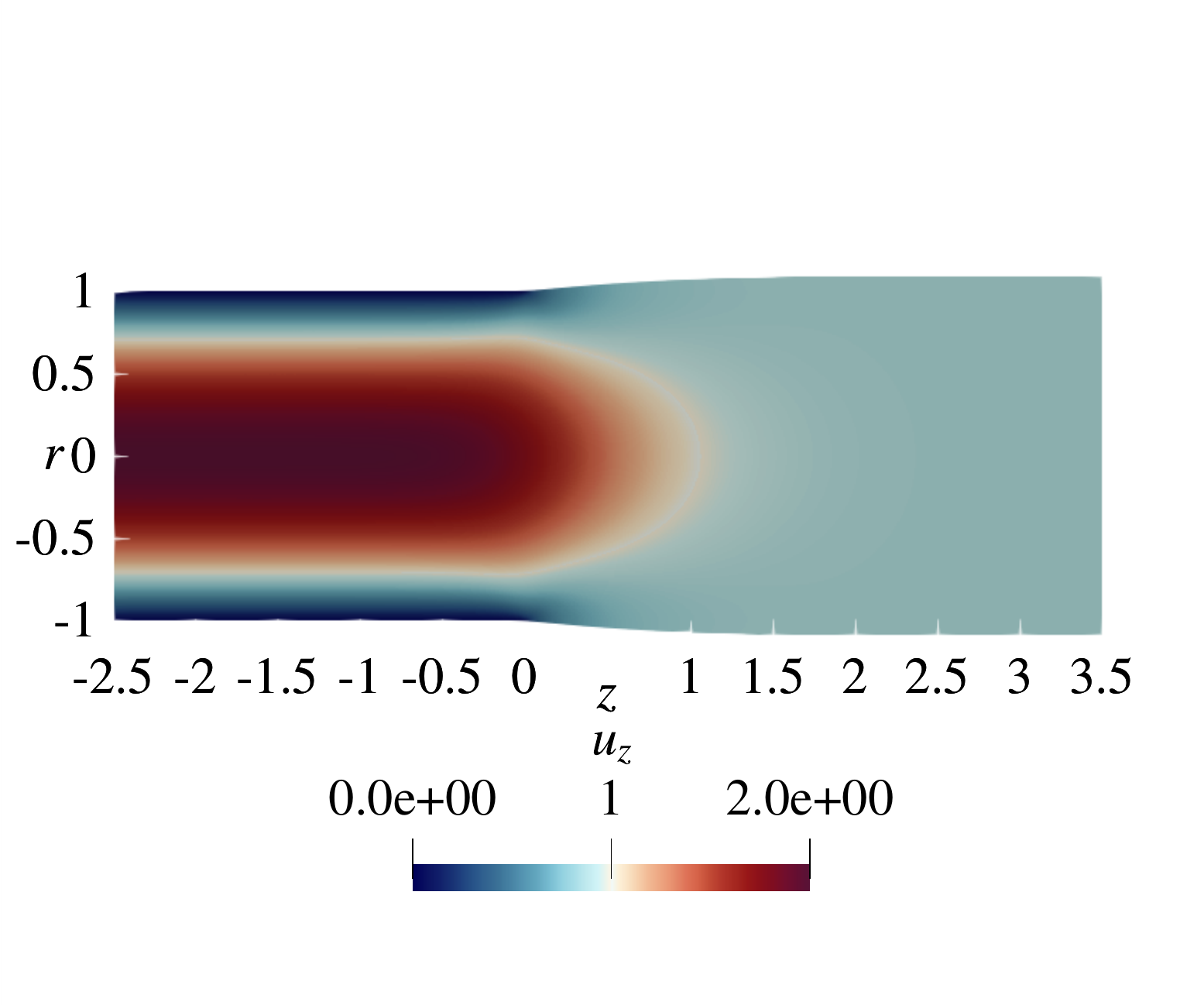}}
    \subfigure[]{\includegraphics[width=0.495\linewidth]{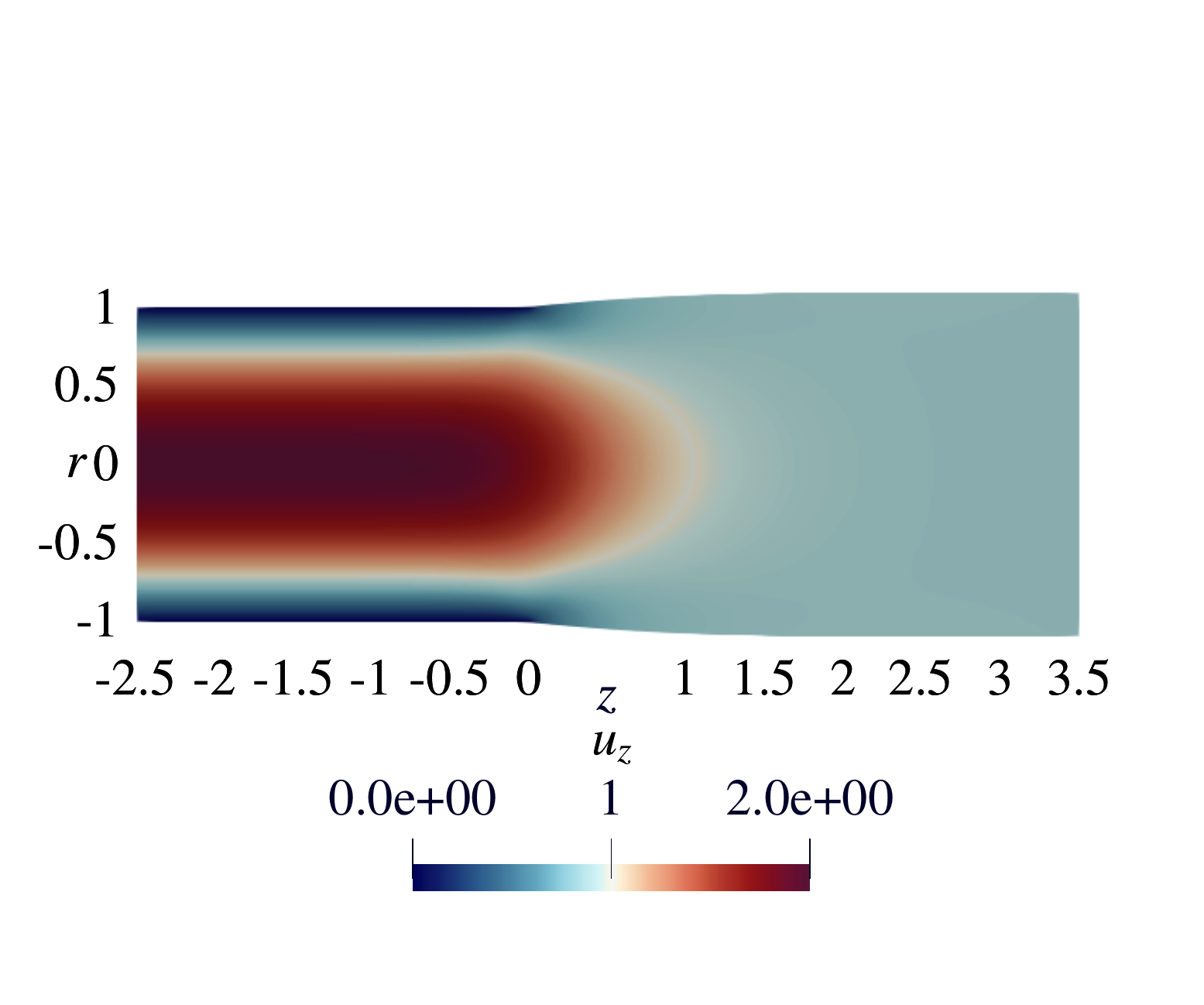}}
    \caption{Die swell simulation result $u_z$ (a) mirrored axisymmetric result (b) 3D Cartesian result of slice through the plane defined by $x = 0$, for parameters $\Rey = 2.5$, $\Oh = \sqrt{3}$  }
    \label{fig:DieSwellRe5u_z}
\end{figure}

\begin{figure}
    \centering
    \subfigure[]{\includegraphics[width=0.495\linewidth]{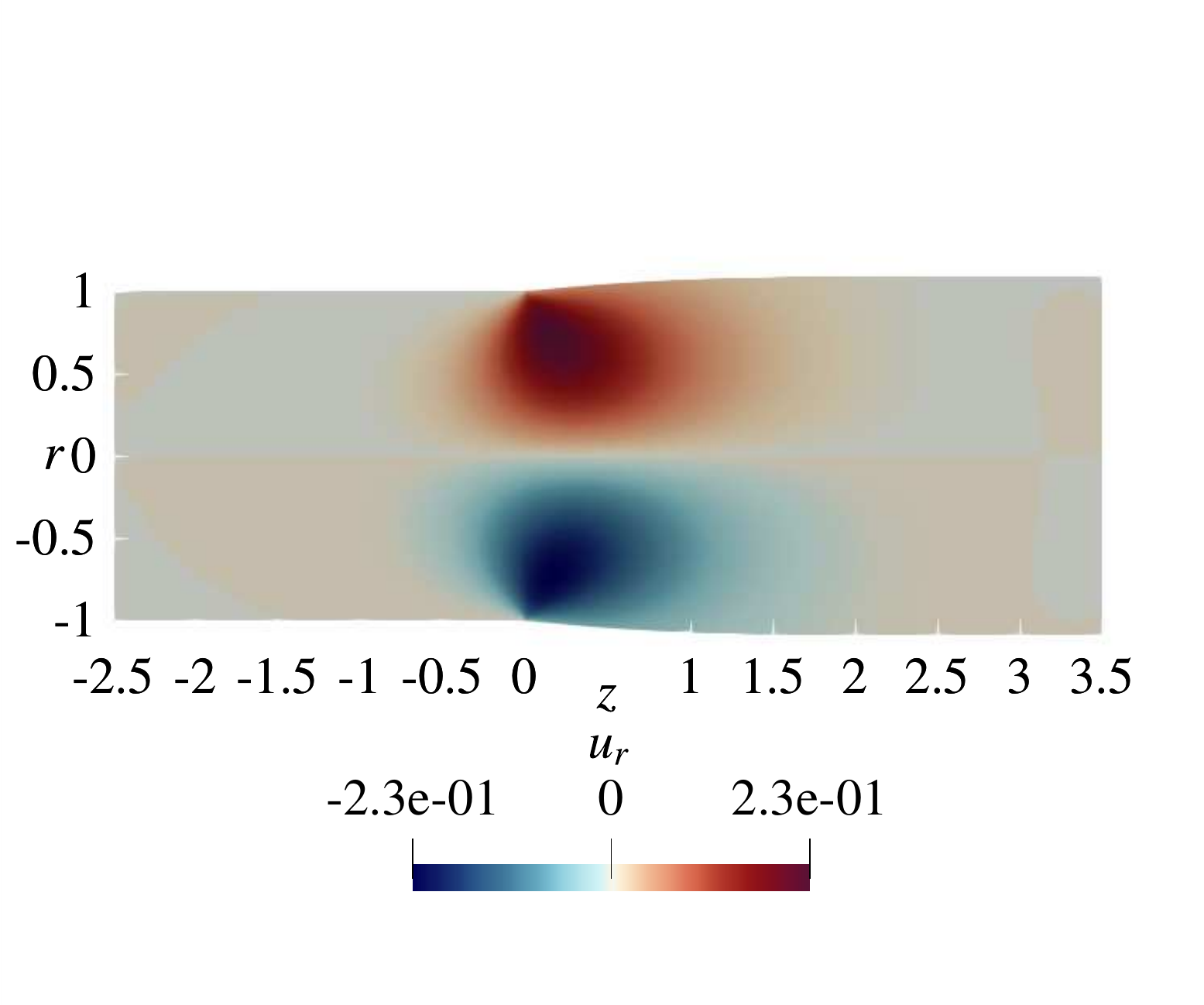}}
    \subfigure[]{\includegraphics[width=0.495\linewidth]{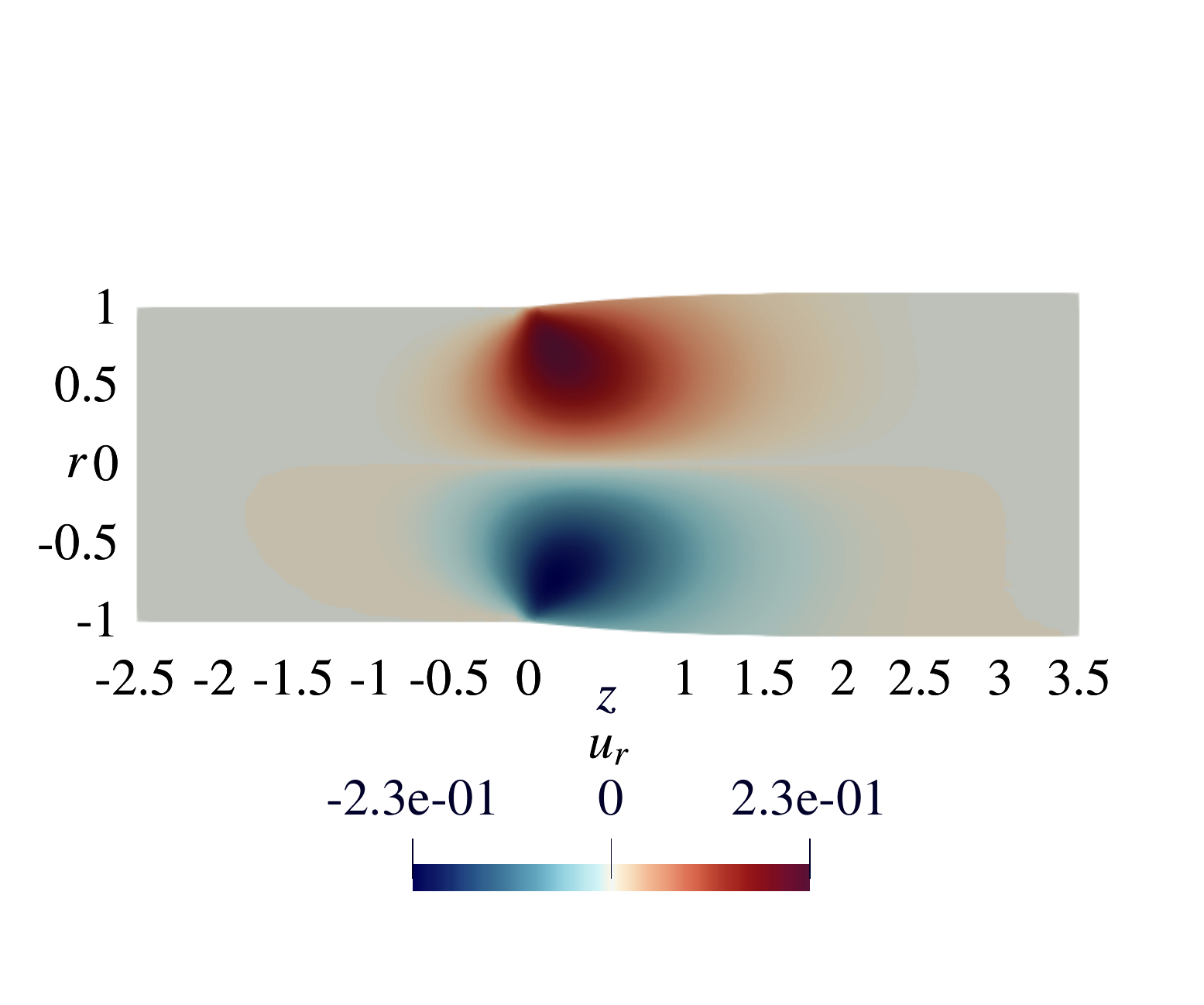}}
    \caption{Die swell simulation result $u_r$ (a) mirrored axisymmetric result (b) 3D Cartesian result of slice through the plane defined by $x = 0$, for parameters $\Rey = 2.5$, $\Oh = \sqrt{3}$  }
    \label{fig:DieSwellRe5u_r}
\end{figure}

\begin{figure}
    \centering
    \subfigure[]{\includegraphics[width=0.495\linewidth]{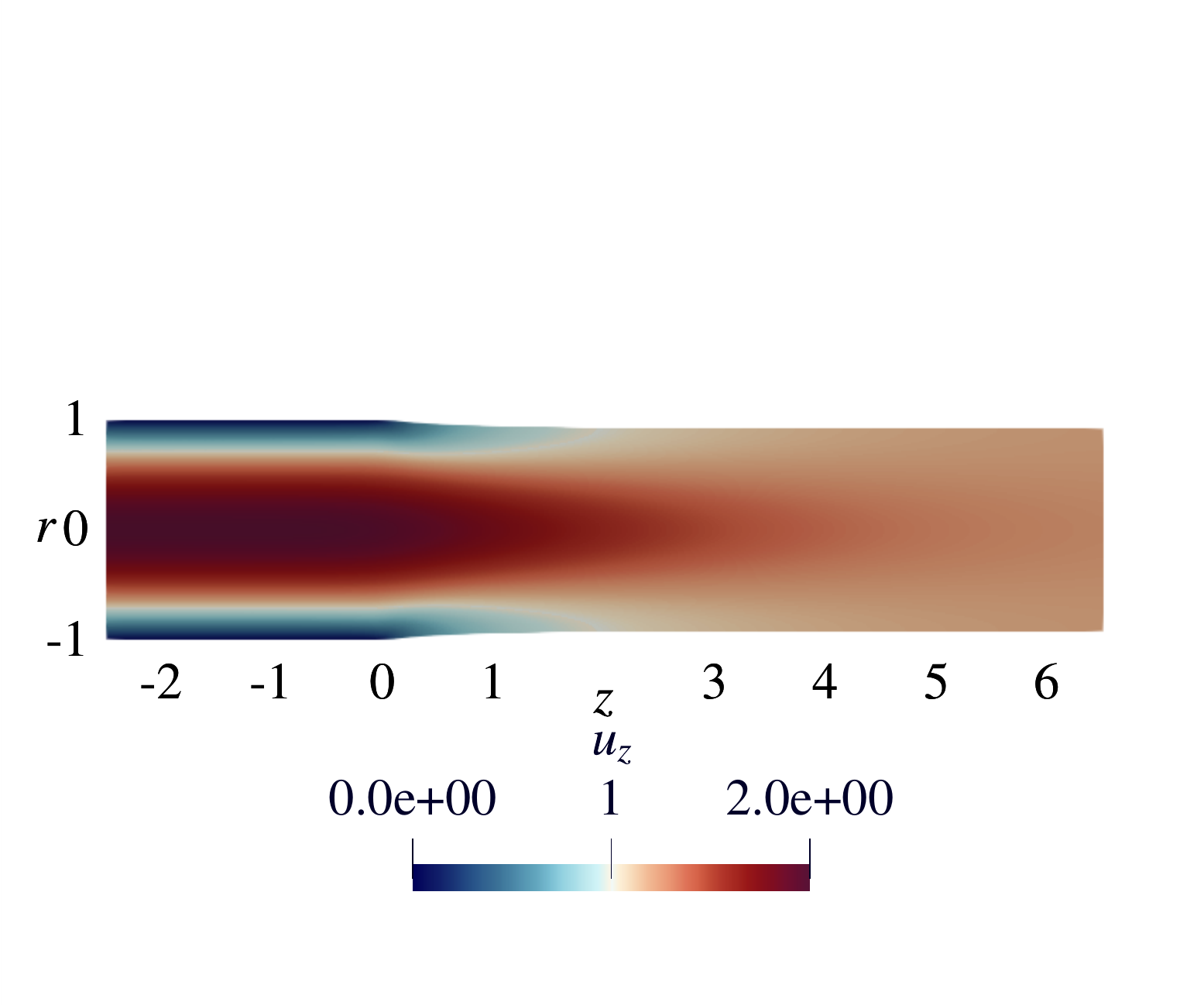}}
    \subfigure[]{\includegraphics[width=0.495\linewidth]{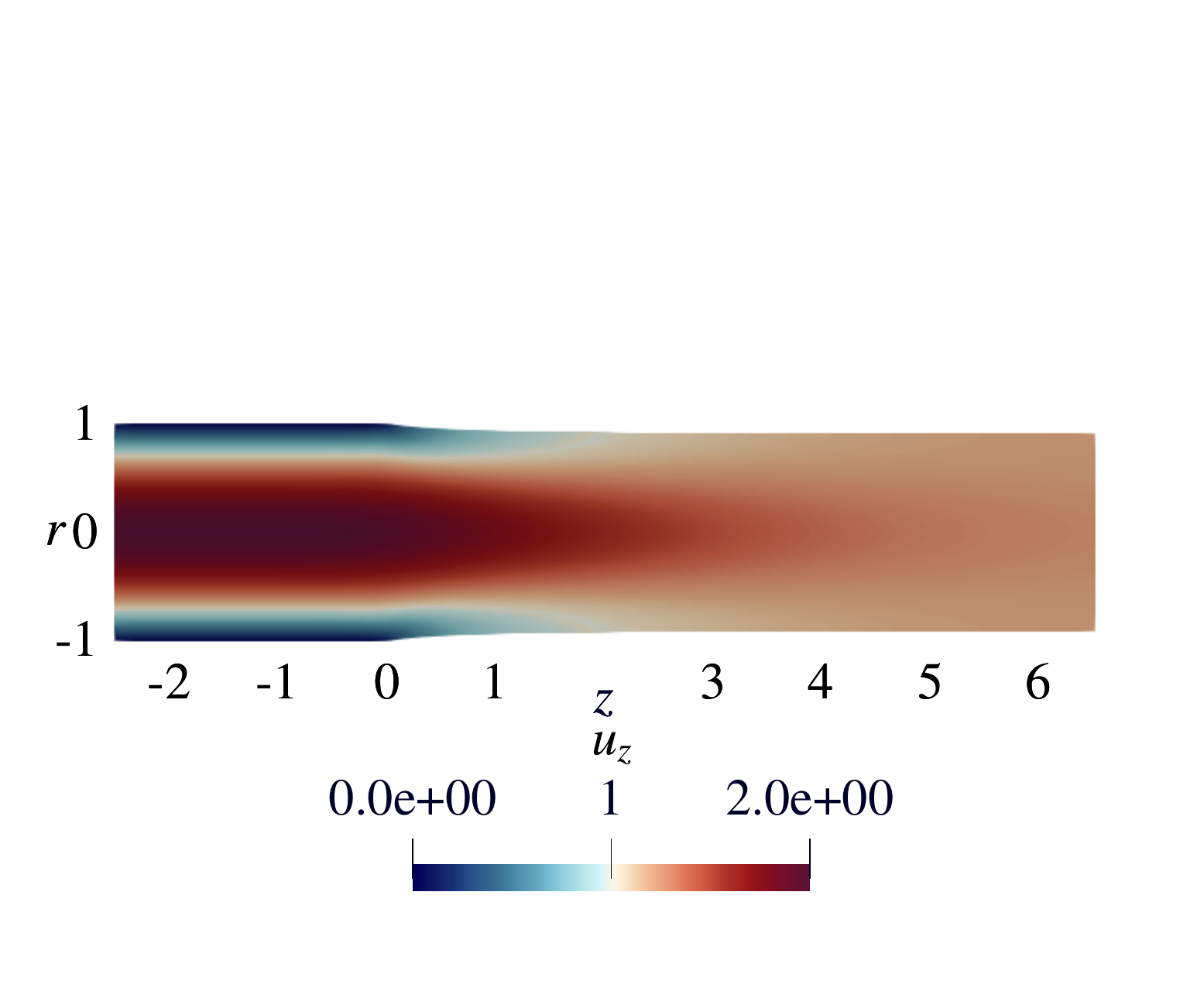}}
    \caption{Die swell simulation result $u_z$ (a) mirrored axisymmetric result (b) 3D Cartesian result of slice through the plane defined by $x = 0$, for parameters $\Rey = 25$, $\Oh = \sqrt{3}$.  }
    \label{fig:DieSwellRe50u_z}
\end{figure}

\begin{figure}
    \centering
    \subfigure[]{\includegraphics[width=0.495\linewidth]{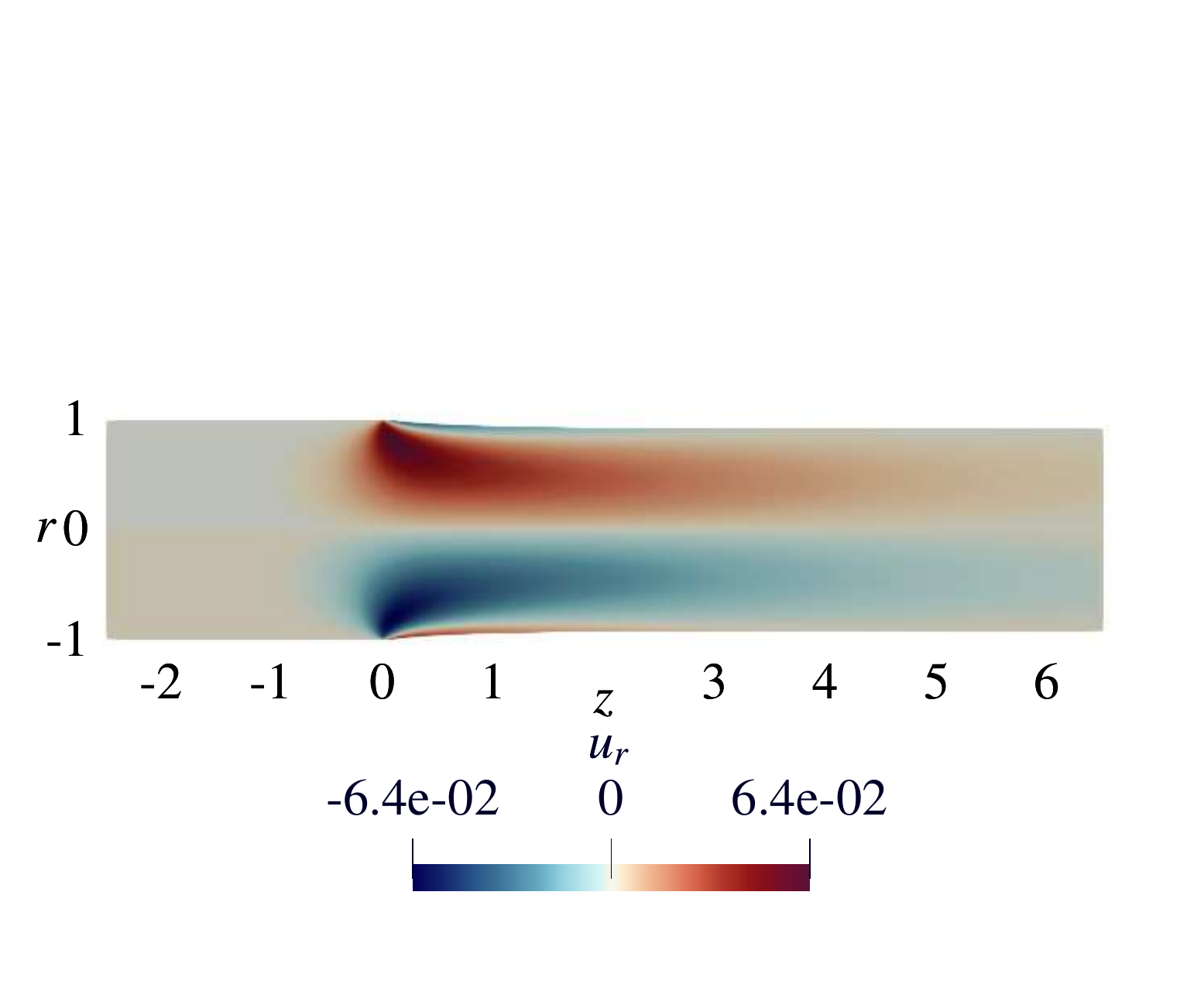}}
    \subfigure[]{\includegraphics[width=0.495\linewidth]{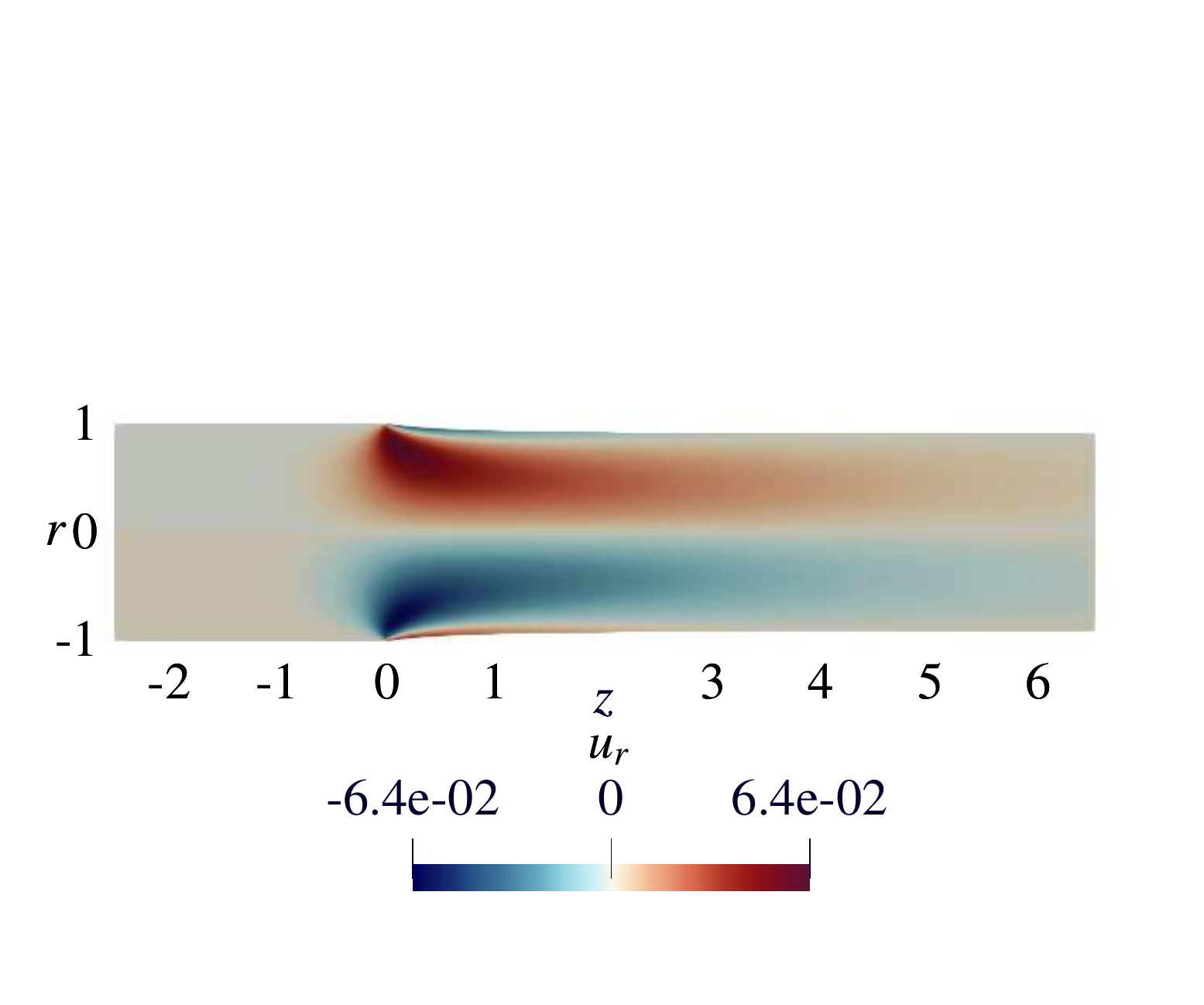}}
    \caption{Die swell simulation result $u_r$ (a) mirrored axisymmetric result (b) 3D Cartesian result of slice through the plane defined by $x = 0$, for parameters $\Rey = 25$, $\Oh = \sqrt{3}$  }
    \label{fig:DieSwellRe50u_r}
\end{figure}

In Figures \ref{fig:DieSwellRe5u_z} and \ref{fig:DieSwellRe5u_r}, the axisymmetric and Cartesian simulations are compared for $\Rey = 2.5$, $\Oh = \sqrt{3}$. Specifically, in Figure \ref{fig:DieSwellRe5u_z}, the axial component of velocity is depicted for both simulations while in Figure \ref{fig:DieSwellRe5u_r}, the radial component is considered. Qualitative discrepancies between the two velocity distributions are indiscernible. Moreover, in both, a substantial expansion of the jet profile from its pipe-outlet radius is predicted; in the axisymmetric simulation, the predicted jet-radius expansion factor, known as the swell ratio $r_\infty$, is 9.0\% while in the Cartesian simulation, it is 9.4\%.

In Figures \ref{fig:DieSwellRe50u_z} and \ref{fig:DieSwellRe50u_r}, a similar comparison is made between axisymmetric and Cartesian results for $\Rey = 25$, $\Oh = \sqrt{3}$. In similar fashion to the preceding simulation, differences are imperceptible between velocity distributions. As for the jet shape, both simulations predict a behaviour opposite to the previous case, namely, a jet contraction of 8.9 \%. 

\begin{figure}
    \centering
    \includegraphics[width=0.95\linewidth]{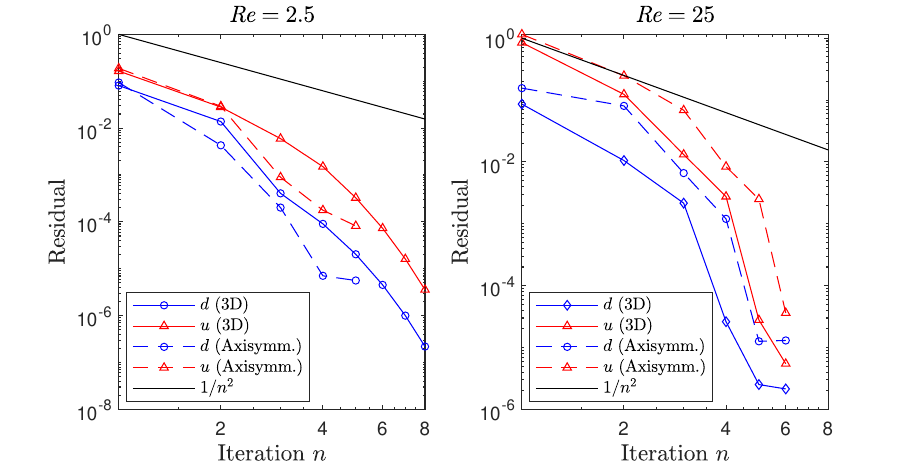}
    \caption{Residual evolution across iterations in axisymmetric and Cartesian simulations for $\Rey = 2.5$ (left) and $\Rey = 25$ (right), with $\textit{Oh} = \sqrt{3}$}
    \label{fig:DieSwellResiduals}
\end{figure}
In Figure \ref{fig:DieSwellResiduals}, the displacement and velocity residuals for both axisymmetric and Cartesian simulations are depicted with respect to the number of iterations. We observe that, in the two Reynolds instances, residuals for axisymmetric and Cartesian simulations decay at a rate that is faster than quadratic and meet the tolerance requirement after a similar number of iterations. The Newton-Raphson algorithm thus verifies its theoretical convergence rate in these numerical examples.

\begin{figure}
    \centering
    \includegraphics{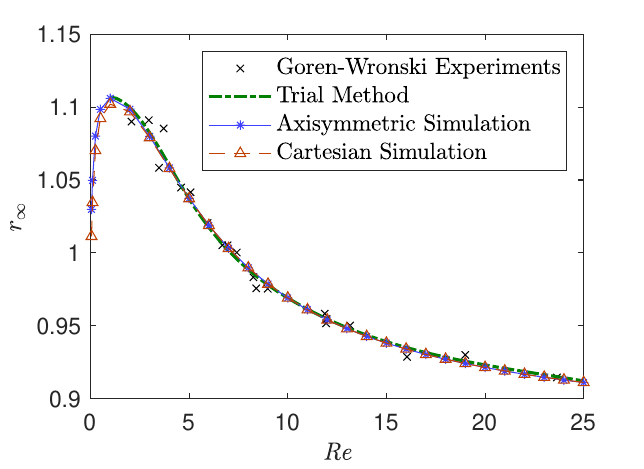}
    \caption{Experimental results of \citet{goren_shape_1966} ($\Oh \approx 1.83 $), results; simulation results of the Trial Method, reproduced with the method of \citet{reddy_finite_1978}; and simulation results in axisymmetric and 3D Cartesian implementations of die-swell ratio for varying $\Rey$ and $\textit{Oh} = \sqrt{3} \approx 1.73$. Cartesian simulations are conducted for $1.4\text{e}6$ DOFs for $\Rey<1$ and $1.1\text{e}6$ DOFs for $1 \leq \Rey$; axisymmetric simulations for $\sim 1\text{e}1$ DOFs}
    \label{fig:SwellRatio}
\end{figure}
In Figure \ref{fig:SwellRatio}, a quantitative comparison to the literature is made by plotting the predicted swell ratios against $\Rey$. Again, the Ohnesorge number is fixed at $\textit{Oh} = \sqrt{3}$. Simulations predict jet swelling for low $\Rey$, and contraction beyond $\Rey \sim 7.5$. Terminal radius predictions are within 0.6\% of one another with the difference being largest for very low $\Rey$. Moreover, the numerical results boast good agreement with the experimental results of \citet{goren_shape_1966} as well as with the simulation results from our own implementation of the Trial Method of \citet{reddy_finite_1978}. This confirms that present numerical results predict with precision swelling and contraction for Reynolds-number values varying from roughly 2 to 25.  Although systematic exploration of the influence of  $\Rey$ was stopped at a value of 25, an additional axisymmetric simulation was conducted for  $\Rey$ equal to 75; this yielded a die-swell ratio of 0.8854. Additional simulations were also conducted for $0.05 \leq \Rey \leq 2$.

\begin{figure}
    \centering
    \includegraphics{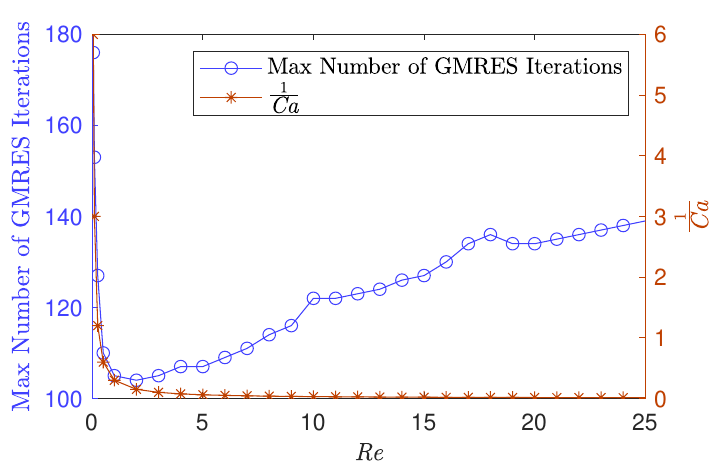}
    \caption{Maximum (out of all Newton iterations) number of GMRES-iterations required for termination in 3D Cartesian simulations of the die-swell problem; results are reported with respect to the inter-related Reynolds and Capillary numbers (for $\Oh = \sqrt{3}$).}
    \label{fig:GMRESIterations}
\end{figure}

Finally, in Table \ref{Table:DieSwell}, we display the computational performance of the 3D Cartesian simulation in the solution of the die-swell problem for varying Reynolds numbers. Times of computation include the computation of an initial guess, that is, the solution to a Stokes flow problem where the Young-Laplace BC has been omitted. This linear, saddle-point problem is solved with a preconditioned, minimum residual algorithm. All simulations were run on one processor in serial for 1'122'762 DOFs and all but one terminated within two hours of simulation time. Moreover, in Figure \ref{fig:GMRESIterations}, a plot of the maximum (out of all Newton iterations) number of GMRES iterations that were required for termination is displayed with respect to $\Rey$ and $\frac{1}{\Ca}$. The number of iterations is largest when the Reynolds and Capillary numbers are low. After dropping to its minimum when $\Rey$ and $\Ca$ are increased, the number of iterations steadily increases for growing Reynolds number.

\begin{table}
\centering
\caption{Assessment of computational performance of TLM on the Cartesian die-swell problem with respect to the inter-related Reynolds and Capillary numbers (for $\Oh = \sqrt{3}$). Computational times include the solution of a Stokes flow problem that served to find the initial guess. All simulations were carried out for $1'122'762 $ DOFs, a Newton tolerance of $10^{-5}$, and run in serial on one processor of the 12th Gen Intel(R) Core(TM) i7-1260P 2.10 GHz.}
    \begin{tabular}{ c c c c c c } \label{Table:DieSwell}
         & & Computation & \# Newton & Avg \# GMRES & Max \# GMRES \\
        $\Rey$ & $\Ca$ & time (hrs)& Iterations & Iterations & Iterations \\
        0.05 & 6.00 & 1.86 & 8 & 79.3 & 176 \\
        0.1 & 3 & 1.79 & 8 & 74.8 & 153 \\
        0.25 & 1.200 & 1.69 & 8 & 60.5 & 127 \\
        0.5 & 0.600 & 1.37 & 6 &  65.3 & 110  \\
        1 & 0.300 & 1.20 & 5 & 64.8 & 105 \\
        2 & 0.150 & 1.16 & 5 & 60.6 & 104 \\
        3 & 0.100 & 1.16 & 5 & 58.2 & 105 \\
        4 & 0.075 & 1.10 & 5 & 56.4 & 107 \\
        5 & 0.060 & 1.01 & 4 & 63.8 & 107 \\
        6 & 0.050 & 1.10 & 5 & 55.6 & 109 \\
        7 & 0.043 & 1.16 & 5 & 58.8 & 111 \\
        8 & 0.038 & 1.14 & 5 & 61.0 & 114 \\
        9 & 0.033 & 1.14 & 5 & 66.4 & 116 \\
        10 & 0.015 & 1.23 & 5 & 69.0 & 122 \\
        11 & 0.027 & 1.32 & 6 & 58.7 & 122 \\
        12 & 0.025 & 1.31 & 6 & 60.2 & 123 \\
        13 & 0.023 & 1.31 & 6 & 62.2 & 124 \\
        14 & 0.021 & 1.50 & 6 & 64.5 & 126 \\
        15 & 0.020 & 1.36 & 6 & 67.3 & 127 \\
        16 & 0.019 & 1.34 & 6 & 71.7 & 130 \\
        17 & 0.018 & 2.70 & 15 & 46.9 & 134 \\
        18 & 0.017 & 1.38 & 6 & 75.8 & 136 \\
        19 & 0.016 & 1.42 & 6 & 76.8 & 134 \\
        20 & 0.015 & 1.48 & 6 & 78.3 & 134 \\
        21 & 0.014 & 1.44 & 6 & 80.5 & 135 \\
        22 & 0.014 & 1.50 & 6 & 82.2 & 136 \\
        23 & 0.013 & 1.43 & 6 & 84.2 & 137 \\
        24 & 0.013 & 1.63 & 6 & 85.8 & 138 \\
        25 & 0.012 & 1.41 & 6 & 86.5 & 139 \\
    \end{tabular} 
\end{table}

\subsection{Thermo-capillary Flow} \label{Sec:ThermocapillarySimulations}

\begin{figure}
    \centering
    \includegraphics{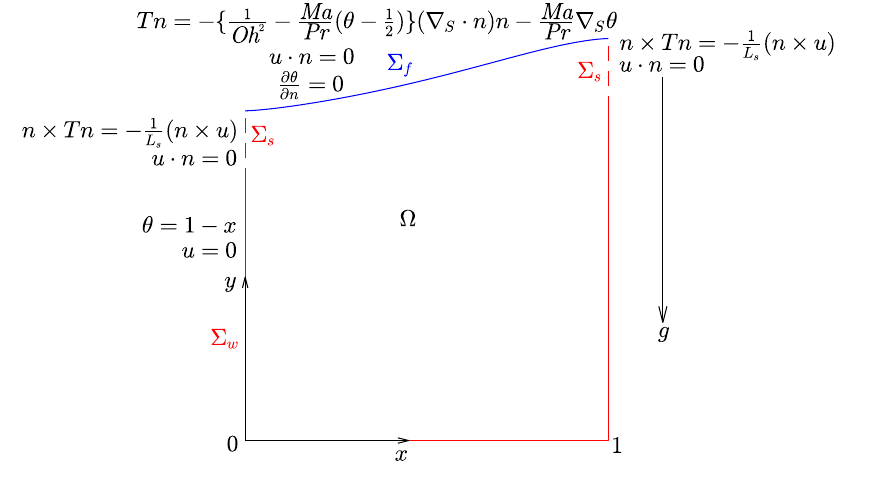}
    \caption{Domain of thermo-capillary container and boundary conditions. Note that the length of the slip regions is not to scale.}
    \label{fig:ThermocapillaryDomain}
\end{figure}

We now address thermo-capillary-driven flow of fluid in a motionless container, as depicted in Figure \ref{fig:ThermocapillaryDomain} and as originally studied by \citet{cuvelier_thermocapillary_1986}. A two-dimensional container with a fixed volume of fluid and an adiabatic free surface on top is subject to a (non-dimensionalised) temperature distribution $\theta = 1 - x$ on its walls. The density of the fluid decreases in the hotter region and increases in the cooler one; the ensuing density gradients affect pressure on the free surface, thus inducing deformation. In turn, they lead to clockwise, Grashof circulation patterns. Moreover, surface tension is taken to decrease as temperature goes up. Tension forces due to capillarity are therefore larger on the cold container side and, as a result, a force directed from hot (left) to cold (right) regions reigns on the free surface. This induces clockwise Marangoni circulation patterns, which, in turn, deform the free surface.

Grashof convection is modeled using the Boussinesq approximation. For surface tension, the Boussinesq approximation is not used in the Young-Laplace law, as it is not required to facilitate computation. As for the contact line, although it is static in steady flows, it must slide along the container surface during the transient so as to reach its steady state configuration. In order to capture this physical mechanism, a small slip region of non-dimensionalised length $L_s = 0.01$ with a Navier-slip BC is included at both contact lines \citep{baer_finite_2000}. As opposed to models without a slip region, this also enables one to impose the contact angle equation weakly. We note that, in this regard, our approach differs from that of \citet{cuvelier_thermocapillary_1986}, to which we refer the reader for details on the non-dimensionalisation. 

The physical system is modeled by the following set of equations.
\begin{equation} \label{Eq:ThermocapillaryStrongForm}
    \begin{cases}
        \text{Find $(\boldsymbol{u},p,\theta,\Sigma_f)\in V\times Q\times\Theta\times S$ such that }\\
        (\u\cdot \bnabla)\u = -\bnabla p + \bnabla^2 \u + \Gr (\theta - \frac{1}{2}) \boldsymbol{e}_y - \frac{\Bo}{\Oh^2}\boldsymbol{e}_y, \text{ in $\Omega$} \\
        \bnabla \cdot \u = 0,\text{ in $\Omega$} \\
        \u\cdot \bnabla\theta - \frac{1}{\Pr}\bnabla^2 \theta = 0,  \text{ in $\Omega$} \\
        \u\cdot \n = 0, \T \n = -\{\frac{1}{\Oh^2} - \frac{\Ma}{\Pr}(\theta-\frac{1}{2})\}(\bnabla_S\cdot \n)\n - \frac{\Ma}{\Pr}\bnabla_S\theta, \frac{\partial \theta}{\partial n} = 0,\text{ on $\Sigma_f$} \\
        \u\cdot\n = 0, \, \sr\cdot (\T\n_r) = - \frac{1}{L_s}(\u\cdot\sr), \,\theta = 1 - x, \text{ on $\Sigma_s$} \\
        \u = 0,\theta = 1 - x, \text{ on $\Sigma_w$} \\
        \int_\Omega dV = 1 \\
        \sf\cdot\sr = \cos(\phi) \\
    \end{cases}
\end{equation}
where $\theta$ and $\Theta$ denote temperature and its function space; $\phi$ denotes the contact angle; $L_s$ is the non-dimensionalised slip region length; $\Gr$, $\Bo$, $\Oh$, $\Pr$, $\Ma$ denote the Grashof, Bond, Ohnesorge, Prandtl and Marangoni numbers and are defined below. 
\begin{align}
\Gr =  \frac{\omega g \rho^2 L^3 \delta T}{\mu^2} && \Pr = \frac{c_p \mu}{k} && \Oh = \frac{\mu}{\sqrt{\rho \gamma L}} && \Bo = \frac{g\rho L^2}{\gamma} && \Ma = \frac{\zeta \gamma c_p \rho L\delta T}{\mu k}
\end{align}
where  $\omega$ is the volume expansion coefficient with respect to temperature; $g$ is the gravitational acceleration; $L$ is the base of the initially square simulation area (see Figure \ref{fig:ThermocapillaryDomain}); $\delta T$ is a characteristic temperature difference between the heated and cooled sides; $c_p$ is specific heat; $k$ is thermal conductivity; and $\zeta$ is the temperature coefficient of surface tension. The weak form of Equation (\ref{Eq:ThermocapillaryStrongForm}) follows.

\begin{equation} \label{Eq:ThermocapillaryWeakForm}
    \begin{cases}
        \text{Find $(\boldsymbol{u},p,\theta,\Sigma_f)\in V\times Q\times\Theta\times S$ such that }\\
        \begin{split}\int_\Omega\{ \v\cdot (\u\cdot \bnabla)\u &+ \T:\bnabla \v - \Gr \, \theta \, v_y \}dV + \int_{\Sigma_s} \frac{1}{L_s}u_y\,v_y\,dS \\
        - \frac{\Ma}{\Pr}\int_{\Sigma_f} &\theta(\bnabla_S\cdot \v)\,dS + \frac{\Ma}{\Pr}\, \cos(\phi)\int_{\partial \Sigma_f} \theta \,v_y\,dl \\
        = &- (\frac{\Gr}{2} + \frac{\Bo}{\Oh^2}) \int_\Omega v_y\,dV \\
        &- ( \frac{1}{\Oh^2} + \frac{\Ma}{2\Pr} ) \int_{\Sigma_f} (\bnabla_S\cdot \v) dS \\
        &+ ( \frac{1}{\Oh^2} + \frac{\Ma}{2\Pr} )\,\cos(\phi) \int_{\partial \Sigma_f} v_y\,dl, \,\forall \v\in V
        \end{split} \\
        \\
        \int_{\Omega} (\bnabla \cdot \u)q dV = 0,\text{ $\forall q\in Q$} \\
        \\
        \int_\Omega ( \u\cdot \bnabla\theta + \frac{1}{\Pr}\bnabla\theta \cdot \bnabla\eta ) dV = 0, \,\forall \eta\in \Theta \\
        \\
        \int_{\Sigma_f} \chi(\u\cdot\n)dS = 0,\,\forall \chi\in M \\
        \\
        \int_\Omega dV = 1 \\
    \end{cases}
\end{equation}
The linearisation of the momentum equation is identical to that of the Navier-Stokes equations, albeit with three additional temperature terms. The terms involving gravity are linearised trivially using Relation (\ref{Eq:VolumeIntegrals}). The two terms ensuing from Marangoni flow can be linearised by directly applying Relation (\ref{Eq:SurfaceDivergenceLinearizationFinal}) from Appendix \ref{appB} and Result (\ref{Eq:LinearizedDynamicEquationResult}) for the contact-angle term. The energy equation may be linearised trivially with Relation (\ref{Eq:VolumeIntegrals}). 

We solve for the normal component of displacement $\alpha := \d\cdot\n$ on $\Sigma_{f,0}$, and its tangent component $\beta = \d\cdot\sf$ on $\Sigma_{f,0}$ so as to accommodate contact angles different from $90^\circ$. Thus, $\alpha$ and $\beta$ are the normal and tangent components of displacement along the free surface. The problem is now stated in weak form as follows.

\begin{equation} \label{Eq:LinearizedThermocapillaryWeakForm}
    \begin{cases}
        \text{Find $(\boldsymbol{u},p,\theta,\alpha,\beta,\lambda)\in V_h\times Q_h\times\Theta_h\times M_{\alpha,h} \times \mathbb{R}^2\times \mathbb{R}$ such that }\\
        \begin{split}
        \int_{\Omega_0}\{ \v\cdot (\u_0\cdot \bnabla)\u &+ \T:\bnabla \v - \Gr \, \theta \, v_y \}dV  + \int_{\Sigma_{s,0}} \frac{1}{L_s}u_y\,v_y\,dS \\
        +\int_{\Sigma_{f,0}}\alpha \{ \v\cdot (&\u_0\cdot \bnabla_S)\u_0 + \T_0:\bnabla_S \v - \Gr \, (\theta_0 - \frac{1}{2}) \, v_y + \frac{\Bo}{\Oh^2}\,v_y \}dS \\
        + \int_{\partial \Sigma_{s,0}}&\frac{1}{L_s} u_{0,y}\,v_y \{ \alpha\, \sin(\phi) + \beta\,\cos(\phi)\}\,dl \\ 
        + \int_{\Sigma_{f,0}} &\{- \frac{\Ma}{\Pr} \theta(\bnabla_S\cdot \v) + \{\frac{1}{\Oh^2}- \frac{\Ma}{\Pr} (\theta_0-\frac{1}{2})\} (\bnabla_S\v\cdot \n)\cdot \bnabla_S\alpha \}\,dS \\
        + \frac{\Ma}{\Pr}\, &\cos(\phi)\int_{\partial \Sigma_{f,0}} \theta \,v_y\,dl \\
        = & - \int_{\Omega_0} (\frac{\Gr}{2} + \frac{\Bo}{\Oh^2}) v_y \,dV \\
        &- ( \frac{1}{\Oh^2} + \frac{\Ma}{2\Pr} ) \int_{\Sigma_{f,0}} (\bnabla_S\cdot \v) dS \\
        &+ ( \frac{1}{\Oh^2} + \frac{\Ma}{2\Pr} )\,\cos(\phi) \int_{\partial \Sigma_{f,0}} v_y\,dl, \, \forall \v\in V_h
        \end{split} \\
        \\
        \int_{\Omega_0} (\bnabla \cdot \u)q dV = 0,\text{ $\forall q\in Q_h$} \\
        \\
        \int_{\Omega_0} \{ \eta(\u_0\cdot \bnabla\theta) + \frac{1}{\Pr}\bnabla\theta \cdot \bnabla\eta \} dV = 0, \,\forall \eta\in \Theta_h \\
        \\
        \int_{\Sigma_f} \{\chi(\u\cdot\n) + \alpha(\u_0\cdot \bnabla_S)\chi + \lambda \chi \} dS = 0,\,\forall \chi\in M_{\alpha,h} \\
        \\
        \alpha \,\cos(\phi) - \beta \,\sin(\phi) = 0, \text{ on $\partial \Sigma_{f,0}$} \\
        \\
        \int_{\partial \Sigma_{f,0}} \alpha \,dS = 1 - \int_{\Omega_0}dV \\
    \end{cases}
\end{equation}
We observe that, while $\alpha$ is present on $\Sigma_{f,0}$, $\beta$ appears in the weak form only in integrals over $\partial \Sigma_{f,0}$, a discrete set of two points in space. As such, it is only necessary to solve for $\beta$ at these two points, i.e. $\beta\in \mathbb{R}^2$. $\beta$ is then extended to $\Sigma_{f,0}$ in the strain-minimisation step (Equation (\ref{Eq:DisplacementProjectionBeta})). $\lambda$ is a Lagrange multiplier of $\alpha$ that is tied to the volume constraint. 

We carry out Picard iteration rather than Newton iteration over the advection terms in the Boussinesq equations \citep{rebholz_picard-newton_2024} so as to decouple $\u$ from the energy equation. While $\alpha$, $\beta$, $\lambda$ and $p$ retain an exact first-order expansion in this case, such is no longer the case for $\u$ and $\theta$ and this may cause the method to converge linearly. On the other hand, the initial guess may be poorer in Picard iteration, as shown by \citet{rebholz_picard-newton_2024}. Last but not least, the decoupling of $\theta$ from $(\u,p,\alpha,\beta,\lambda)$ allows the use of the preconditioner given in Section \ref{Sec:Preconditioner} and, in turn, of iterative solution, hence increasing the potential system size and speed of solution.

The domain is discretised into a mesh of linear, triangular elements. Taylor-Hood elements are used for velocity and pressure \citep{taylor_numerical_1973}. Quadratic Lagrange elements are implemented for $\alpha$ and $\theta$. The residual tolerance is fixed to be $\epsilon = 10^{-4}$ on variables $\u$, $\theta$, $\alpha$ and $\beta$. 

\begin{figure}
    \centering
    \subfigure[]{\includegraphics[width=0.495\linewidth]{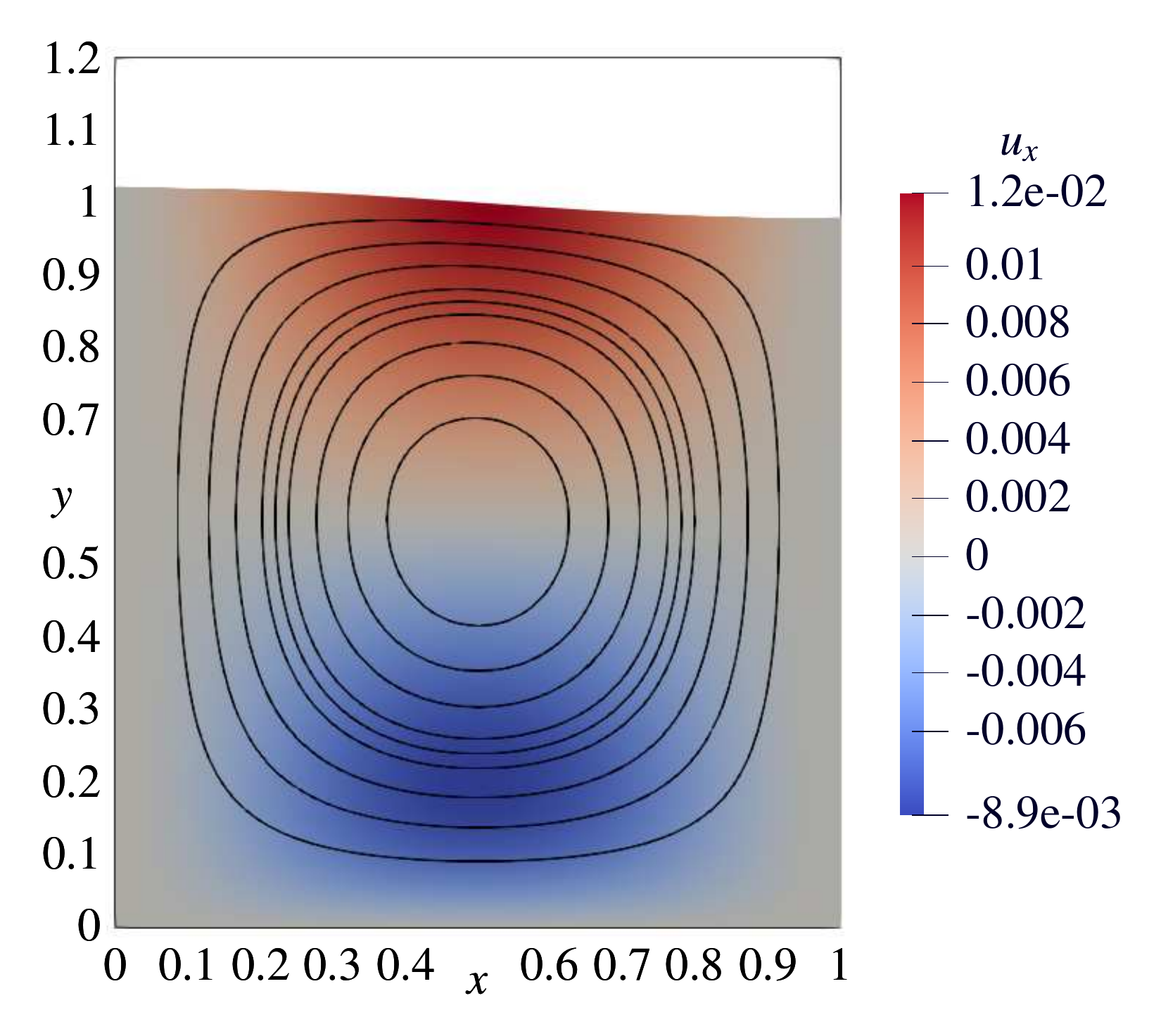}}
    \subfigure[]{\includegraphics[width=0.495\linewidth]{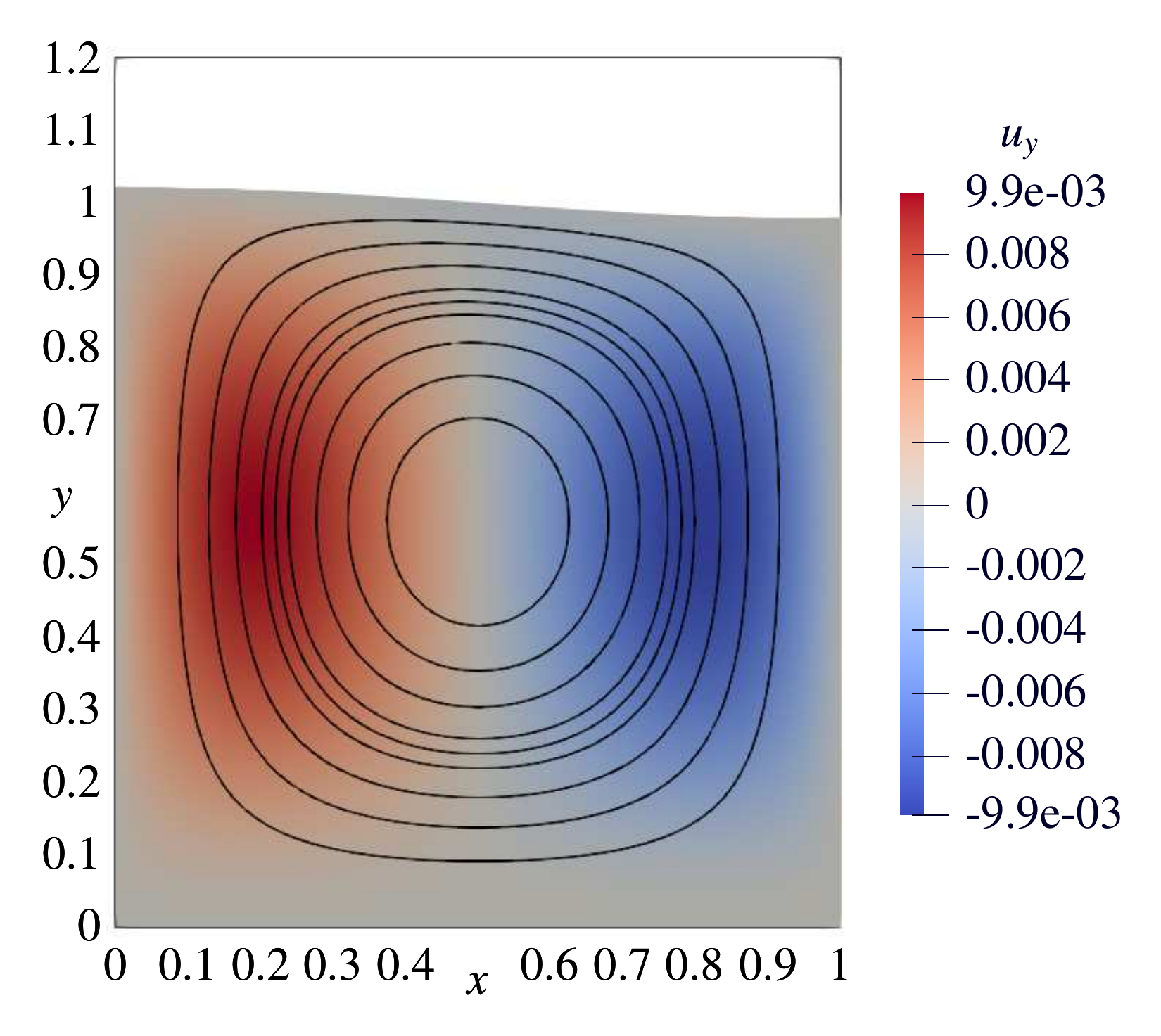}}
    \caption{Grashof-induced velocity distribution with streamlines (black) (a) $u_x$ (b) $u_y$; for parameters $\Gr = 2$, $\Pr = 0.73$, $\Ma = 0$, $\Bo = 0$, $\Oh = 1$  }
    \label{fig:Grashof2}
\end{figure}

\begin{figure}
    \centering
    \subfigure[]{\includegraphics[width=0.495\linewidth]{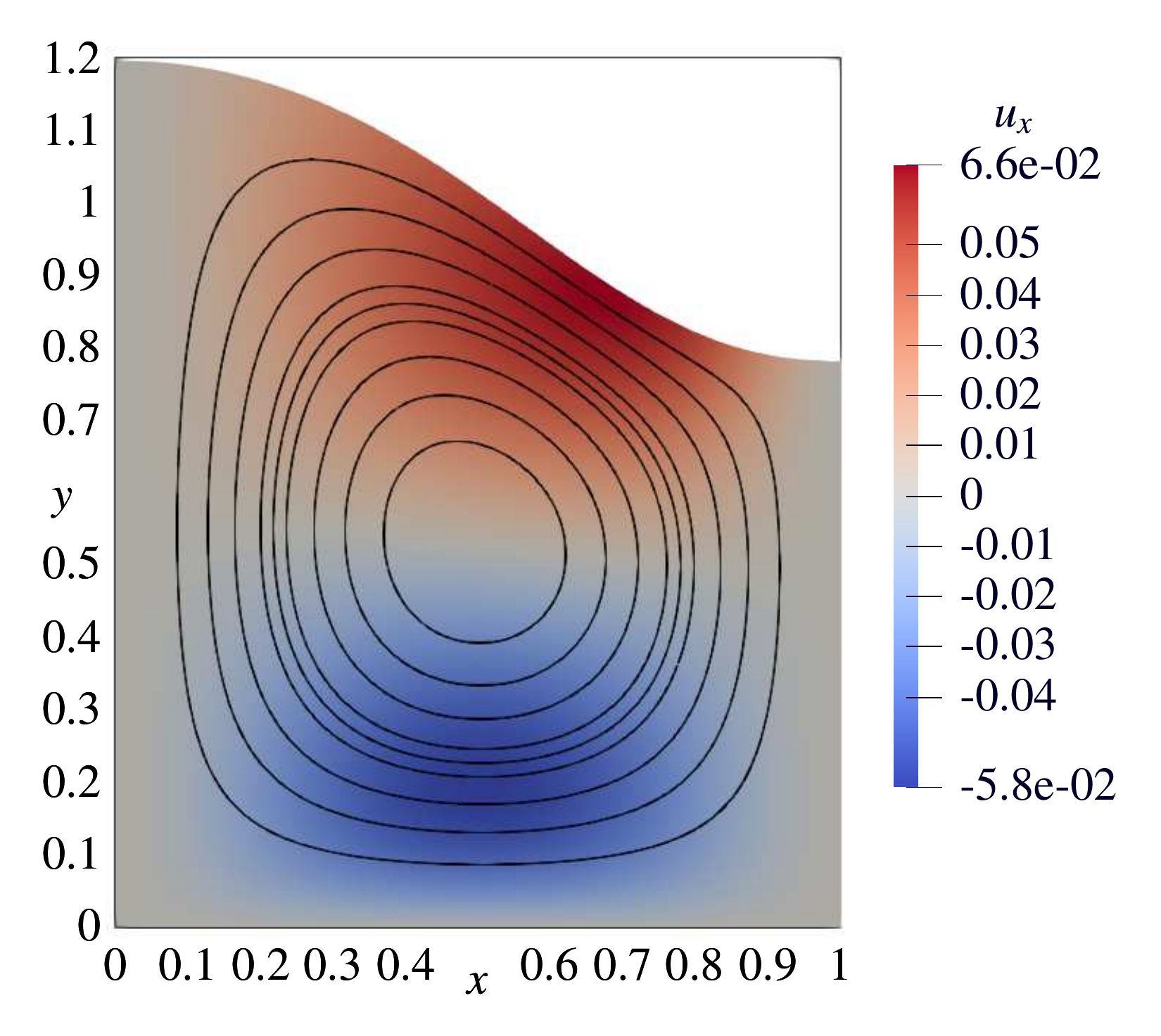}}
    \subfigure[]{\includegraphics[width=0.495\linewidth]{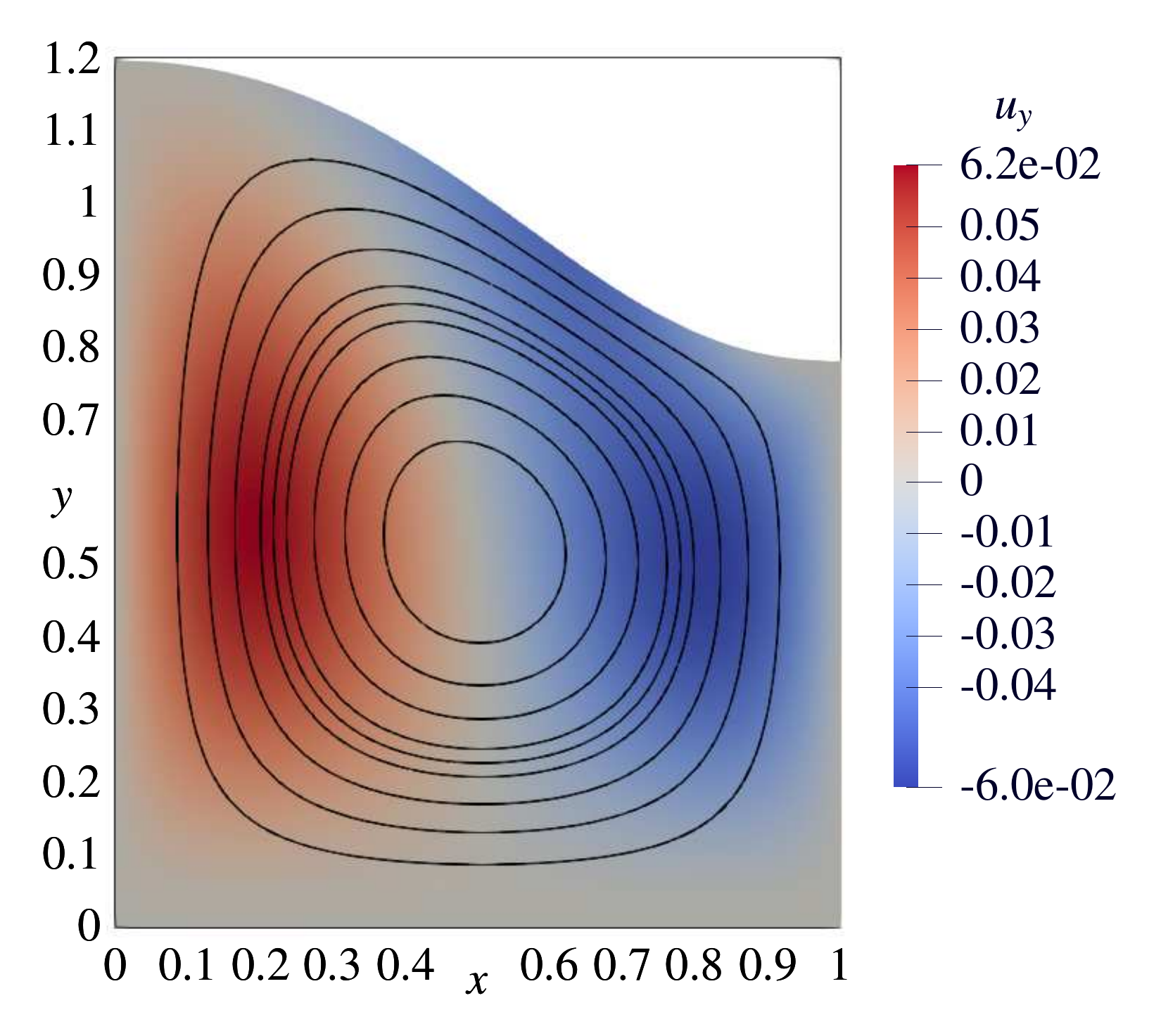}}
    \caption{Grashof-induced velocity distribution with streamlines (black) (a) $u_x$ (b) $u_y$; for parameters $\Gr = 14$, $\Pr = 0.73$, $\Ma = 0$, $\Bo = 0$, $\Oh = 1$  }
    \label{fig:Grashof14}
\end{figure}

To compare to results in the study by \citet{cuvelier_thermocapillary_1986}, we first consider simulations carried out with a contact angle $\phi = \frac{\pi}{2}$. In Figures \ref{fig:Grashof2} and \ref{fig:Grashof14}, the velocity distribution for convection induced solely by buoyancy is displayed for Grashof numbers $\Gr = 2$ (Figure \ref{fig:Grashof2}) and $\Gr = 14$ (Figure \ref{fig:Grashof14}). Numerical continuation was required for $\Gr = 14$. In both cases, the free surface rises on the hot wall and sinks on the other side while clockwise Grashof convection takes place. Circulation is more vigorous for $\Gr = 14$ and engenders significant deformation, with a peak deformation of roughly 20\%. The streamlines, quasi symmetric for $\Gr = 2$, are significantly altered in the vicinity of the free surface due to the latter's deformation for $\Gr = 14$, thus further breaking symmetry.

\begin{figure}
    \centering
    \includegraphics[width = 0.75\linewidth]{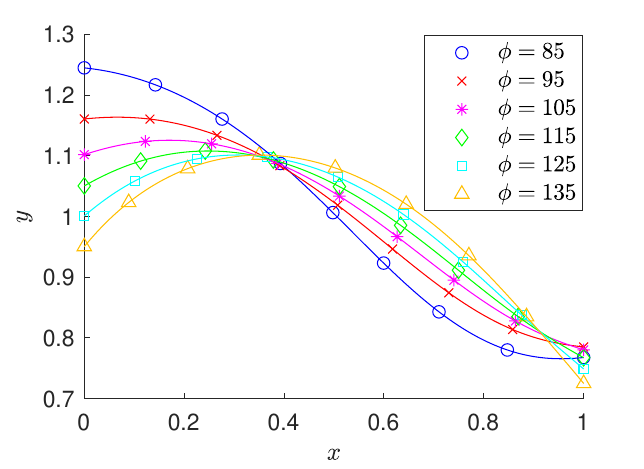}
    \caption{Free surface predictions for varying contact angle $\phi$ in Grashof-induced convection with parameters $\Pr = 0.73$, $\Gr = 14$, $\Bo = 0$, $\Oh = 1$  }
    \label{fig:Grashof14Angles}
\end{figure}

Exploring now the influence of $\phi$, Figure \ref{fig:Grashof14Angles} shows, for $\Gr = 14$, the free-surface dependence on the contact angle $\phi$. Numerical continuation starting from the solution for $\phi = 90^\circ$ is used in order to simulate Grashof convection under different wetting conditions. We observe that the amount of surface in contact with the container reduces as the contact angle increases; this is consistent with the physics of wetting. When performing numerical continuation, the farther the contact angle was from $90^\circ$, the smaller the angle increment had to be. At both ends of the spectrum of angles displayed in Figure \ref{fig:Grashof14Angles}, numerical continuation became intractable as the angle increment became prohibitively small.  

\begin{figure}
    \centering
    \subfigure[]{\includegraphics[width=0.495\linewidth]{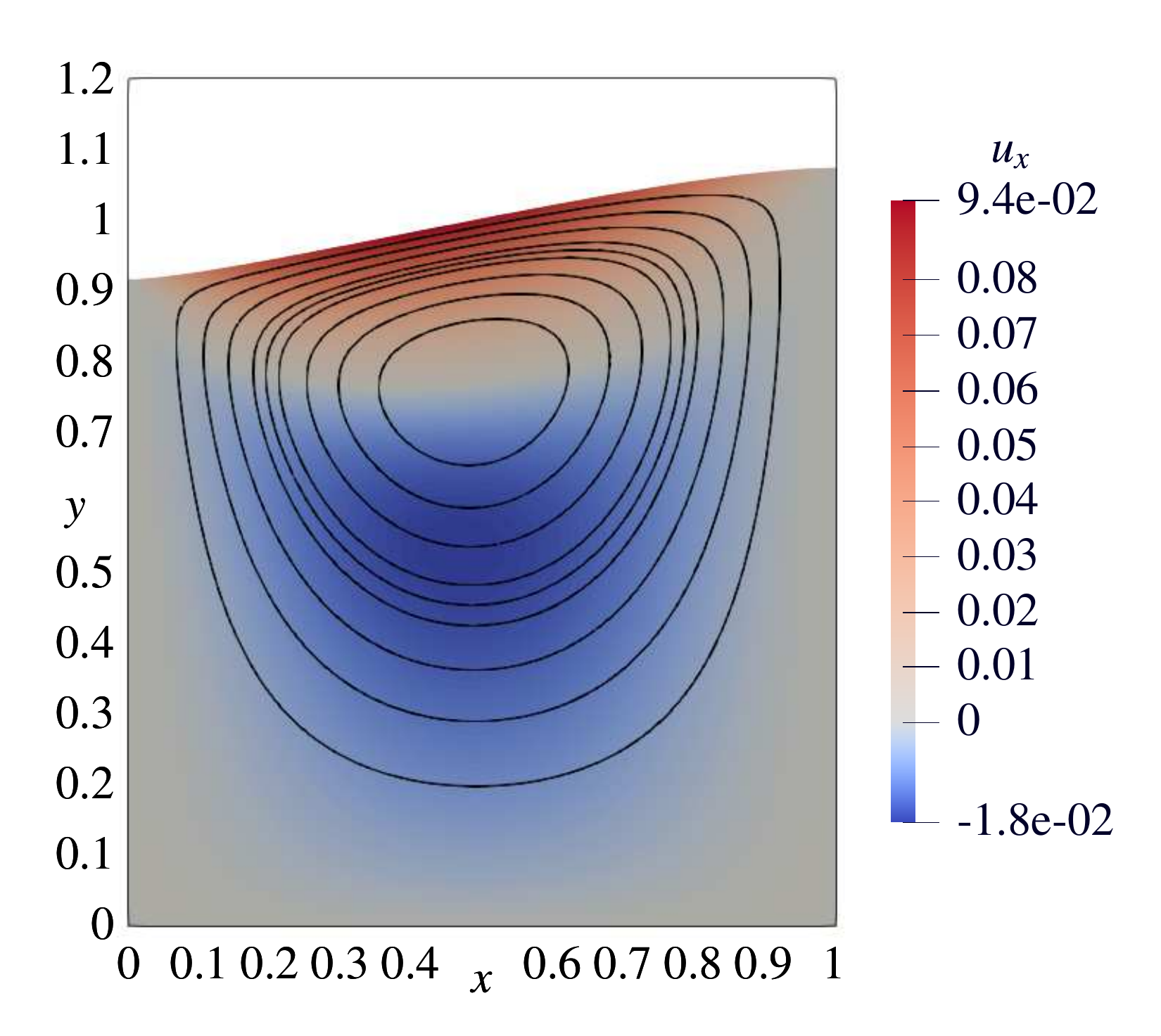}}
    \subfigure[]{\includegraphics[width=0.495\linewidth]{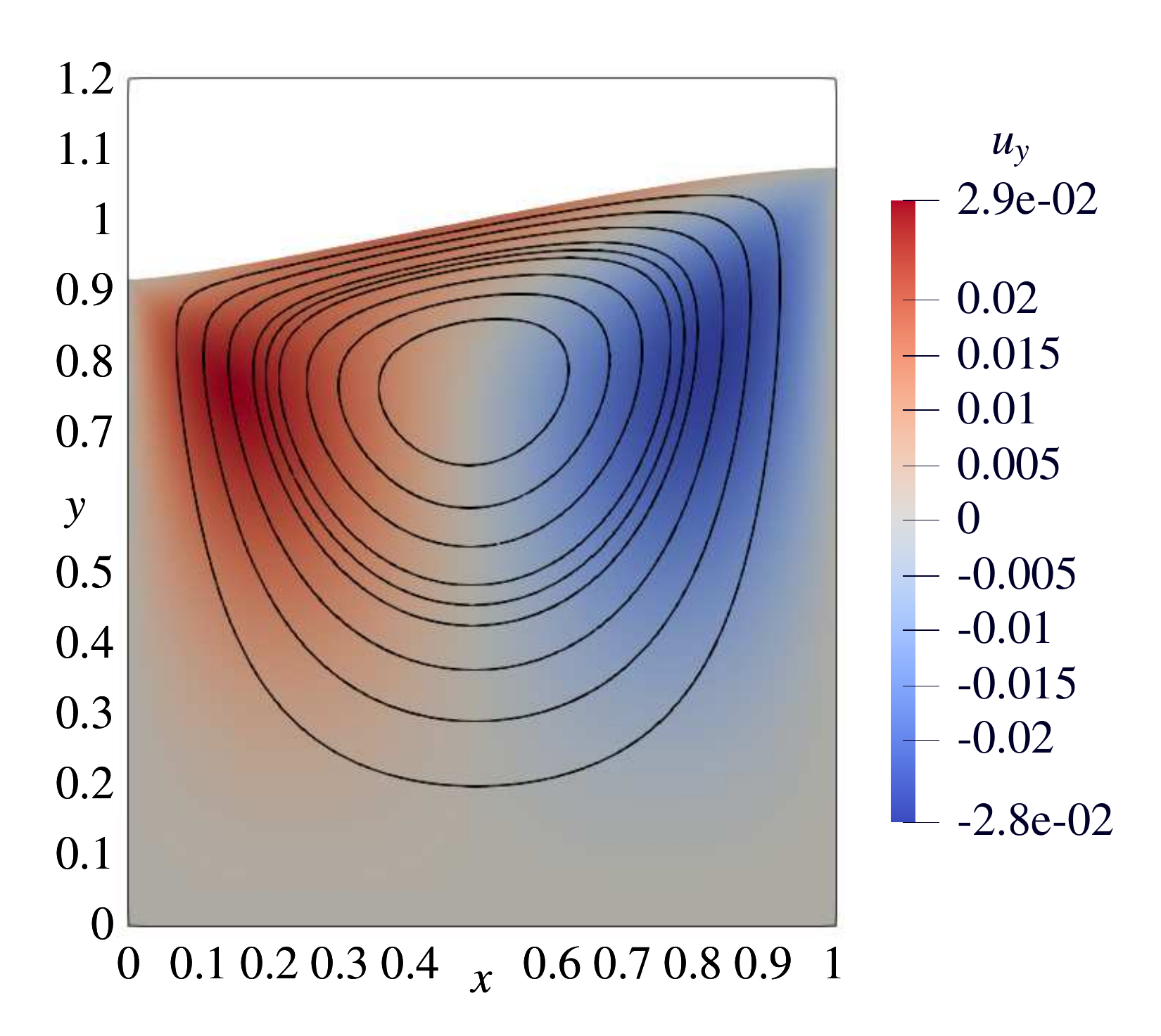}}
    \caption{Marangoni-induced velocity distribution with streamlines (black) (a) $u_x$ (b) $u_y$; for parameters $\Gr = 0$, $\Pr = 0.73$, $\Ma = 0.5$, $\Bo = 0$, $\Oh = 1$  }
    \label{fig:Marangoni0.5}
\end{figure}

\begin{figure}
    \centering
    \subfigure[]{\includegraphics[width=0.495\linewidth]{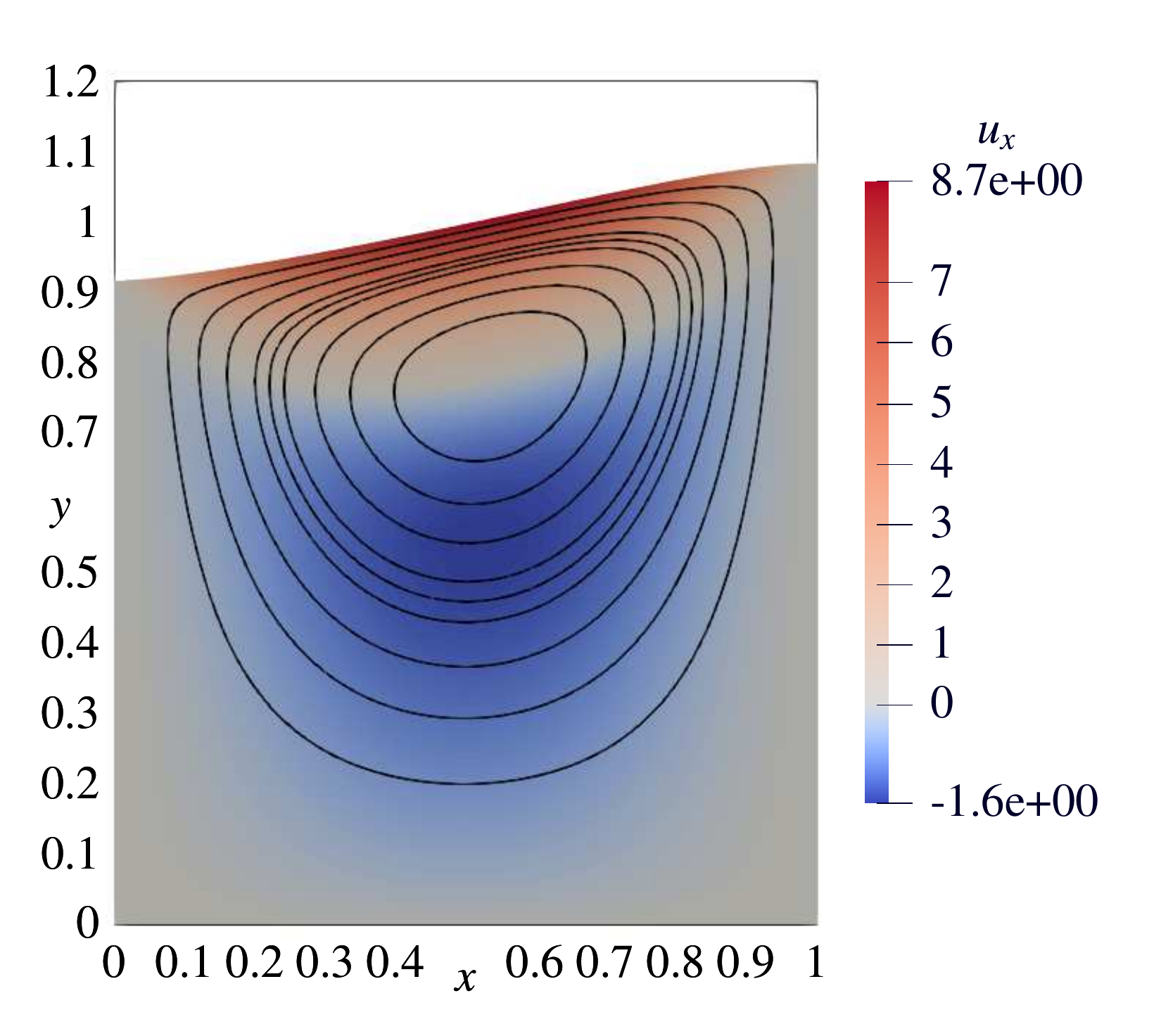}}
    \subfigure[]{\includegraphics[width=0.495\linewidth]{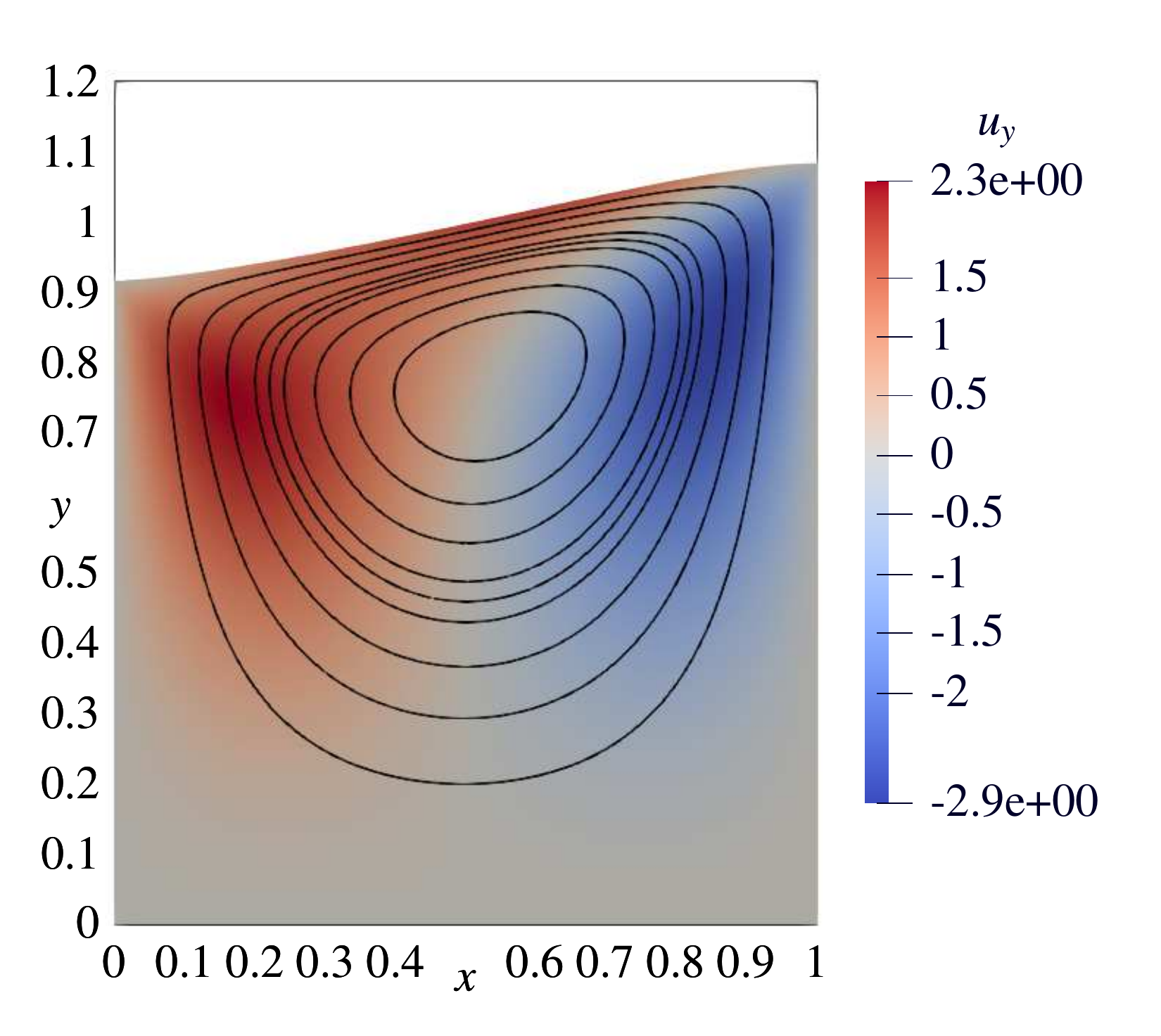}}
    \caption{Marangoni-induced velocity distribution with streamlines (black) (a) $u_x$ (b) $u_y$; for parameters $\Gr = 0$, $\Pr = 0.73$, $\Ma = 50$, $\Bo = 0$, $\Oh = 0.1$  }
    \label{fig:Marangoni50}
\end{figure}

In Figure \ref{fig:Marangoni0.5} where $\Ma = 0.5$, and Figure \ref{fig:Marangoni50} where $\Ma = 50$, emphasis is put on the Marangoni effect. Without effects of buoyancy, i.e. $\Gr = 0$, and when $\Ma > 0$, convection is driven solely by surface-tension gradients. Although circulation is once again clockwise, the free surface now adopts the opposite shape, thus rising on the cold side and falling on the other. Fluid flow is also more localised to the free surface, it being the primary source of momentum. 

\begin{figure}
    \centering
    \includegraphics[]{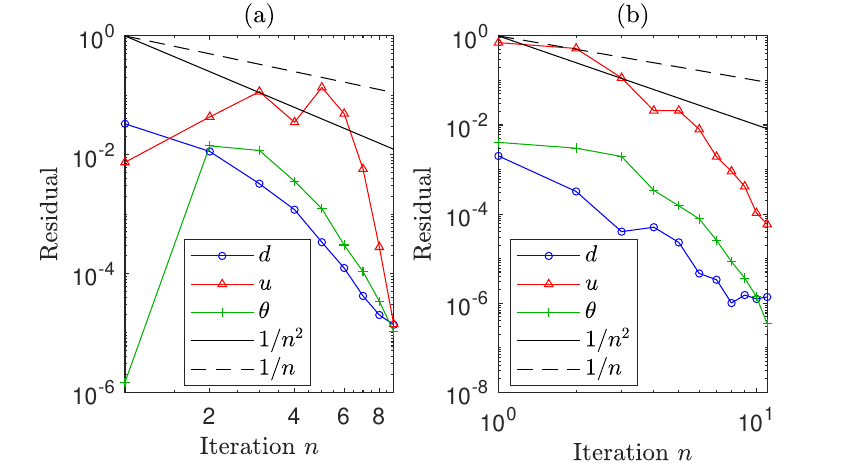}
    \caption{Convergence residuals for (a) $\Oh = 1$, $\Pr = 0.73$, $\phi = 85^\circ$, $\Gr = 14$, $\Ma = 0$ (b) $\Oh = 0.1$, $\Pr = 0.73$, $\phi = 90^\circ$, $\Gr = 1200$, $\Ma = 50$}
    \label{fig:ThermocapillaryResiduals}
\end{figure}

Solution residuals for two distinct cases are depicted in Figure \ref{fig:ThermocapillaryResiduals}, including an example of Grashof convection for contact angle $\phi = 85^\circ$ and a mixed convection problem. While residuals tend to decrease rather slowly across the first few iterations, their rate of convergence drastically accelerates across final iterations, even exceeding quadratic decay despite the fact that a linear scheme is employed to iterate over nonlinear advection terms. In both cases, velocity was the last unknown to meet the tolerance requirement.

In Table \ref{Table:Thermocapillary}, we compare our predictions of the height of each contact line to those of \citet{cuvelier_thermocapillary_1986}. We observe that in surface-tension-driven flow, the discrepancy is roughly $10^{-2}$ while for buoyancy-driven flow, it is one order of magnitude smaller.

\begin{figure}
    \centering
    \subfigure[]{\includegraphics[width=0.495\linewidth]{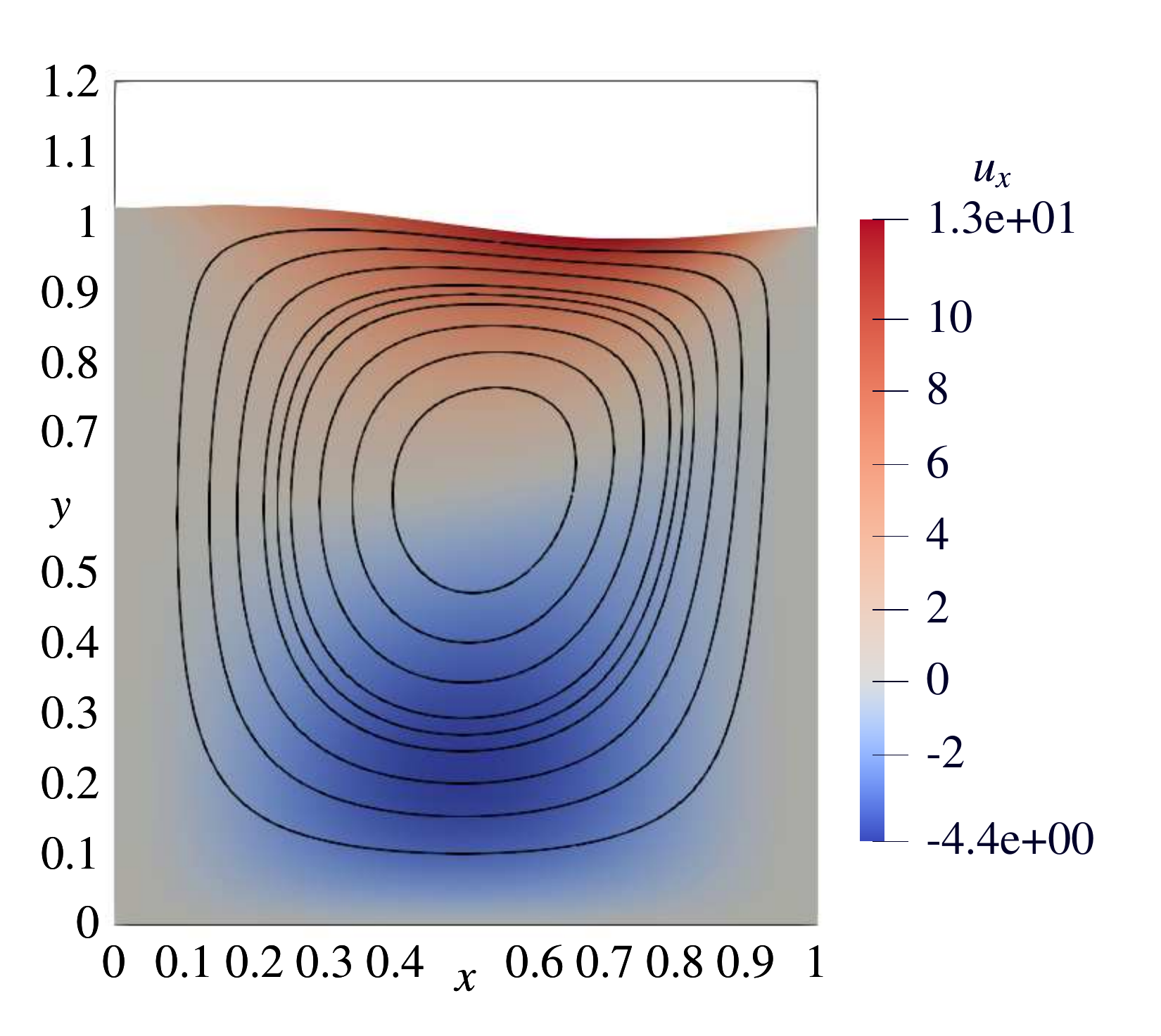}}
    \subfigure[]{\includegraphics[width=0.495\linewidth]{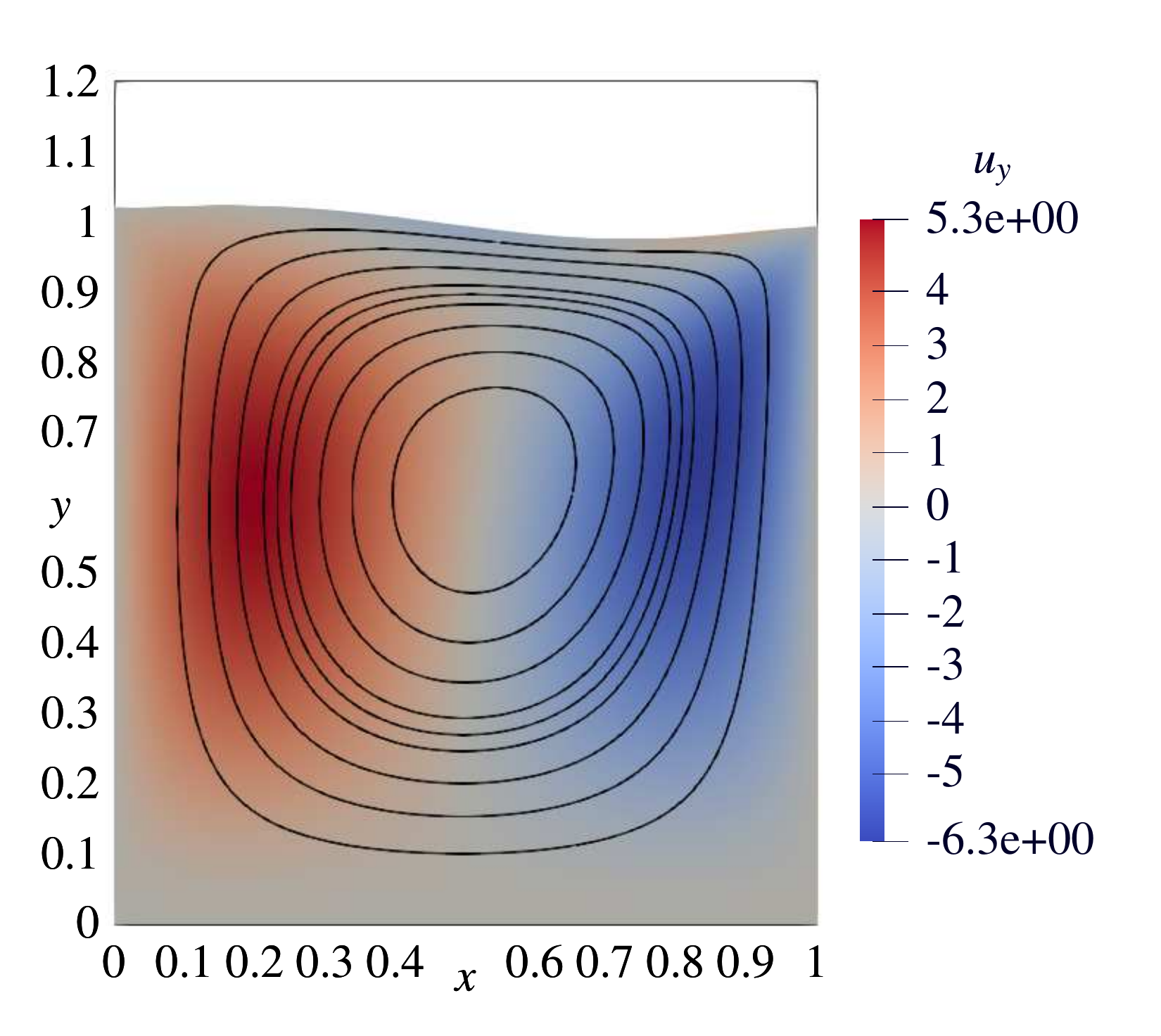}}
    \caption{Grashof and Marangoni-driven velocity distribution with streamlines (black) (a) $u_x$ (b) $u_y$; for parameters $\Gr = 900$, $\Pr = 0.73$, $\Ma = 50$, $\Bo = 0$, $\Oh = 0.1$  }
    \label{fig:Gr900Marangoni50}
\end{figure}

\begin{table}
\centering
\caption{Absolute difference between left and right contact-line heights predicted by TLM and those of \citet{cuvelier_thermocapillary_1986}, $|\Delta h_l| $ and $|\Delta h_r|$, for $\Pr = 0.73$, $\Bo = 0$ and $\phi = \frac{\pi}{2}$}
    \begin{tabular}{ c c c c c c } \label{Table:Thermocapillary}
        $\Gr$ & $\Ma$ & $\Oh$ & $\Bo$ & $ |\Delta h_l|$ & $ |\Delta h_r|$ \\
        2 & 0 & 1 & 0 & 0.000 & 0.001 \\
        14 & 0 & 1 & 0 & 0.008 & 0.009 \\
        0 & 0.5 & 1 & 0 & 0.028 & 0.011 \\
        1200 & 0 & 0.1 & 0 & 0.003 & 0.006 \\
        1200 & 0 & 0.1 & 1 & 0.003 & 0.003 \\
        1200 & 0 & 0.1 & 1000 & 0.000 & 0.000 \\
        0 & 50 & 0.01 & 0 & 0.001 & 0.000 \\
        0 & 50 & 0.1 & 0 & 0.025 & 0.007 \\
    \end{tabular} 
\end{table}
Finally, in Figure \ref{fig:Gr900Marangoni50}, the results for the velocity distribution of a mixed convection problem where $\Gr = 900$ and $\Ma = 50$ are displayed. Above, it was pointed out that, while buoyancy effects tend to raise the hot side of the free surface, surface-tension effects do the opposite; Figure \ref{fig:Gr900Marangoni50} depicts the interplay between these two convection mechanisms. While surface tension drives motion from hot to cold, thus causing the free surface to rise to the right, buoyancy drives the interface up on the left-hand side and tends to lower it on the cold side. The opposing behaviours of these convection mechanisms lead to the interface undulation present in Figure \ref{fig:Gr900Marangoni50}.

\section{Discussion}

The linearised problem (\ref{Eq:LinearizedWeakFormReduced}) derived in Section \ref{Sec:Linearisation} provides a weak form that includes fluid variables $\u$, $p$, $p_a$ as well as displacement $\d$. Unlike earlier methods that contain the latter in volume integrals and therefore require a non-physical equation governing $\d$ in the bulk, the present formulation is significantly more efficient due to the fact that $\d$ is present in the weak form only on the free surface; specifically, only through its normal component $\alpha = \d\cdot \n$. Moreover, on $\partial \Sigma_{f,0}$, $\alpha$ and tangent component $\beta = \d\cdot\sr$ are the only components to appear at first order. Thus, effectively, considering that $\sr$ is the normal unit vector of curve $\partial \Sigma_{f,0}$ along rigid surface $\Sigma_{r,0}$, the first-order dependence of any surface or curve on displacement is only on its normal vector. We believe that this reduction of free-surface-flow problems is optimal in the sense that it reduces the presence of $\d$ in the governing equations to its physical essence. 

The implications of this reduction are three-fold: 

(i) The size of the linear system at each Newton iteration is nearly halved and this significantly increases the limits of modeling precision while speeding up computation. 

(ii) An approximation of the ideal preconditioner in Equation (\ref{Eq:IdealPreconditioner}) is possible. To be more specific, direct inversion of the lumped block
\begin{equation} \label{Eq:LumpedBlock}
\begin{pmatrix}
    \H + \E\hat{\boldsymbol{S}}\C & \E\hat{\boldsymbol{S}} \boldsymbol{L} & \boldsymbol{N} + \E \hat{\boldsymbol{S}} \boldsymbol{Z} \\
    \boldsymbol{J} & \boldsymbol{W} & 0 \\
    \K & 0 & \G
\end{pmatrix}
\end{equation}
is tractable in Step 1 of Algorithm (\ref{Eq:PreconditionerInversionAlgorithm}) since the unknown, $\d$, possesses DOFs only on the free surface and not in the bulk. Otherwise, this system would be enlarged by an additional system of bulk equations for the 3D displacement vector field (e.g. strain-minimisation Problem (\ref{Eq:DisplacementStrainMinimization})), which would rapidly become prohibitively large for direct solution. Exact inversion would still be possible, but only with inner-iterations that would make GMRES prohibitively slow. An alternative approach could be to use an algebraic preconditioner for the enlarged block system but this would further worsen the approximation, leaving then no guarantee that the approximation would remain sufficiently accurate for iterative solution.

(iii) Unknowns with homogeneous Neumann BCs along the free surface are decoupled from displacement. Indeed, given $\d$'s presence only on $\Sigma_{f,0}$, in the event of the BC in question, $\d$ completely vanishes at first order from the weak form. This is the precise reason why the temperature equation in thermo-capillary Problem (\ref{Eq:LinearizedThermocapillaryWeakForm}) is independent of $\alpha$. In that particular case, when combined with a Picard iteration scheme, one succeeds in completely decoupling temperature from $\u$, $p$, $\alpha$ and $\beta$, albeit at the price of linear convergence.

Given these three attributes, simulation of cylindrical die swell, modeled in 3D Cartesian coordinates, is made possible on one processor, as demonstrated by the results in Section \ref{Sec:DieSwellSimulations}. The comparison of 3D and axisymmetric TLM results, to both Trial-Method simulations and experiments in Figure \ref{fig:SwellRatio}, validates the extension from 2D to 3D of the TLM carried out in Section \ref{Sec:Linearisation}. 

Agreement is verified over a spectrum of parameters that include the interplay between advection, viscosity and capillarity. These three effects each give rise to a specific behaviour, including jet contraction in the case of inertia-dominated flow, for $10\lesssim\Rey$; jet swelling for viscous flow, when $2.5 \lesssim \Rey  \lesssim 5$; and conforming of the interface to the minimal surface, for $\Rey \lesssim 2.5$. 

In the inertia-dominated case, at the pipe outlet, momentum is heavily concentrated at the centre of the channel. Since viscosity-driven diffusion of momentum to the free surface is weak, rather than being accelerated by viscous diffusion, the free surface contracts as it is dragged by the high-momentum centre of the jet and is then accelerated by virtue of mass conservation. We note in passing that the swell ratio that is reached at $\Rey = 25$ is far from the $\frac{\sqrt{3}}{2}$ limit that one derives analytically by assuming conservation of kinetic energy \citep{goren_shape_1966} free of surface tension contributions (justified in the present simulation series by $\Ca \overset{\Rey\rightarrow \infty}{\longrightarrow}\infty$ at fixed $\Oh$): the transition from Poiseuille to slug flow  at $\Rey \leq 25$ comes with significant energy dissipation at the die exit.  At higher Reynolds numbers this dissipation apparently vanishes, given that the die-swell ratio of 0.8854 predicted for $\Rey = 75$ comes very near the inviscid $\frac{\sqrt{3}}{2}= 0.8660$ limit.

For highly-viscous flow, the pipe's wall induces shear stress or, in other words, vorticity in the fluid. At the stick-slip transition, this results in an outward-pushing stress on the free surface, driving the fluid to swell as it exits the die. This is counteracted by the inertia-induced propensity to contract, which eventually wins out when $10\lesssim\Rey$.

As for the high-capillarity limit, we point out that, in the numerical examples explored in Section \ref{Sec:DieSwellSimulations}, $\Ca \propto \Rey$ and, therefore, $\Rey \rightarrow 0$ implies $\Ca \rightarrow 0$. The low-Reynolds limit may then be interpreted as the limit where the mean cross-sectional velocity $U \rightarrow 0$. Surface-tension forces dominate in this limit and insist on the free surface remaining a cylinder of radius $r = 1$. This behaviour is visible in Figure \ref{fig:SwellRatio} for $\Rey \lesssim 2.5$ where viscous swelling is observed to decline for decreasing $\Rey$. We stress however that, while simulation is capable of pinning the interface on the pipe corner, in experiments, for low Reynolds numbers, the triple line tends to slide away from the pipe corner, thus wetting the pipe edge \citep{goren_shape_1966}. In this limit, the steady flows predicted by simulation may be unstable and therefore not observed in practice.

Minor differences between 3D and axisymmetric swell-ratio predictions are also to be noted in Figure \ref{fig:SwellRatio} for low Reynolds numbers. We believe that these differences arise due to the relative coarseness of the 3D mesh and that they would be eradicated by mesh refinement. Considering the very limited amount of computational resources used here and that meshes were refined locally at the stick-slip transition by the user rather than via a criterion based on rigorous error estimation, one can imagine several straightforward improvements of calculations. Local, \textit{ad-hoc} mesh refinement based on, for instance, the Zienkiewicz-Zhu estimator \citep{picasso_anisotropic_2003}, could serve to better allocate resources to more sensitive domain regions. In addition, parallelisation may enable the distribution of memory across several processors in order to increase the number of DOFs. As highlighted by Table \ref{Table:DieSwell}, computational time is not a severe issue and, moreover, iterative solution, which is enabled by the Preconditioner given in Section \ref{Sec:Preconditioner}, ensures that it scales moderately with the number of DOFs, when used in the appropriate range of parameters. 

In Figure \ref{fig:GMRESIterations}, one observes that for low Reynolds numbers ($\Rey \lesssim 5$), the number of GMRES iterations is at its largest, and is positively correlated with $\frac{1}{\Ca}$, i.e. surface tension forces. This occurs when the approximations made in the ordinary Navier-Stokes blocks of Preconditioner (\ref{Eq:ApproximatePreconditioner}) are known to be adequate \citep{elman_finite_2014}, and suggests that performance degradation of the preconditioner may be attributed to approximations made in lumped Block (\ref{Eq:LumpedBlock}). For $ 5\lesssim \Rey$, in Figure \ref{fig:GMRESIterations}, one observes a steady increase in the number of iterations, which is consistent with the literature on Navier-Stokes preconditioning \citep{elman_finite_2014}. Indeed, such a dependency on $\Rey$ is known to occur when pressure's Schur complement is approximated by the mass matrix, whose choice is effective in the event of highly-viscous flow, but can degrade as $\Rey$ increases. A better approximation, for larger $\Rey$, would have been the pressure convection-diffusion preconditioner or the least-squares commutator preconditioner \citep{elman_finite_2014}. This also suggests that any adverse effects of approximations that were made in lumped Block (\ref{Eq:LumpedBlock}), have decayed for larger $\Rey$ and that the approximation is suitable in this range.

As for the residual evolution in Figure \ref{fig:DieSwellResiduals}, the die-swell convergence results are unequivocally quadratic, and verify the theoretical convergence rate of Newton-Raphson methods that comes out of the rigorous linearisation carried out in Section \ref{Sec:Linearisation}. The linearisation thus guarantees faster convergence than artificial transient simulations or Trial Methods.

The thermo-capillary flow in Section \ref{Sec:ThermocapillarySimulations} is a multi-physics example, where a homogeneous Neumann BC on temperature along the free interface may be leveraged, together with a Picard iteration, to decouple it from $\u$, $p$ and $\alpha$. Although this violates the exactness of the linearisation in Section \ref{Sec:Linearisation}, it permits the use of Preconditioner (\ref{Eq:ApproximatePreconditioner}). In this example, fluid flow is induced by both Grashof and Marangoni convection. Both effects cause eddies within the enclosed fluid domain and deform the free interface. The ensuing circulation patterns, although clockwise in both cases, have opposite effects on the shape of the free surface in Figures \ref{fig:Grashof14}, \ref{fig:Marangoni50} and \ref{fig:Gr900Marangoni50}, as buoyancy-driven convection elevates the hot container side and surface-tension-driven convection does the converse. 

Table \ref{Table:Thermocapillary} demonstrates agreement with the results of \citet{cuvelier_thermocapillary_1986}. We believe that any difference may be attributed to the comparative coarseness of their meshes, which were limited by the computational power at the time (1986). Indeed, differences are less marked in buoyancy-driven situations than in surface-tension driven simulations where flow patterns are more localised to the free surface and, thus require higher resolution. Together with modern computational power, iterative solution, enabled by Preconditioner (\ref{Eq:ApproximatePreconditioner}), is most likely responsible for the difference between the two simulation sets. 

Simulations in Section \ref{Sec:ThermocapillarySimulations} explore the effect of the contact angle on interface shapes in Grashof convection, as an extension of the effects explored by \citet{cuvelier_thermocapillary_1986}. Although no prior solution exists that can be confronted with present results, their correctness is corroborated by convergence of the free surface to the imposed contact angle, despite the fact that the initial guess of these flows possessed a different $\phi$. This suggests correctness of the contact-line BCs and, hence, of the linearisation in Section \ref{Sec:Linearisation}. We would also like to point out that the artificial slip region enabled imposing these BCs weakly. Even though this BC may appear to be non-physical in steady, free-surface flows since the contact line is static, all numerical simulations converged to solutions that verified the no-slip condition along the slip region. 

\section{Conclusion}

The method presented here, an extension of the original TLM that was proposed by \citet{kruyt_total_1988}, endows computer simulations with a significantly enhanced ability to simulate 3D free-surface flows with the sharp-interface approach. It is predicated on a rigorous first-order expansion of the nonlinear FEM weak form that ensures second-order convergence of the iterative method and results in a substantial reduction in the computational size of interfacial-flow problems with respect to existing exact-Newton methods. Specifically, it produces a monolithic linear system that is only slightly more than half the size of state-of-the-art Newton methods, while fully encompassing surface-tension effects and covering moving contact lines with arbitrary contact angles in 3D. Pairing this method with Preconditioner (\ref{Eq:ApproximatePreconditioner}) renders iterative solution robust in the appropriate range of dimensionless parameters, albeit without tainting the equation system with stabilisation terms. It thus surmounts the conditioning issues typically encountered by exact-Newton methods in monolithic system inversion, while still permitting the use of LBB-compliant elements.

The resulting significant size reduction and novel iterative-solution capabilities enables higher-resolution solution of free-surface flows without any compromise on convergence rate or accuracy. Its use to solve the cylindrical die-swell problem shows agreement with experiment and prior simulation with no adjustable parameter employed, despite the fact that maximum computational power has not yet been harnessed. Not only were simulations run on one processor and in serial, but they were also carried out on meshes that were not tailored by rigorous error estimation. All but one simulation converged within two hours, while including over one million DOFs. Parallelisation and adaptive meshing should thus significantly further the limits of this extended TLM's computational power. 

In Section \ref{Sec:ThermocapillarySimulations}, the method is employed to solve a multi-physics problem entailing both thermal transport and interfacial flows, matching earlier calculations by \citet{cuvelier_thermocapillary_1986}. The present TLM is thus shown to be accurate in the modeling of buoyancy and surface-tension-driven convection and is further applied to arbitrary contact-angle flows.

Only steady, free-surface flow problems were solved numerically here; however, the present method reaches beyond steady simulations. Indeed, the Jacobian of nonlinear weak form (\ref{Eq:WeakForm}), as derived in Section \ref{Sec:Linearisation} is also directly applicable to growth rate calculations in linear stability analysis of interfacial flows. Given that the linearisation was carried out for time-dependent domains in the derivation presented here, it may be applied to transient problems as well, without any restriction on the time-integration method. In addition, the derivation methodology is applicable to any kind of FBP and could serve to reduce the computational footprint across a much broader range of physics than were addressed here. The scope for implementation of the present method thus likely goes well beyond what is exposed in the present contribution.


\backsection[Acknowledgements]{We would like to acknowledge Professor François Gallaire of EPFL for fruitful discussions and the \href{https://www.epfl.ch/research/facilities/scitas/}{SCITAS} platform of EPFL for computational resources.}

\backsection[Funding]{ This research project is sponsored by an Advanced Grant from the Swiss National Science Foundation, Grant No. TMAG-2-2O932A.}

\backsection[Declaration of interests]{The authors report no conflict of interest.}


\backsection[Author ORCIDs]{T. Benkley https://orcid.org/0009-0001-7611-2530; S.Deparis https://orcid.org/0000-0002-2832-6630; P. Ricci https://orcid.org/0000-0003-3117-2238; A. Mortensen https://orcid.org/0000-0002-8267-2008}


\appendix

\section{Kinematic Equation Identity}\label{appA}

Let us prove that, for two sufficiently smooth vector fields $\u,\,\v$ and a sufficiently smooth scalar field $\mu$,
\begin{equation} \label{Eq:AppendixIdentity}
\begin{split}
    \mu \n \cdot  \{(\v \cdot \bnabla)\u - (\u\cdot \bnabla_S )\v\} = &\bnabla_S\cdot (-\mu \n\times(\u\times \v)) + (\v\cdot \n)(\u\cdot \bnabla_S)\mu \\
    &- (\u\cdot \n)(\v\cdot \bnabla_S)\mu - \mu(\u\cdot \n) (\bnabla_S\cdot \v) + \mu (\v\cdot \n)(\bnabla\cdot \u)
\end{split}
\end{equation}
on $\Sigma_f$ where $\n$ is the outward-pointing unit normal vector on $\Sigma_f$.

Using $\bnabla = \bnabla_S + \n\frac{\partial}{\partial n} $, we decompose the left-hand side of Identity (\ref{Eq:AppendixIdentity}).
\begin{equation} \label{Eq:AppendixStep1}
    \begin{split}
        \mu \n \cdot  \{(\v \cdot \bnabla)\u - (\u\cdot \bnabla_S )\v\} &= \mu \n \cdot  \{(\v \cdot \bnabla_S)\u - (\u\cdot \bnabla_S )\v\} + \mu (\v\cdot n)(\n\cdot \frac{\partial \u}{\partial n})
    \end{split}
\end{equation}
Now, we apply the following vector calculus identity
\begin{equation} \label{Eq:AppendixStep}
\begin{split}
    \mu \, \n\cdot (\a\cdot \bnabla_S)\b = &\mu \{\n\cdot (\a\cdot \bnabla_S)\b - \a\cdot (\n\cdot \bnabla_S)\b \} \\
    = &\mu (\a\times \n) \cdot (\bnabla_S\times \b) \\
    = &\mu \{ \b\cdot (\bnabla_S\times (\a\times \n)) + \bnabla_S\cdot (\b\times(\a\times\n)) \} \\
    = &\bnabla_S \cdot (\mu \b\times(\a\times\n)) - (\b\times(\a\times \n))\cdot \bnabla_S\mu + \mu \, \b \cdot (\bnabla_S\times(\a\times \n)) \\
    = &\bnabla_S \cdot (\mu \b\times(\a\times\n)) - (\b\times(\a\times \n))\cdot \bnabla_S\mu \\
    &+ \mu \b \cdot (\a (\bnabla_S\cdot \n) - \n(\bnabla_S\cdot \a) - (\a\cdot \bnabla_S)\n )
\end{split}
\end{equation}
which holds for any pair of, sufficiently smooth fields $\a$, $\b$. When applied to the first two terms in Equation (\ref{Eq:AppendixStep1}), it gives rise to
\begin{equation}
    \begin{split}
        \mu \n \cdot  \{(\v \cdot \bnabla_S)\u &- (\u\cdot \bnabla_S )\v\} \\
        = &\bnabla_S\cdot\{ \mu ( \u\times(\v\times\n) - \v\times(\u\times\n) ) \} - \{\u\times(\v\times\n) - \v\times(\u\times\n) \}\cdot\bnabla_S\mu \\
        &+ \mu\{ (\v\cdot\n)(\bnabla_S\cdot\u) - (\u\cdot\n)(\bnabla_S\cdot\v) + \v\cdot(\u\cdot\bnabla_S)\n - \u\cdot(\v\cdot\bnabla_S)\n \} \\
        = & \bnabla_S \cdot (-\mu\, \n\times(\u\times \v) ) + (\v\cdot\n)(\u\cdot\bnabla_S)\mu - (\u\cdot\n)(\v\cdot \bnabla_S)\mu \\
        &- \mu(\u\cdot\n)(\bnabla_S\cdot \v) + \mu(\v\cdot \n)(\bnabla_S\cdot \u) + \v\cdot (\u\cdot\bnabla_S)\n - \u\cdot(\v\cdot\bnabla_S)\n
    \end{split}
\end{equation}
In the next step, the subsequent identity is used.
\begin{equation}
    \v\cdot (\u\cdot\bnabla_S)\n - \u\cdot(\v\cdot\bnabla_S)\n = (\u\times\v)\cdot (\bnabla_S\times \n) = 0
\end{equation}
It is identically zero since the surface curl of a unit normal vector vanishes everywhere on the relevant surface \citep{weatherburn_differential_1927}. Finally, plugging in the above steps proves the statement.
\begin{equation}
    \begin{split}
        \mu \n \cdot  \{(\v \cdot \bnabla)\u &- (\u\cdot \bnabla_S )\v\} \\
        = &\bnabla_S \cdot (-\mu \n\times(\u\times \v) ) + (\v\cdot\n)(\u\cdot\bnabla_S)\mu - (\u\cdot\n)(\v\cdot \bnabla_S)\mu \\
        &- \mu(\u\cdot\n)(\bnabla_S\cdot \v) + \mu(\v\cdot \n)(\bnabla\cdot \u)
    \end{split}
\end{equation} 

\section{Surface Divergence Integral Linearisation} \label{appB}

In this Section, The Einstein notation convection is adopted for subscripts.

The linearisation of the product between a scalar field $\psi$ and the surface divergence of a vector field $\v$, integrated over a surface is carried out. 

In the first step, the surface gradient is linearised.
\begin{equation}
    \begin{split}
        (\delta_{ij} - n_i&n_j) \frac{\partial}{\partial x_j} \approx ( \delta_{ij} - \{ n_i-(\delta_{ik}-n_in_k)\frac{\partial d_m}{\partial X_k}n_m\} \{n_j-(\delta_{jp}-n_jn_p)\frac{\partial d_q}{\partial X_p}n_q) \} \frac{\partial X_l}{\partial x_j} \frac{\partial}{\partial X_l}
    \end{split}
\end{equation}
where the rules of linearisation in Equation (\ref{Eq:LinearizationRelations}) are used. Additionally, by virtue of a Neumann series, the inverse of the deformation gradient approximates as 

\begin{equation}
    F_{lj}^{-1} = \frac{\partial X_l}{\partial x_j} \approx \delta_{lj} - \frac{\partial d_l}{\partial X_j}.
\end{equation}
Inserting this relation and removing higher-order terms yields the next equation.
\begin{equation}
    \begin{split}
        (\delta_{ij} - n_in_j) \frac{\partial}{\partial x_j}  \approx \{&\delta_{ij} - n_in_j - (\delta_{mj}-n_mn_j) (\delta_{ik}-n_in_k)\frac{\partial d_m}{\partial X_k} \\
        &+ (\delta_{jk}-n_jn_k)\frac{\partial d_m}{\partial X_k} n_mn_i\} \frac{\partial}{\partial X_j}
    \end{split}
\end{equation}
The dot product is taken with a vector field which is linearised as $v_i \approx v_i + d_k\frac{\partial v_i}{\partial X_k}$ and engenders
\begin{equation} \label{Eq:LinearizedSurfaceDivergence}
    \begin{split}
        \bnabla_S \cdot \v \approx &\bnabla_S \cdot \v + \bnabla_S\cdot \{(\d\cdot \bnabla)\v\} - (\bnabla_S \d)^T:\bnabla_S\v + (\bnabla_S \d\cdot \n)\cdot (\bnabla_S \v \cdot \n) \\
        = &\bnabla_S \cdot \v + \bnabla_S\cdot \{ (\d\cdot \n) \frac{\partial \v}{\partial n} \} + \underbrace{\bnabla_S\cdot \{(\d\cdot \bnabla_S)\v\}}_{=:A} - (\bnabla_S \d)^T:\bnabla_S\v \\
        &+ (\bnabla_S \d\cdot \n)\cdot (\bnabla_S \v \cdot \n)
    \end{split}
\end{equation}
$A$ is expanded in the subsequent step:
\begin{equation}
    \begin{split}
        A = &(\delta_{ij}-n_in_j)\frac{\partial }{\partial X_j}(d_l(\delta_{kl}-n_kn_l)\frac{\partial v_i}{\partial X_k})  \\
        = &(\delta_{ij}-n_in_j)\{\frac{\partial d_l}{\partial X_j}(\delta_{kl}-n_kn_l) \frac{\partial v_i}{\partial X_k} + d_l (\delta_{kl}-n_kn_l) \frac{\partial^2 v_i}{\partial X_j \partial X_k} - d_l \frac{\partial (n_kn_l)}{\partial X_j} \frac{\partial v_i}{\partial X_k} \} \\
        = &(\bnabla_S \d)^T:\bnabla_S\v + (\delta_{ij}-n_in_j) d_l \{ - (n_l\frac{\partial n_k}{\partial X_j} + n_k\frac{\partial n_l}{\partial X_j})\frac{\partial v_i}{\partial X_k} + (\delta_{kl}-n_kn_l)\frac{\partial^2v_i}{\partial X_j\partial X_k}\} \\
        = &(\bnabla_S \d)^T:\bnabla_S\v - (\d\cdot \n)(\bnabla_S\n)^T:\bnabla_S\v - \d\cdot(\frac{\partial \v}{\partial n}\cdot \bnabla_S)\n \\
        &+  d_l(\delta_{ij}-n_in_j)(\delta_{kl}-n_kn_l)\frac{\partial^2v_i}{\partial X_j\partial X_k}
    \end{split}
\end{equation}
We recall the linearisation of surface measures (see Equation (\ref{Eq:LinearizationRelations})) which gives
\begin{equation}
    \int_{\Sigma_f} .\,dS \approx \int_{\Sigma_{f,0}} .\,(1+\bnabla_S\cdot\d)dS.
\end{equation}
Then, plugging $A$ into Equation (\ref{Eq:LinearizedSurfaceDivergence}) yields
\begin{equation} \label{Eq:SurfaceDivergenceLinearizationStep}
    \begin{split}
        \int_{\Sigma_f} \psi(\bnabla_S\cdot \v) dS \approx &\int_{\Sigma_{f,0}} \{ \psi(\bnabla_S \cdot \v) + \psi\bnabla_S\cdot \{ (\d\cdot \n) \frac{\partial \v}{\partial n} \} - \psi(\d\cdot \n)(\bnabla_S\n)^T:\bnabla_S\v \\
        &- \psi\,\d\cdot(\frac{\partial \v}{\partial n}\cdot \bnabla_S)\n +  \psi\,d_l(\delta_{ij}-n_in_j)(\delta_{kl}-n_kn_l)\frac{\partial^2v_i}{\partial X_j\partial X_k} \\
        &+ \psi(\bnabla_S \d\cdot \n)\cdot (\bnabla_S \v \cdot \n) +  \psi \underbrace{(\bnabla_S\cdot \v)(\bnabla_S\cdot \d)}_{=:B} \\
        &+ (\bnabla_S\cdot\v)(\d\cdot\bnabla \psi)\} dS
    \end{split}
\end{equation}
$B$ is expanded:
\begin{equation}
    \begin{split}
        B = &\bnabla_S \cdot \{ \d(\bnabla_S\cdot \v) \} - (\d\cdot \bnabla_S)(\bnabla_S\cdot \v) \\
        = &\bnabla_S \cdot \{ \d(\bnabla_S\cdot \v) \} - d_l(\delta_{kl} - n_kn_l)\frac{\partial }{\partial X_k}\{ (\delta_{ij}-n_in_j)\frac{\partial v_i}{\partial X_j} \} \\
        = &\bnabla_S \cdot \{ \d(\bnabla_S\cdot \v) \} + d_l(\delta_{kl} - n_kn_l)\{ \frac{\partial (n_in_j)}{\partial X_k}\frac{\partial v_i}{\partial X_j} - (\delta_{ij}-n_in_j)\frac{\partial^2 v_i}{\partial X_j\partial X_k} \} \\
        = &\bnabla_S \cdot \{ \d(\bnabla_S\cdot \v) \} + \frac{\partial \v}{\partial n}\cdot(\d\cdot \bnabla_S)\n + (\bnabla_S\v\cdot\n)\cdot(\d\cdot \bnabla_S)\n \\
        &- d_l(\delta_{kl} - n_kn_l)(\delta_{ij}-n_in_j)\frac{\partial^2 v_i}{\partial X_j\partial X_k} 
    \end{split}
\end{equation}
Now, $B$ is plugged back into Relation (\ref{Eq:SurfaceDivergenceLinearizationStep}):
\begin{equation}
    \begin{split}
        \int_{\Sigma_f} \psi(\bnabla_S\cdot \v) dS \approx &\int_{\Sigma_{f,0}} \{ \psi (\bnabla_S \cdot \v) + \underbrace{\psi\, \bnabla_S\cdot \{ (\d\cdot \n) \frac{\partial \v}{\partial n} \}}_{=:G} - \psi (\d\cdot \n)(\bnabla_S\n)^T:\bnabla_S\v \\
        &\underbrace{- \psi\,\d\cdot(\frac{\partial \v}{\partial n}\cdot \bnabla_S)\n}_{=:C_1}
        + \underbrace{\psi (\bnabla_S \d\cdot \n)\cdot (\bnabla_S \v \cdot \n)}_{=:D_2} + \underbrace{\psi \bnabla_S \cdot \{ \d(\bnabla_S\cdot \v) \}}_{=:E_1} \\
        &+ \underbrace{\psi \frac{\partial \v}{\partial n}\cdot(\d\cdot \bnabla_S)\n}_{=:C_2} + \underbrace{\psi (\bnabla_S\v\cdot\n)\cdot(\d\cdot \bnabla_S)\n}_{=:D_1} + \underbrace{(\bnabla_S\cdot\v)(\d\cdot\bnabla\psi)}_{=:E_2} \} dS
    \end{split}
\end{equation}
Using the fact that $\bnabla_S\times \n = 0$ \citep{weatherburn_differential_1927}, the following integrands cancel out.
\begin{equation}
    C_1 + C_2 = - \psi\,\d\cdot(\frac{\partial \v}{\partial n}\cdot \bnabla_S)\n + \psi \frac{\partial \v}{\partial n}\cdot(\d\cdot \bnabla_S)\n = \psi (\d\times \frac{\partial \v}{\partial n})\cdot (\bnabla_S\times \n) = 0
\end{equation}
Again, using the irrotational property of the normal vector,
\begin{equation}
    D_1 = \psi (\bnabla_S\v\cdot \n) \cdot (\d\cdot \bnabla_S)\n = \psi (\bnabla_S\v\cdot \n)\cdot(\bnabla_S\n \cdot \d),
\end{equation}
which in turn implies that the sum of the two following integrands satisfies
\begin{equation}
\begin{split}
    D_1+D_2 = \psi (\bnabla_S \d\cdot \n)\cdot (\bnabla_S \v \cdot \n)+ \psi(\bnabla_S \v\cdot \n) \cdot ( \d \cdot \bnabla_S)\n = \psi (\bnabla_S\v\cdot \n)\cdot \bnabla_S(\d\cdot \n).
\end{split}
\end{equation}
Additionally, we employ relations
\begin{equation}
    E_1 + E_2 = \bnabla_S\cdot \{ \psi \d(\bnabla_S\cdot \v) \} + (\d\cdot \n)(\bnabla_S\cdot \v)\frac{\partial \psi}{\partial n}
\end{equation}
and 
\begin{equation}
    G = \bnabla_S\cdot \{ \psi(\d\cdot\n)\frac{\partial \v}{\partial n} \} - (\d\cdot\n)(\frac{\partial\v}{\partial n}\cdot \psi).
\end{equation}
These computations lead to the following intermediate result.
\begin{equation} \label{Eq:SurfaceDivergenceLinearizationIntermediate}
    \begin{split}
        \int_{\Sigma_f} \psi(\bnabla_S\cdot \v) dS \approx &\int_{\Sigma_{f,0}} \{ \psi(\bnabla_S \cdot \v) + \underbrace{\bnabla_S\cdot\{ \psi(\d\cdot\n)\frac{\partial \v}{\partial n} + \psi \d (\bnabla_S\cdot \v ) \}}_{=:H} \\
        &- \psi(\d\cdot \n)(\bnabla_S\n)^T:\bnabla_S\v + \psi(\bnabla_S\v\cdot \n)\cdot \bnabla_S(\d\cdot \n) \\
        & + (\bnabla_S\cdot\v)(\d\cdot\n)\frac{\partial\psi}{\partial n} - (\d\cdot\n)(\frac{\partial\v}{\partial n}\cdot\bnabla_S\psi)\} dS 
    \end{split}
\end{equation}
We now apply the surface divergence theorem (\ref{Eq:DivergenceTheorem}) to the surface integral of $H$.
\begin{equation} 
    \begin{split}
        \int_{\Sigma_{f,0}}H\,dS = \int_{\Sigma_{f,0}} &\bnabla_S\cdot \{ \psi(\d\cdot\n)\frac{\partial \v}{\partial n} + \psi \d (\bnabla_S\cdot \v ) \} dS \\
        = &\int_{\Sigma_{f,0}} \psi (\d\cdot\n)(\bnabla_S\cdot \n )\{\frac{\partial \v}{\partial n}\cdot \n + \bnabla_S\cdot\v \}  dS \\
        &+ \int_{\partial \Sigma_{f,0}} \psi \{(\d\cdot \n_f)(\sf\cdot \frac{\partial \v}{\partial n_f})  + (\d\cdot \sf) (\bnabla_S \cdot \v) \}dl
    \end{split}
\end{equation}
Finally, the resulting expression is given by
\begin{equation} \label{Eq:SurfaceDivergenceLinearizationFinal}
    \begin{split}
        \int_{\Sigma_f} \psi(\bnabla_S\cdot \v) dS \approx &\int_{\Sigma_{f,0}} \{ \psi(\bnabla_S \cdot \v) + \psi(\d\cdot \n) (\n\cdot \frac{\partial \v}{\partial n} + \bnabla_S\cdot \v )(\bnabla_S\cdot \n) \\
        &- \psi(\d\cdot \n)(\bnabla_S\n)^T:\bnabla_S\v + \psi(\bnabla_S\v\cdot \n)\cdot \bnabla_S(\d\cdot \n) \\
        & + (\bnabla_S\cdot\v)(\d\cdot\n)\frac{\partial\psi}{\partial n} - (\d\cdot\n)(\frac{\partial\v}{\partial n}\cdot\bnabla_S\psi)\} dS \\
        & + \int_{\partial \Sigma_{f,0}} \psi \{(\d\cdot \n_f)(\sf\cdot \frac{\partial \v}{\partial n_f})  + (\d\cdot \sf) (\bnabla_S \cdot \v) \}dl
    \end{split}
\end{equation}

\bibliographystyle{jfm}
\bibliography{TLM_Extension}

\begin{thebibliography}{67}
\expandafter\ifx\csname natexlab\endcsname\relax\def\natexlab#1{#1}\fi
\def\au#1{#1} \def\ed#1{#1} \def\yr#1{#1}\def\at#1{#1}\def\jt#1{\textit{#1}} \def\bt#1{#1}\def\bvol#1{\textbf{#1}} \def\vol#1{#1} \def\pg#1{#1} \def\publ#1{#1}\def\arxiv#1{#1}\def\org#1{#1}\def\st#1{\textit{#1}}

\bibitem[Abubakar \& Matar(2022)]{abubakar_linear_2022}
{\sc \au{Abubakar, H.A.} \& \au{Matar, O.K.}} \yr{2022}  \at{Linear stability analysis of {Taylor} bubble motion in downward flowing liquids in vertical tubes}.  \jt{J. Fluid Mech.}  \bvol{941},  \pg{A2}.

\bibitem[Anthony {\em et~al.\/}(2023)Anthony, Wee, Garg, Thete, Kamat, Wagoner, Wilkes, Notz, Chen, Suryo, Sambath, Panditaratne, Liao \& Basaran]{anthony_sharp_2023}
{\sc \au{Anthony, Christopher~R.}, \au{Wee, Hansol}, \au{Garg, Vishrut}, \au{Thete, Sumeet~S.}, \au{Kamat, Pritish~M.}, \au{Wagoner, Brayden~W.}, \au{Wilkes, Edward~D.}, \au{Notz, Patrick~K.}, \au{Chen, Alvin~U.}, \au{Suryo, Ronald}, \au{Sambath, Krishnaraj}, \au{Panditaratne, Jayanta~C.}, \au{Liao, Ying-Chih} \& \au{Basaran, Osman~A.}} \yr{2023}  \at{Sharp {Interface} {Methods} for {Simulation} and {Analysis} of {Free} {Surface} {Flows} with {Singularities}: {Breakup} and {Coalescence}}.  \jt{Annu. Rev. Fluid Mech.}  \bvol{55}~(1),  \pg{707--747}.

\bibitem[Baer {\em et~al.\/}(2000)Baer, Cairncross, Schunk, Rao \& Sackinger]{baer_finite_2000}
{\sc \au{Baer, Thomas~A.}, \au{Cairncross, Richard~A.}, \au{Schunk, P.~Randall}, \au{Rao, Rekha~R.} \& \au{Sackinger, Phillip~A.}} \yr{2000}  \at{A finite element method for free surface flows of incompressible fluids in three dimensions. {Part} {II}. {Dynamic} wetting lines}.  \jt{Int. J. Numer. Meth. Fluids}  \bvol{33}~(3),  \pg{405--427}.

\bibitem[Balay {\em et~al.\/}(2015{\natexlab{{\em a\/}}})Balay, Abhyankar, Adams, Brown, Brune, Buschelman, Dalcin, Eijkhout, Gropp, Kaushik, Knepley, McInnes, Rupp, Smith, Zampini \& Zhang]{petsc-user-ref}
{\sc \au{Balay, Satish}, \au{Abhyankar, Shrirang}, \au{Adams, Mark~F.}, \au{Brown, Jed}, \au{Brune, Peter}, \au{Buschelman, Kris}, \au{Dalcin, Lisandro}, \au{Eijkhout, Victor}, \au{Gropp, William~D.}, \au{Kaushik, Dinesh}, \au{Knepley, Matthew~G.}, \au{McInnes, Lois~Curfman}, \au{Rupp, Karl}, \au{Smith, Barry~F.}, \au{Zampini, Stefano} \& \au{Zhang, Hong}} \yr{2015{\natexlab{{\em a\/}}}}  \bt{{PETS}c users manual}. {\em Tech. Rep.\/} ANL-95/11 - Revision 3.6.  \org{Argonne National Laboratory}.

\bibitem[Balay {\em et~al.\/}(2015{\natexlab{{\em b\/}}})Balay, Abhyankar, Adams, Brown, Brune, Buschelman, Dalcin, Eijkhout, Gropp, Kaushik, Knepley, McInnes, Rupp, Smith, Zampini \& Zhang]{petsc-web-page}
{\sc \au{Balay, Satish}, \au{Abhyankar, Shrirang}, \au{Adams, Mark~F.}, \au{Brown, Jed}, \au{Brune, Peter}, \au{Buschelman, Kris}, \au{Dalcin, Lisandro}, \au{Eijkhout, Victor}, \au{Gropp, William~D.}, \au{Kaushik, Dinesh}, \au{Knepley, Matthew~G.}, \au{McInnes, Lois~Curfman}, \au{Rupp, Karl}, \au{Smith, Barry~F.}, \au{Zampini, Stefano} \& \au{Zhang, Hong}} \yr{2015{\natexlab{{\em b\/}}}} {PETS}c {W}eb page. \url{http://www.mcs.anl.gov/petsc}.

\bibitem[Belhachmi {\em et~al.\/}(2006)Belhachmi, Bernardi \& Deparis]{belhachmi_weighted_2006}
{\sc \au{Belhachmi, Zakaria}, \au{Bernardi, Christine} \& \au{Deparis, Simone}} \yr{2006}  \at{Weighted {Clément} operator and application to the finite element discretization of the axisymmetric {Stokes} problem}.  \jt{Numer. Math.}  \bvol{105}~(2),  \pg{217--247}.

\bibitem[Brackbill {\em et~al.\/}(1992)Brackbill, Kothe \& Zemach]{brackbill_continuum_1992}
{\sc \au{Brackbill, J.U}, \au{Kothe, D.B} \& \au{Zemach, C}} \yr{1992}  \at{A continuum method for modeling surface tension}.  \jt{Journal of Computational Physics}  \bvol{100}~(2),  \pg{335--354}.

\bibitem[Brennen(2006)]{brennen_fundamentals_2006}
{\sc \au{Brennen, Christopher~E.}} \yr{2006} {\em Fundamentals of multiphase flow\/}, 1st edn.  \publ{Cambridge: Cambridge University Press}.

\bibitem[Brezzi(1974)]{brezzi_existence_1974}
{\sc \au{Brezzi, F.}} \yr{1974}  \at{On the existence, uniqueness and approximation of saddle-point problems arising from lagrangian multipliers}.  \jt{R.A.I.R.O. Analyse Numérique}  \bvol{8}~(R2),  \pg{129--151}.

\bibitem[Cahn \& Hilliard(1958)]{cahn_free_1958}
{\sc \au{Cahn, John~W.} \& \au{Hilliard, John~E.}} \yr{1958}  \at{Free {Energy} of a {Nonuniform} {System}. {I}. {Interfacial} {Free} {Energy}}.  \jt{The Journal of Chemical Physics}  \bvol{28}~(2),  \pg{258--267}.

\bibitem[Cairncross {\em et~al.\/}(2000)Cairncross, Schunk, Baer, Rao \& Sackinger]{cairncross_finite_2000}
{\sc \au{Cairncross, Richard~A.}, \au{Schunk, P.~Randall}, \au{Baer, Thomas~A.}, \au{Rao, Rekha~R.} \& \au{Sackinger, Phillip~A.}} \yr{2000}  \at{A finite element method for free surface flows of incompressible fluids in three dimensions. {Part} {I}. {Boundary} fitted mesh motion}.  \jt{Int. J. Numer. Meth. Fluids}  \bvol{33}~(3),  \pg{375--403}.

\bibitem[Chowdhury {\em et~al.\/}(2022)Chowdhury, Yadaiah, Prakash, Ramakrishna, Dixit, Gupta \& Buddhi]{chowdhury_laser_2022}
{\sc \au{Chowdhury, Sohini}, \au{Yadaiah, N.}, \au{Prakash, Chander}, \au{Ramakrishna, Seeram}, \au{Dixit, Saurav}, \au{Gupta, Lovi~Raj} \& \au{Buddhi, Dharam}} \yr{2022}  \at{Laser powder bed fusion: a state-of-the-art review of the technology, materials, properties \& defects, and numerical modelling}.  \jt{Journal of Materials Research and Technology}  \bvol{20},  \pg{2109--2172}.

\bibitem[Christodoulou \& Scriven(1992)]{christodoulou_discretization_1992}
{\sc \au{Christodoulou, K.N.} \& \au{Scriven, L.E.}} \yr{1992}  \at{Discretization of free surface flows and other moving boundary problems}.  \jt{Journal of Computational Physics}  \bvol{99}~(1),  \pg{39--55}.

\bibitem[Crank(1987)]{crank_free_1987}
{\sc \au{Crank, J.}} \yr{1987} {\em Free and {Moving} {Boundary} {Problems}\/}. {\em Oxford science publications\/} 1.  \publ{Clarendon Press}.

\bibitem[Crosetto {\em et~al.\/}(2011)Crosetto, Deparis, Fourestey \& Quarteroni]{crosetto_parallel_2011}
{\sc \au{Crosetto, Paolo}, \au{Deparis, Simone}, \au{Fourestey, Gilles} \& \au{Quarteroni, Alfio}} \yr{2011}  \at{Parallel {Algorithms} for {Fluid}-{Structure} {Interaction} {Problems} in {Haemodynamics}}.  \jt{SIAM J. Sci. Comput.}  \bvol{33}~(4),  \pg{1598--1622}.

\bibitem[Cryer(1976)]{cryer_technical_nodate}
{\sc \au{Cryer, Colin~W}} \yr{1976} {\em A survey of trial free-boundary methods for the numerical solution of free boundary problems\/}.  \publ{University of Wisconsin, Mathematics Research Center}.

\bibitem[Cuvelier \& Driessen(1986)]{cuvelier_thermocapillary_1986}
{\sc \au{Cuvelier, C.} \& \au{Driessen, J.~M.}} \yr{1986}  \at{Thermocapillary free boundaries in crystal growth}.  \jt{J. Fluid Mech.}  \bvol{169}~(-1),  \pg{1}.

\bibitem[Cuvelier \& Schulkes(1990)]{cuvelier_numerical_1990}
{\sc \au{Cuvelier, C.} \& \au{Schulkes, R. M. S.~M.}} \yr{1990}  \at{Some {Numerical} {Methods} for the {Computation} of {Capillary} {Free} {Boundaries} {Governed} by the {Navier}–{Stokes} {Equations}}.  \jt{SIAM Rev.}  \bvol{32}~(3),  \pg{355--423}.

\bibitem[Daversin-Catty {\em et~al.\/}(2021)Daversin-Catty, Richardson, Ellingsrud \& Rognes]{daversin-catty_abstractions_2021}
{\sc \au{Daversin-Catty, Cécile}, \au{Richardson, Chris~N.}, \au{Ellingsrud, Ada~J.} \& \au{Rognes, Marie~E.}} \yr{2021}  \at{Abstractions and {Automated} {Algorithms} for {Mixed} {Domain} {Finite} {Element} {Methods}}.  \jt{ACM Trans. Math. Softw.}  \bvol{47}~(4),  \pg{1--36}.

\bibitem[Deparis(2004)]{deparis_numerical_2004}
{\sc \au{Deparis, Simone}} \yr{2004}  \at{Numerical analysis of axisymmetric flows and methods for fluid-structure interaction arising in blood flow simulation}. PhD thesis, Lausanne, EPFL.

\bibitem[Di~Martino {\em et~al.\/}(2001)Di~Martino, Guadagni, Fumero, Ballerini, Spirito, Biglioli \& Redaelli]{di_martino_fluidstructure_2001}
{\sc \au{Di~Martino, E.S.}, \au{Guadagni, G.}, \au{Fumero, A.}, \au{Ballerini, G.}, \au{Spirito, R.}, \au{Biglioli, P.} \& \au{Redaelli, A.}} \yr{2001}  \at{Fluid–structure interaction within realistic three-dimensional models of the aneurysmatic aorta as a guidance to assess the risk of rupture of the aneurysm}.  \jt{Medical Engineering \& Physics}  \bvol{23}~(9),  \pg{647--655}.

\bibitem[Dowell \& Hall(2000)]{dowell_modeling_2000}
{\sc \au{Dowell, Earl~H} \& \au{Hall, Kenneth~C}} \yr{2000}  \at{{MODELING} {OF} {FLUID}-{STRUCTURE} {INTERACTION}}.  \jt{Annual Reviews of Fluid Mechanics} .

\bibitem[Elgohary \& Attia(2017)]{elgohary_3d-modeling_2017}
{\sc \au{Elgohary, Mohamed} \& \au{Attia, Walia}} \yr{2017}  \at{{3D}-{MODELING} {OF} {LONG} {SPAN} {BRIDGES} {AND} {FLUID}-{STRUCTURE} {INTERACTION} {DOMAINS}}.  \jt{Journal of Al-Azhar University Engineering Sector}  \bvol{12}~(45),  \pg{1267--1284}.

\bibitem[Elman {\em et~al.\/}(2014)Elman, Silvester \& Wathen]{elman_finite_2014}
{\sc \au{Elman, Howard}, \au{Silvester, David} \& \au{Wathen, Andy}} \yr{2014} {\em Finite {Elements} and {Fast} {Iterative} {Solvers}: with {Applications} in {Incompressible} {Fluid} {Dynamics}\/}.  \publ{Oxford University Press}.

\bibitem[Ettouney \& Brown(1983)]{ettouney_finite-element_1983}
{\sc \au{Ettouney, Hisham~M} \& \au{Brown, Robert~A}} \yr{1983}  \at{Finite-element methods for steady solidification problems}.  \jt{Journal of Computational Physics}  \bvol{49}~(1),  \pg{118--150}.

\bibitem[Fernández \& Moubachir(2005)]{fernandez_newton_2005}
{\sc \au{Fernández, Miguel~Ángel} \& \au{Moubachir, Marwan}} \yr{2005}  \at{A {Newton} method using exact jacobians for solving fluid–structure coupling}.  \jt{Computers \& Structures}  \bvol{83}~(2-3),  \pg{127--142}, publisher: Elsevier BV.

\bibitem[Fraggedakis {\em et~al.\/}(2017)Fraggedakis, Papaioannou, Dimakopoulos \& Tsamopoulos]{fraggedakis_discretization_2017}
{\sc \au{Fraggedakis, D.}, \au{Papaioannou, J.}, \au{Dimakopoulos, Y.} \& \au{Tsamopoulos, J.}} \yr{2017}  \at{Discretization of three-dimensional free surface flows and moving boundary problems via elliptic grid methods based on variational principles}.  \jt{Journal of Computational Physics}  \bvol{344},  \pg{127--150}.

\bibitem[Gerbeau {\em et~al.\/}(2004)Gerbeau, Lelièvre \& Le~Bris]{gerbeau_modeling_2004}
{\sc \au{Gerbeau, J.-F.}, \au{Lelièvre, T.} \& \au{Le~Bris, C.}} \yr{2004}  \at{Modeling and simulation of the industrial production of aluminium: the nonlinear approach}.  \jt{Computers \& Fluids}  \bvol{33}~(5-6),  \pg{801--814}.

\bibitem[Gerritsma \& Phillips(2000)]{gerritsma_spectral_2000}
{\sc \au{Gerritsma, M.I.} \& \au{Phillips, T.N.}} \yr{2000}  \at{Spectral {Element} {Methods} for {Axisymmetric} {Stokes} {Problems}}.  \jt{Journal of Computational Physics}  \bvol{164}~(1),  \pg{81--103}.

\bibitem[Girault \& Raviart(1986)]{girault_finite_1986}
{\sc \au{Girault, Vivette} \& \au{Raviart, Pierre-Arnaud}} \yr{1986} {\em Finite {Element} {Methods} for {Navier}-{Stokes} {Equations}\/},  \st{Springer {Series} in {Computational} {Mathematics}},  \vol{vol.~5}.  \publ{Berlin, Heidelberg: Springer Berlin Heidelberg}.

\bibitem[Goren \& Wronski(1966)]{goren_shape_1966}
{\sc \au{Goren, Simon~L.} \& \au{Wronski, Stanislaw}} \yr{1966}  \at{The shape of low-speed capillary jets of {Newtonian} liquids}.  \jt{J. Fluid Mech.}  \bvol{25}~(1),  \pg{185--198}.

\bibitem[Hirt \& Nichols(1981)]{hirt_volume_1981}
{\sc \au{Hirt, C.W} \& \au{Nichols, B.D}} \yr{1981}  \at{Volume of fluid ({VOF}) method for the dynamics of free boundaries}.  \jt{Journal of Computational Physics}  \bvol{39}~(1),  \pg{201--225}.

\bibitem[Hou {\em et~al.\/}(2012)Hou, Wang \& Layton]{hou_numerical_2012}
{\sc \au{Hou, Gene}, \au{Wang, Jin} \& \au{Layton, Anita}} \yr{2012}  \at{Numerical {Methods} for {Fluid}-{Structure} {Interaction} — {A} {Review}}.  \jt{Commun. comput. phys.}  \bvol{12}~(2),  \pg{337--377}.

\bibitem[Katopodes(2019)]{katopodes_free-surface_2019}
{\sc \au{Katopodes, Nikolaos~D.}} \yr{2019} {\em Free-surface flow: environmental fluid mechanics\/}.  \publ{Oxford [England] ; Cambridge, MA: Butterworth-Heinemann}.

\bibitem[Kim(2012)]{kim_phase-field_2012}
{\sc \au{Kim, Junseok}} \yr{2012}  \at{Phase-{Field} {Models} for {Multi}-{Component} {Fluid} {Flows}}.  \jt{Commun. comput. phys.}  \bvol{12}~(3),  \pg{613--661}.

\bibitem[Kruyt {\em et~al.\/}(1988)Kruyt, Cuvelier, Segal \& Van Der~Zanden]{kruyt_total_1988}
{\sc \au{Kruyt, N.~P.}, \au{Cuvelier, C.}, \au{Segal, A.} \& \au{Van Der~Zanden, J.}} \yr{1988}  \at{A total linearization method for solving viscous free boundary flow problems by the finite element method}.  \jt{Int. J. Numer. Meth. Fluids}  \bvol{8}~(3),  \pg{351--363}.

\bibitem[Kurz {\em et~al.\/}(2023)Kurz, Fisher \& Rappaz]{kurz_fundamentals_2023}
{\sc \au{Kurz, Wilfried}, \au{Fisher, David} \& \au{Rappaz, Michel}} \yr{2023} {\em Fundamentals of {Solidification}\/}, 5th edn.  \publ{Baech: Trans Tech Publications Ltd}.

\bibitem[LaCamera {\em et~al.\/}(1992)LaCamera, Ziegler \& Kozarek]{lacamera1992magnetohydrodynamics}
{\sc \au{LaCamera, AF}, \au{Ziegler, DP} \& \au{Kozarek, RL}} \yr{1992} {\em Magnetohydrodynamics in the Hall-H{\'e}roult process, an overview\/}.  \publ{TMS Seattle}.

\bibitem[Lake(1988)]{osti_5112525}
{\sc \au{Lake, L~W}} \yr{1988} {\em Enhanced oil recovery\/}.  \publ{Old Tappan, NJ; Prentice Hall Inc.}

\bibitem[Logg \& Wells(2010)]{logg_dolfin_2010}
{\sc \au{Logg, Anders} \& \au{Wells, Garth~N.}} \yr{2010}  \at{{DOLFIN}: {Automated} finite element computing}.  \jt{ACM Trans. Math. Softw.}  \bvol{37}~(2),  \pg{1--28}.

\bibitem[Lucy(1977)]{lucy_numerical_1977}
{\sc \au{Lucy, L.~B.}} \yr{1977}  \at{A numerical approach to the testing of the fission hypothesis}.  \jt{The Astronomical Journal}  \bvol{82},  \pg{1013}.

\bibitem[Mohan \& Tomar(2024)]{mohan_volume_2024}
{\sc \au{Mohan, Ananthan} \& \au{Tomar, Gaurav}} \yr{2024}  \at{Volume of {Fluid} {Method}: {A} {Brief} {Review}}.  \jt{J Indian Inst Sci}  \bvol{104}~(1),  \pg{229--248}.

\bibitem[Monaghan(2005)]{monaghan_smoothed_2005}
{\sc \au{Monaghan, J~J}} \yr{2005}  \at{Smoothed particle hydrodynamics}.  \jt{Rep. Prog. Phys.}  \bvol{68}~(8),  \pg{1703--1759}.

\bibitem[Osher \& Sethian(1988)]{osher_fronts_1988}
{\sc \au{Osher, Stanley} \& \au{Sethian, James~A}} \yr{1988}  \at{Fronts propagating with curvature-dependent speed: {Algorithms} based on {Hamilton}-{Jacobi} formulations}.  \jt{Journal of Computational Physics}  \bvol{79}~(1),  \pg{12--49}.

\bibitem[Peskir \& Shiryaev(2006)]{sirjaev_optimal_2006}
{\sc \au{Peskir} \& \au{Shiryaev}} \yr{2006} {\em Optimal {Stopping} and {Free}-{Boundary} {Problems}\/}. {\em Lectures in {Mathematics}. {ETH} {Zürich} {Ser}\/} 1.  \publ{Basel: Springer Basel AG}.

\bibitem[Picasso(2003)]{picasso_anisotropic_2003}
{\sc \au{Picasso, M.}} \yr{2003}  \at{An {Anisotropic} {Error} {Indicator} {Based} on {Zienkiewicz}--{Zhu} {Error} {Estimator}: {Application} to {Elliptic} and {Parabolic} {Problems}}.  \jt{SIAM J. Sci. Comput.}  \bvol{24}~(4),  \pg{1328--1355}.

\bibitem[Popinet(2009)]{popinet_accurate_2009}
{\sc \au{Popinet, Stéphane}} \yr{2009}  \at{An accurate adaptive solver for surface-tension-driven interfacial flows}.  \jt{Journal of Computational Physics}  \bvol{228}~(16),  \pg{5838--5866}.

\bibitem[Rao {\em et~al.\/}(2016)Rao, Schunk, Hariprasad, Ortiz, Tjiptowidjojo \& Secor]{rao_3d_nodate}
{\sc \au{Rao, Rekha~R.}, \au{Schunk, Peter~Randall}, \au{Hariprasad, Daniel}, \au{Ortiz, Weston}, \au{Tjiptowidjojo, Kris} \& \au{Secor, Robert}} \yr{2016} 3d viscoelastic flow with free and moving boundaries using a stabilized finite element method. \url{https://www.osti.gov/biblio/1372203}.

\bibitem[Rapuch(2005)]{rapuch_american_2005}
{\sc \au{Rapuch, Gregory}} \yr{2005}  \at{American options and the free boundary exercise region: a {PDE} approach}.  \jt{Interfaces Free Bound.}  \bvol{7}~(1),  \pg{79--98}.

\bibitem[Rebholz \& Hawkins(2024)]{rebholz_picard-newton_2024}
{\sc \au{Rebholz, L.} \& \au{Hawkins, E.}} \yr{2024} The {Picard}-{Newton} iteration for the {Boussinesq} equations.  \bt{In {\em 16th {World} {Congress} on {Computational} {Mechanics} and 4th {Pan} {American} {Congress} on {Computational} {Mechanics}\/}}.  \publ{CIMNE}.

\bibitem[Reddy \& Tanner(1978)]{reddy_finite_1978}
{\sc \au{Reddy, K.R.} \& \au{Tanner, R.I.}} \yr{1978}  \at{Finite element solution of viscous jet flows with surface tension}.  \jt{Computers \& Fluids}  \bvol{6}~(2),  \pg{83--91}.

\bibitem[Ruschak(1980)]{ruschak_method_1980}
{\sc \au{Ruschak, Kenneth~J.}} \yr{1980}  \at{A method for incorporating free boundaries with surface tension in finite element fluid‐flow simulators}.  \jt{Numerical Meth Engineering}  \bvol{15}~(5),  \pg{639--648}.

\bibitem[Sackinger {\em et~al.\/}(1996)Sackinger, Schunk \& Rao]{sackinger_newtonraphson_1996}
{\sc \au{Sackinger, P.A.}, \au{Schunk, P.R.} \& \au{Rao, R.R.}} \yr{1996}  \at{A {Newton}–{Raphson} {Pseudo}-{Solid} {Domain} {Mapping} {Technique} for {Free} and {Moving} {Boundary} {Problems}: {A} {Finite} {Element} {Implementation}}.  \jt{Journal of Computational Physics}  \bvol{125}~(1),  \pg{83--103}.

\bibitem[Saito \& Scriven(1981)]{saito_study_1981}
{\sc \au{Saito, H.} \& \au{Scriven, L.E.}} \yr{1981}  \at{Study of coating flow by the finite element method}.  \jt{Journal of Computational Physics}  \bvol{42}~(1),  \pg{53--76}.

\bibitem[Scardovelli \& Zaleski(1999)]{scardovelli_direct_1999}
{\sc \au{Scardovelli, Ruben} \& \au{Zaleski, Stéphane}} \yr{1999}  \at{{DIRECT} {NUMERICAL} {SIMULATION} {OF} {FREE}-{SURFACE} {AND} {INTERFACIAL} {FLOW}}.  \jt{Annu. Rev. Fluid Mech.}  \bvol{31}~(1),  \pg{567--603}.

\bibitem[Schunk {\em et~al.\/}(2013)Schunk, Rao, Chen, Labreche, Sun, Hopkins, Moffat, Roach, Hopkins, Notz, Roberts, Sackinger, Subia, Wilkes, Baer, Noble \& Secor]{schunk_goma_2013}
{\sc \au{Schunk, Peter}, \au{Rao, Rekha}, \au{Chen, Ken}, \au{Labreche, Duane}, \au{Sun, Amy}, \au{Hopkins, Matthew}, \au{Moffat, Harry}, \au{Roach, Robert}, \au{Hopkins, Polly}, \au{Notz, Patrick}, \au{Roberts, Scott}, \au{Sackinger, Philip}, \au{Subia, Samuel}, \au{Wilkes, Edward}, \au{Baer, Thomas}, \au{Noble, David} \& \au{Secor, Robert}} \yr{2013}  \bt{{GOMA} 6.0 - {A} {Full}-{Newton} {Finite} {Element} {Program} for {Free} and {Moving} {Boundary} {Problems} with {Coupled} {Fluid}/ {Solid} {Momentum}, {Energy}, {Mass}, and {Chemical} {Species} {Transport}: {User}’s {Guide}}. {\em Tech. Rep.\/} SAND--2013-1844, 1089869, 456348.  \org{Sandia National Laboratories}.

\bibitem[Schunk {\em et~al.\/}(2002)Schunk, Heroux, Rao, Baer, Subia \& Sun]{schunk_iterative_2002}
{\sc \au{Schunk, P~Randall}, \au{Heroux, Michael~A}, \au{Rao, Rekha~R}, \au{Baer, Thomas~A}, \au{Subia, Samuel~R} \& \au{Sun, Amy Cha-Tien}} \yr{2002}  \bt{Iterative {Solvers} and {Preconditioners} for {Fully}-{Coupled} {Finite} {Element} {Formulations} of {Incompressible} {Fluid} {Mechanics} and {Related} {Transport} {Problems}}. {\em Tech. Rep.\/} SAND2001-3512, 793401.  \org{Sandia National Laboratories}.

\bibitem[Sharma(2015)]{sharma_level_2015}
{\sc \au{Sharma, Atul}} \yr{2015}  \at{Level set method for computational multi-fluid dynamics: {A} review on developments, applications and analysis}.  \jt{Sadhana}  \bvol{40}~(3),  \pg{627--652}.

\bibitem[Silliman \& Scriven(1980)]{silliman_separating_1980}
{\sc \au{Silliman, William~J} \& \au{Scriven, L.E}} \yr{1980}  \at{Separating how near a static contact line: {Slip} at a wall and shape of a free surface}.  \jt{Journal of Computational Physics}  \bvol{34}~(3),  \pg{287--313}.

\bibitem[Slikkerveer {\em et~al.\/}(1996)Slikkerveer, Van~Lohuizen \& O'Brien]{slikkerveer_implicit_1996}
{\sc \au{Slikkerveer, P.~J.}, \au{Van~Lohuizen, E.~P.} \& \au{O'Brien, S. B.~G.}} \yr{1996}  \at{{AN} {IMPLICIT} {SURFACE} {TENSION} {ALGORITHM} {FOR} {PICARD} {SOLVERS} {OF} {SURFACE}-{TENSION}-{DOMINATED} {FREE} {AND} {MOVING} {BOUNDARY} {PROBLEMS}}.  \jt{Int. J. Numer. Meth. Fluids}  \bvol{22}~(9),  \pg{851--865}.

\bibitem[Stone {\em et~al.\/}(2004)Stone, Stroock \& Ajdari]{stone_engineering_2004}
{\sc \au{Stone, H.A.}, \au{Stroock, A.D.} \& \au{Ajdari, A.}} \yr{2004}  \at{Engineering {Flows} in {Small} {Devices}: {Microfluidics} {Toward} a {Lab}-on-a-{Chip}}.  \jt{Annu. Rev. Fluid Mech.}  \bvol{36}~(1),  \pg{381--411}.

\bibitem[Taylor \& Hood(1973)]{taylor_numerical_1973}
{\sc \au{Taylor, C.} \& \au{Hood, P.}} \yr{1973}  \at{A numerical solution of the {Navier}-{Stokes} equations using the finite element technique}.  \jt{Computers \& Fluids}  \bvol{1}~(1),  \pg{73--100}.

\bibitem[Tenderini(2024)]{tenderini_reduced_2024}
{\sc \au{Tenderini, Riccardo}} \yr{2024}  \at{Reduced order models to address temporal complexity and geometrical variability in hemodynamics}. PhD thesis, Lausanne, EPFL.

\bibitem[Thompson {\em et~al.\/}(1982)Thompson, Warsi \& Wayne~Mastin]{thompson_boundary-fitted_1982}
{\sc \au{Thompson, Joe~F}, \au{Warsi, Zahir~U.A} \& \au{Wayne~Mastin, C}} \yr{1982}  \at{Boundary-fitted coordinate systems for numerical solution of partial differential equations—{A} review}.  \jt{Journal of Computational Physics}  \bvol{47}~(1),  \pg{1--108}.

\bibitem[Viswanathan {\em et~al.\/}(2008)Viswanathan, Apelian, Donahue, DasGupta, Gywn, Jorstad, Monroe, Sahoo, Prucha \& Twarog]{viswanathan_casting_2008}
{\sc \au{Viswanathan, Srinath}, \au{Apelian, Diran}, \au{Donahue, Raymond~J.}, \au{DasGupta, Babu}, \au{Gywn, Michael}, \au{Jorstad, John~L.}, \au{Monroe, Raymond~W.}, \au{Sahoo, Mahi}, \au{Prucha, Thomas~E.} \& \au{Twarog, Daniel}}, ed. \yr{2008} {\em Casting\/}.  \publ{ASM International}.

\bibitem[Weatherburn(1927)]{weatherburn_differential_1927}
{\sc \au{Weatherburn, C.E.}} \yr{1927} {\em Differential {Geometry} of {Three} {Dimensions}\/}. {\em Differential {Geometry} of {Three} {Dimensions}\/} Bd. 1.  \publ{Cambridge University Press}.

\bibitem[Wijshoff(2018)]{wijshoff_drop_2018}
{\sc \au{Wijshoff, Herman}} \yr{2018}  \at{Drop dynamics in the inkjet printing process}.  \jt{Current Opinion in Colloid \& Interface Science}  \bvol{36},  \pg{20--27}.

\end{thebibliography}



\end{document}